\documentclass[11pt]{article}
\usepackage{amssymb}
\usepackage{amsmath}
\usepackage{epsfig}
\newlength{\bredde}
\def\slash#1{\settowidth{\bredde}{$#1$}\ifmmode\,\raisebox{.15ex}{/}
\hspace*{-\bredde} #1\else$\,\raisebox{.15ex}{/}\hspace*{-\bredde} #1$\fi}
\newcommand{\be}{\begin{equation}}
\newcommand{\ee}{\end{equation}}
\newcommand{\bea}{\begin{eqnarray}}
\newcommand{\eea}{\end{eqnarray}}
\newcommand{\nn}{\nonumber}
\newcommand{\Dirac}{\rlap{\hspace{-.8mm} \slash} D}
\newcommand{\al}{\alpha}

\newcommand{\sig}{\sigma}

\newcommand{\eps}{\epsilon}
\newcommand{\sect}[1]{\setcounter{equation}{0}\section{#1}}

\def\Tr{{\mbox{Tr}}}
\def\re{{\Re\mbox{e}}}
\def\im{{\Im\mbox{m}}}
\def\dc{d{\cal C}}
\def\dd{d{\cal D}}

\begin{document}
\topmargin -1.4cm
\oddsidemargin -0.8cm
\evensidemargin -0.8cm
\title{\Large{{\bf 
The Complex Laguerre Symplectic 
Ensemble\\ of Non-Hermitian Matrices 
}}}

\vspace{1.5cm}

\author{~\\{\sc G. Akemann}\\~\\
Department of Mathematical Sciences\\
School of Information Systems, Computing and Mathematics\\
Brunel University West London\\ 
Uxbridge UB8 3PH, United Kingdom
}
\date{}
\maketitle
\vfill
\begin{abstract}
We solve the complex extension of the chiral Gaussian 
Symplectic Ensemble, defined as a Gaussian two-matrix model of
chiral non-Hermitian quaternion real matrices. This leads to the appearance of
Laguerre polynomials in the complex plane and we prove their orthogonality.
Alternatively, a complex eigenvalue representation of this ensemble 
is given for general weight functions. All $k$-point correlation functions 
of complex eigenvalues are given in terms of the corresponding 
skew orthogonal polynomials in the complex plane for finite-$N$,
where $N$ is the matrix size or number of eigenvalues, respectively.
We also allow for an arbitrary number of complex conjugate pairs of 
characteristic polynomials in the weight function, corresponding to 
massive quark flavours in applications to field theory. Explicit expressions 
are given in the large-$N$ limit at both weak and strong non-Hermiticity for 
the weight of the Gaussian two-matrix model. This model can be mapped to 
the complex Dirac operator spectrum with non-vanishing chemical potential. 
It belongs to the symmetry class of either the adjoint representation or 
two colours in the fundamental representation using staggered lattice
fermions. 

\end{abstract}
\vfill

\thispagestyle{empty}
\newpage

\renewcommand{\thefootnote}{\arabic{footnote}}
\setcounter{footnote}{0}

\sect{Introduction}\label{intro}

The first complex random matrix models date back to Ginibre \cite{Gin}
where the three classical Wigner-Dyson ensembles were generalised. 
In recent years we have seen a revival of complex matrix 
models, due to new applications as well as new insights and
solutions. Todays interest in such models ranges from  Quantum Chromodynamics
(QCD) with chemical potential \cite{Steph}, 
fractional Quantum-Hall effect \cite{PdF}, Coulomb plasma \cite{FJ} and  
growth processes \cite{Zabrodin} to
scattering in Quantum chaos reviewed in \cite{FS}. New insights include
the discovery of the regime of weak non-Hermiticity \cite{FKS}, interpolating
between correlations of real eigenvalues of the Wigner-Dyson ensembles 
and those correlations 
computed by Ginibre. Furthermore, we know that there are many more
complex matrix model symmetry classes than
those with real eigenvalues \cite{Denis}.  

The aim of this work is to solve such a new complex matrix 
model, by generalising a chiral version of one of the Wigner-Dyson ensembles.
Although this is
motivated by a specific application to QCD we hope that our result will be
useful in other areas as well, because of the wide range of
applicability of matrix models.
A major problem in QCD is that the Dirac operator becomes
non-Hermitian when introducing a chemical potential for the quarks. In general
this spoils the positivity of the action and thus its interpretation as a
Boltzmann weight when performing numerical simulations of QCD on an Euclidean
space-time lattice. 

A way out of this dilemma is to study properties of field theories
similar to QCD, where Dirac eigenvalues come in complex conjugate pairs. 
There are two symmetry classes 
different from QCD, with the Dirac operator having an
orthogonal or symplectic symmetry compared to unitary. This feature holds at
both zero \cite{Jac3fold} and non-zero chemical potential \cite{HOV}.   
In this work we will construct a complex 
matrix model with symplectic symmetry and solve it analytically. 
It has the virtue to be testable in lattice simulations including dynamical
fermions. These are absent in the so-called quenched approximation,
suppressing the Dirac operator determinants, and such matrix model predictions
\cite{A02,SplitVerb2} have already been
confirmed in quenched lattice simulations of QCD \cite{AW} and the symplectic
symmetry class \cite{ABMLP},
where preliminary results of this work were used.
Let us emphasise that unquenched matrix model predictions for QCD 
with chemical potential do exist 
\cite{James,AOSV}.

The most important ingredient of matrix models is the way they incorporate
the repulsion between eigenvalues or levels. The form of the repulsion term is
dictated by the Jacobian when changing from matrices to eigenvalues. For real
eigenvalues it is given by the absolute value of a Vandermonde determinant 
of the eigenvalues 
to the power of the Dyson index $\beta=1,2$, or $4$, labelling the number of
independent matrix elements in the Orthogonal, Unitary or Symplectic Ensembles 
respectively (OE, UE or SE). The same is true for the squared singular values
of the so-called chiral ensembles (chOE, chUE or chSE) which have an additional
block structure. When employing the method of orthogonal polynomials 
\cite{Mehta} all
eigenvalue correlations of the corresponding Gaussian ensembles (GUE etc.) 
can be obtained using
Hermite polynomials, or Laguerre polynomials in the chiral classes.

What is known about the level repulsion of complex eigenvalues and the
corresponding correlations functions? For the complex GUE one obtains 
again the
modulus squared of the Vandermonde determinant 
of complex eigenvalues \cite{Gin}. For the
corresponding chiral ensemble it was conjectured in \cite{A02} to have  
the same Jacobian of squared complex eigenvalues, which was later obtained from
a two-matrix model realisation \cite{James}\footnote{
Chiral complex models with a single matrix were first defined in
\cite{Steph,HOV} 
without having an eigenvalue representation.}.
While for $\beta=4$ a Vandermonde to the power 4 can be constructed using
normal matrices \cite{H2,Zabrodin}, 
the Jacobian of the complex GSE of generic non-Hermitian quaternion real
matrices is more complicated \cite{Gin}. It shows an additional 
level repulsion from the real line as pointed out in \cite{EK} (see also
\cite{KE}), where 
this ensemble was solved for an arbitrary weight function. 
A similar feature was observed numerically in \cite{HOV}, 
where the level repulsion for all three 
complex chiral ensembles were compared. 

Our task here is to derive the corresponding Jacobian for the
complex chGSE using a two-matrix model of  non-Hermitian quaternion real
matrices, in analogy to the approach \cite{James}. 
Because of the QCD application we allow for exact
zero-eigenvalues, using rectangular matrices.
The result for the Jacobian turns out to be again 
that of the non-chiral complex GSE of squared variables (generalised to
rectangular matrices), hence having an
additional level repulsion from the imaginary axis.

We solve our model introducing orthogonal polynomials in the complex
plane, where we encounter complex Laguerre instead of Hermite polynomials
\cite{PdF,FKS98,EK} in non-chiral ensembles. 
Other techniques include supersymmetry,
e.g. in \cite{Efetov,FKS98,KE}, the replica method \cite{EK2,SplitVerb2} or a
fermionic mapping \cite{Hastings} which we do not discuss here.
Our computation completes the task of solving the complex extensions of the 
chiral and non-chiral Wigner-Dyson $\beta=2$ and 4 ensembles. 
The difficulty with the remaining $\beta=1$ ensembles is that they have a
finite fraction of real eigenvalues, with the corresponding
probabilities to be determined. 
Only the spectral density at weak non-Hermiticity of the complex GOE is know
so far \cite{Efetov}.

The article is organised as follows. In the following section \ref{finiteNev} 
the complex chSE is defined for a general
weight function, and all correlation functions are 
given at finite-$N$ in terms of the corresponding skew orthogonal polynomials.
The subsections \ref{quenchedrho} and \ref{Rmass} distinguish between weights
without and with explicit insertions of characteristic polynomials or mass
terms respectively, where we have to impose a two-fold mass degeneracy.  
In subsection \ref{skewcomplexL} two examples for weight functions 
and their corresponding skew-orthogonal polynomials 
are given, 
a Ginibre type weight and a Gaussian combined with a $K$-Bessel
function. The latter originates from the Gaussian two-matrix representation
of our model discussed in detail 
in section \ref{Dirac}. The orthogonal
polynomials of the $K$-Bessel
function weight are Laguerre polynomials in the complex plane, and
we prove both their orthogonality and skew orthogonality, with the help of
appendix \ref{proofLOP}\footnote{The Gaussian weight first introduced in 
 \cite{A02} for complex Laguerre polynomials $L_N^\nu$ 
agrees only asymptotically or at
 special values of $\nu$ with
 the correct $K$-Bessel weight. The latter was introduced in
 \cite{James} and conjectured to satisfy orthogonality.}.
In section \ref{weak} we take the large-$N$ limit at weak non-Hermiticity 
for the ensemble of complex skew orthogonal Laguerre polynomials. 
Explicit examples for the quenched and massive spectral density 
as well as for massive partition functions are given in subsections 
\ref{ex:m=0density} and \ref{ex:massdensity} .
Section \ref{strong} is devoted to the strong non-Hermiticity limit, 
giving the same set of examples.
Technical details for the derivation of the limiting kernel at strong
non-Hermiticity are collected in the appendix \ref{Appdiff}.
In section \ref{Dirac} we relate the Gaussian two-matrix model
representation of our complex model 
to the Dirac spectrum with chemical potential. We also provide a
matrix representation of characteristic polynomials. The derivation of
the Jacobian is 
deferred to appendix \ref{Jacobian}. 
Our findings are summarised in section \ref{disc}.


\sect{Correlation functions of the complex chSE 
at finite-$N$}\label{finiteNev}

In this section we define the complex extension of the chSE 
for a general class of weight functions. 
The diagonalisation of the 
matrix model representation in terms of two non-Hermitian quaternion real 
matrices is discussed in more detail in 
section \ref{Dirac} and appendix \ref{Jacobian}. 
This Gaussian matrix model leads to a specific weight 
function containing $K$-Bessel functions.
The reason is the nontrivial decoupling of eigenvalues
and eigenvectors which is not present for ensembles with real eigenvalues. 
The two-matrix model representation 
determines the Jacobian we write down below as the 
interaction term for the eigenvalues.

For the calculation of the eigenvalue correlation functions 
in terms of skew orthogonal
polynomials we proceed in two steps.  
We first treat the so-called quenched case where no explicit 
characteristic polynomials (determinants) are present in the weight function. 
Although the expressions we give for the complex eigenvalue 
correlations in the first step, eq. (\ref{detformula}) below,  
are true for general weight functions, it is easier to include determinants as
mass terms in a second step. Namely we can re-express the corresponding 
massive correlation functions in terms of the quenched ones, following the
same logic as in \cite{AK,A01}.

\subsection{Complex eigenvalue correlation functions}\label{quenchedrho}

The matrix model partition function of the complex Gaussian chSE 
is defined as the following matrix integral 
\be
{\cal Z}_N
\equiv
 \int d\Phi  d\Psi 
\exp\left[-N\Tr (\Phi^\dag \Phi\ +\ \Psi^\dag\Psi)\right] \ ,
\label{Zmatrix}
\ee
where $\Phi$ and $\Psi$ are two matrices with quaternion real elements of
rectangular size $(N+\nu)\times N$ without further symmetry properties.
In order to transform to eigenvalues we parametrise the following linear
combinations as 
\bea
C &\equiv& i \Phi\ + \mu \Psi \ \ =\ U(X+R)V\nn\\
D &\equiv& i \Phi^{\dagger} + \mu \Psi^{\dagger} \ =\ V^\dag(Y+S)U^\dag\ ,
\eea
being equivalent to independent Schur decompositions of $CD$ and
$DC$. $U$ and $V$ are symplectic matrices and the diagonal matrices $X$ and 
$Y$ contain the complex conjugate eigenvalue pairs  $(x_k,x_k^*)$ and
$(y_k,y_k^*)$, respectively. After integrating out the second set we obtain  
in terms of the variables 
\be
z_k^2\ \equiv\ -x_k y_k 
\ee
the following expression for the partition function
\be
{\cal Z}_N\ \equiv\  
\int \prod_{j=1}^N d^2z_j\ w(z_j,z_j^\ast)
\prod_{k>l}^N |z_k^2-z_l^2|^2\ |z_k^2-z_l^{\ast\,2}|^2
\prod_{h=1}^N |z_h^2-z_h^{\ast\,2}|^2\ .
\label{Zev}
\ee
For more details we refer to section \ref{Dirac} and appendix \ref{Jacobian}
for the Jacobian. 
The integrals in $z_k\in C$ extend over the whole complex plane 
with $d^2z\equiv d\re(z)d\im(z)$. Here 
$w(z,z^\ast)$  denotes a {\it general} 
positive definite weight function defined 
in complex plane. It depends on both $z$ and $z^\ast$. 
The only restriction 
we want to impose here is the existence of all positive moments, 
$\int d^2z\ w(z,z^\ast)\, z^{2k}<\infty$, $k\in$ N , 
and similarly for the complex conjugated ones. 
For the Gaussian matrix model eq. (\ref{Zmatrix}) the weight is given by (see
section \ref{Dirac})
\be
w_K^{(2\nu)}(z,z^\ast) \ =\  |z|^{4\nu+2} 
K_{2\nu}\left(\frac{N(1+\mu^2)|z|^2}{2\mu^2}\right)
\exp\left[\frac{N(1-\mu^2)}{4\mu^2}(z^2+z^{*\,2})\right]
\ ,\ \ \mu\in\,(0,1]\ ,
\ee
a second example is given below 
in eq. (\ref{Gweight}) (see also \cite{AV03}).
The general weight $w(z,z^\ast)$ may depend on extra parameters such as 
the non-Hermiticity parameter $\mu>0$, or masses $m_f$ 
as we will discuss later. The former will allow us   
to take limits to either real eigenvalues by letting $\mu\to0$, 
or to maximal non-Hermiticity at $\mu\to 1$, given for example 
by the chiral Ginibre type weight 
$~|z|\exp[-zz^\ast]$. 
Eq. (\ref{Zev}) can  be derived in a different way, inserting
squared arguments into the Jacobian of the non-chiral, complex SE \cite{Gin}.
This is 
in analogy to the construction of the chiral from the non-chiral complex UE
\cite{A02}.

Below we compute all eigenvalue correlations in terms of skew orthogonal 
polynomials in the complex plane. This  
generalises the construction \cite{EK} using 
complex skew orthogonal polynomials for the non-chiral complex SE. There,
complex Hermite polynomials were used for 
the special case of a Gaussian potential. In our chiral model 
with the special weight of the Gaussian 
two-matrix model we encounter complex Laguerre polynomials, 
as expected from the complex chUE (see \cite{A02,James}).

For a general weight function the 
$k$-point correlation functions of complex eigenvalues 
can be defined in the standard way \cite{Mehta}
\be
R_N(z_1,\ldots,z_k)\ \equiv\ \frac{N!}{(N-k)!} 
\frac{1}{{\cal Z}_N}
\int\prod_{j=k+1}^N  d^2z_j
\prod_{i=1}^N w(z_i,z_i^{\ast})
\prod_{n>l}^N |z_n^2-z_l^2|^2|z_n^2-z_l^{\ast\,2}|^2
\prod_{h=1}^N |z_h^2-z_h^{\ast\,2}|^2\ .
\label{defR_k}
\ee
Being real positive functions 
they depend also on the complex conjugated variables $z_1^*,\ldots,z_k^*$.
We suppress this dependence in the following for simplicity.  
In analogy to the non-chiral model \cite{EK} the eigenvalue correlations 
show a depletion of the eigenvalues from the real axis, being
identically zero there. 
This property was found also numerically for a chiral
complex one-matrix model with chemical potential \cite{HOV}. 
Since our Jacobian or interaction term 
is given in terms of squared variables compared to \cite{EK},
there is the same depletion also from the imaginary axis, due to 
$z_h^2-z_h^{\ast\,2}=4i\re(z_h)\im(z_h)$. Such a depletion has been 
already detected in lattice data for the corresponding symmetric class 
\cite{ABMLP}.

Following \cite{Mehta,EK} all $k$-point correlation functions are determined 
through a quaternion determinant \cite{Mehta} of a matrix valued kernel,
\be
R_N(z_1,\ldots,z_k)\ =\ \mbox{Qdet}_{i,j=1,\ldots,k}
\left[{\cal K}_N(z_i,z_j)\right] \ ,
\label{detformula}
\ee
where the kernel is given by
\be
{\cal K}_N(z_1,z_2)\ \equiv\ 
\left[(z_1^{\ast\,2}-z_1^2)(z_2^{\ast\,2}-z_2^2)
w(z_1,z_1^\ast)w(z_2,z_2^{\ast})
\right]^{\frac12}
\left(
\begin{array}{ll}
\kappa_N(z_1^\ast,z_2) &-\kappa_N(z_1^\ast,z_2^\ast)\\
\kappa_N(z_1,z_2)      &-\kappa_N(z_1,z_2^\ast)
\end{array}
\right).
\label{kernel}
\ee
Here we have introduced the pre-kernel $\kappa_N(z_1,z_2^\ast)$,
\footnote{For simplicity we do not give the
more general expression of \cite{EK} here.} 
\be
\kappa_N(z_1,z_2^\ast)\ \equiv\  
\sum_{k=0}^{N-1} \frac{1}{r_k} \left(q_{2k+1}(z_1)q_{2k}(z_2^\ast)-
q_{2k+1}(z_2^\ast)q_{2k}(z_1)\right)  \ .
\label{prekernel}
\ee
Examples for such correlation functions are the spectral density given by 
\be
R_N(z)\ =\ (z^{\ast\,2}-z^2)\ w(z,z^\ast)\ \kappa_N(z,z^\ast)\ ,
\label{rho}
\ee
and the two-point function
\bea
R_N(z,u) &=& (z^{\ast\,2}-z^2)(u^{\ast\,2}-u^2)\,w(z,z^\ast)w(u,u^\ast)\nn\\
&&\times\left( \kappa_N(z,z^\ast) \kappa_N(u,u^\ast) - |\kappa_N(z,u)|^2 
+|\kappa_N(z,u^\ast)|^2 \right)\ .
\label{2point}
\eea
The pre-kernel eq. (\ref{prekernel}) is expressed in terms of skew-orthogonal 
polynomials $q_k(z)$.  
These are polynomials of degree $k$ in the complex variable $z\in C$ squared,
$q_k(z)=z^{2k}+{\cal O}(z^{2k-2})$, where we have chosen the monic
normalisation (because of the form of eq. (\ref{Zev}) we only need
polynomials of squared arguments, see also \cite{A02}).
The skew-orthogonal polynomials 
are defined with respect to the following antisymmetric scalar product
\be
\langle h|g\rangle_S\ \equiv\ \int d^2z\ w(z,z^\ast)
(z^{\ast\,2}-z^2)(h(z)g(z)^\ast-h(z)^\ast g(z)),
\label{scalar}
\ee
by obeying
\bea
\langle q_{2k+1}|q_{2l}\rangle_S &=& -\langle q_{2l}|q_{2k+1}\rangle_S
\ =\ r_k\ \delta_{kl}\nn\ ,\\
\langle q_{2k+1}|q_{2l+1}\rangle_S &=& \, \ \ \langle q_{2l}|q_{2k}\rangle_S \
\ \ \ =\ 0\ .
\label{skewdef}
\eea
For a general weight they enjoy the following representation 
\bea
q_{2N}(z)&=& \left\langle \prod_{l=1}^N(z^2-z_l^2)(z^{2}-z_l^{*\,2})
\right\rangle_{{\cal Z}_N}\ ,
\label{qevenvev}\\
q_{2N+1}(z)&=& 
\left\langle \prod_{l=1}^N(z^2-z_l^2)(z^{2}-z_l^{*\,2})
(\ z^2+\sum_{k=1}^N(z_k^2+z_k^{*\,2})+ c_N\ )
\right\rangle_{{\cal Z}_N}\ ,
\label{qoddvev}
\eea
which holds in complete analogy to \cite{EK}.
Here  $c_N$ is an arbitrary constant that can be set to zero. Due to this
representation we can observe that the polynomials have real coefficients,
as $q_{2N}(z)^*=q_{2N}(z^*)$ and $q_{2N+1}(z)^*=q_{2N+1}(z^*)$.
A similar representation of the skew orthogonal polynomials also holds 
for the SE with real eigenvalues \cite{Bert}.
In section \ref{Dirac} we will 
also give a matrix representation of the $q_{2N}(z)$ 
in terms of a characteristic polynomial.
The expectation values are to be taken with respect to the partition function 
eq. (\ref{Zev}) (not to be confused with the antisymmetric scalar product).
The skew orthogonal polynomials $q_{2N}(z)$ and  $q_{2N+1}(z)$ 
both depend only on the variable $z$.
While eq. (\ref{detformula}) for the $k$-point correlation functions 
holds for general weight functions 
it is convenient to explicitly split off determinants or mass terms from the
weight. We shall do this in the next subsection, allowing us to compute
characteristic polynomials more general than in the example
eq. (\ref{qevenvev}) above.

\subsection{Massive complex correlation functions}
\label{Rmass}

The massive matrix model partition function of the complex chSE 
we wish to compute is given by 
\bea
{\cal Z}^{(4N_f)}_N(\{m_f\}) &\equiv&  
\int\prod_{j=1}^N d^2z_j\ w(z_j,z_j^*) 
\prod_{f=1}^{N_f} |m_f|^{4\nu} 
|z_j^2 + m_f^2|^2  |z_j^{2} + m_f^{\ast\,2}|^2 
\nn\\
&&\times\prod_{k>l}^N |z_k^2-z_l^2|^2\ |z_k^2-z_l^{\ast\,2}|^2
\prod_{h=1}^N |z_h^2-z_h^{\ast\,2}|^2\ ,
\label{Zevm}
\eea
where we allow for complex masses $m_f^2\neq m_f^{*\,2}$ 
is this subsection. 
Compared to the partition function eq. (\ref{Zev}) we have explicitly
introduced mass terms in complex conjugated pairs as extra sources.
The new weight function including the masses is still
positive definite\footnote{This is no longer true for mass terms in the 
complex chUE corresponding the the QCD symmetry class.}.
The mass terms can also be 
written as a determinant of quaternion real non-Hermitian matrices as we will
show in section \ref{Dirac}. Therefore these partition functions can 
be interpreted as characteristic polynomials, and we 
will give explicit formulas for them below.

In the presence of a non-Hermiticity parameter $\mu$
we can make contact with the
chSE by taking the Hermitian limit $\mu\to0$. 
This leads to 4-fold degenerate (real) masses, as indicated by 
the superscript\footnote{Due to Kramer's degeneracy the mass terms 
automatically come in doubly degenerate pairs for the symplectic 
ensemble, see also sect. \ref{Dirac}.}.  
Our aim is to express the partition function as well as all 
complex eigenvalue correlation functions in the 
presence of mass terms through the eigenvalue correlations without 
mass terms at $N_f=0$. For that reason we need 
to impose such a fourfold mass degeneracy as in \cite{AK}. There, 
massive correlation functions were expressed through quenched ones 
for Hermitian self-dual matrices \cite{AK}, 
and we will follow the same strategy here. 
Similar results were also obtained for complex non-Hermitian matrices in the 
non-chiral case $\beta=2$ \cite{A01}, 
where only a twofold degeneracy is needed due to 
the different Jacobian $\prod_{k>l}|z_k^2-z_l^2|^2$.

To start we compute the massive partition function eq. (\ref{Zevm}) 
in terms of quenched correlation functions. 
When writing the mass terms as $z_j^2 + m_f^2=z_j^2 - (im_f)^2$
they can be 
absorbed into a larger Jacobian of $N+N_f$ eigenvalues 
$z_1,\ldots,z_N,im_1,\ldots, im_{N_f}$, after multiplying with constant
factors as 
$\prod_{f=1}^{N_f} |m_f^2-m_f^{\ast\,2}|^2\neq0$. 
The massive partition function is thus proportional 
to a quenched $N_f$-point correlation function at $i$ times the 
complex masses. Filling 
in all explicit factors from the definition eq. (\ref{defR_k}) we thus obtain
\bea
{\cal Z}^{(4N_f)}_N(\{m_f\}) &=& \frac{N!}{(N+N_f)!} 
\frac{\prod_{f=1}^{N_f}|m_f|^{4\nu}
{\cal Z}_{N+N_f}^{(0)}R_{N+N_f}^{(0)}(im_1,\ldots,im_{N_f})
}{\prod_{f=1}^{N_f}w(im_f,-im_f^*)|m_f^2-m_f^{\ast\,2}|^2\ 
\prod_{k>l}^{N_f}|m_k^2-m_l^2|^2|m_k^2-m_l^{\ast\,2}|^2}\ 
\nn\\
&=& \frac{N!\ {\cal Z}_{N+N_f}^{(0)}
\prod_{f=1}^{N_f}|m_f|^{4\nu} }{(N+N_f)!} \nn\\
&&\times\frac{
\mbox{Qdet}_{j,k=1,\ldots,N_f}
\left[\left(\!
\begin{array}{ll}
\kappa_N(-im_j^\ast,im_k) &-\kappa_N(-im_j^\ast,-im_k^\ast)\\
\kappa_N(im_j,im_k)      &-\kappa_N(im_j,-im_k^\ast)
\end{array}
\!\right)\right] 
}{\prod_{f=1}^{N_f}(m_f^2-m_f^{\ast\,2})\ 
\prod_{k>l}^{N_f}|m_k^2-m_l^2|^2|m_k^2-m_l^{\ast\,2}|^2}.\nn\\
\label{Zmrho}
\eea
Dividing by ${\cal Z}_{N+N_f}^{(0)}$ and switching back to parameters 
$u_f=im_f$ this gives an explicit expression 
for products of 
$L$ pairs of characteristic polynomials times their 
complex conjugates
\bea
&&\left\langle \prod_{j=1}^N  \prod_{l=1}^L
(z_j^2-u_l^2)(z_j^{*\,2}-u_l^2)(z_j^2-u_l^{*\,2})(z_j^{*\,2}-u_l^{*\,2})
\right\rangle_{{\cal Z}_N} \nn\\
&=& \frac{N!
\prod_{l=1}^{L}|u_l|^{4\nu} }{(N+L)!}
\frac{
\mbox{Qdet}_{l,k=1,\ldots,L}
\left[\left(\!
\begin{array}{ll}
\kappa_N(u_l^\ast,u_k) &-\kappa_N(u_l^\ast,u_k^\ast)\\
\kappa_N(u_l,u_k)      &-\kappa_N(u_l,u_k^\ast)
\end{array}
\!\right)\right] 
}{\prod_{f=1}^{L}(u_f^{*\,2}-u_f^{2})\ 
\prod_{k>l}^{N_f}|u_k^2-u_l^2|^2|u_k^2-u_l^{\ast\,2}|^2}\ .
\label{genchar}
\eea
It extends the corresponding results \cite{AV03} for the corresponding
$\beta=2$ 
class. 
Next we use the same trick to compute the massive eigenvalue correlation 
functions, 
obtaining
\bea
R^{(4N_f)}_N(z_1,\ldots,z_k) &=& \frac{N!}{(N+N_f)!} 
\frac{{\cal Z}_{N+N_f}^{(0)}\prod_{f=1}^{N_f}|m_f|^{4\nu}}
{{\cal Z}_{N}^{(4N_f)}(\{m_f\})}
\\
&\times&\frac{R_{N+N_f}^{(0)}(z_1,\ldots,z_k,im_1,\ldots,im_{N_f})}{
\prod_{f=1}^{N_f}w(im_f,-im_f^*)|m_f^2-m_f^{\ast\,2}|^2\ 
\prod_{h>l}^{N_f}|m_h^2-m_l^2|^2|m_h^2-m_l^{\ast\,2}|^2}\ .\nn
\eea
Inserting the above relation (\ref{Zmrho}) for the massive partition function 
all pre-factors cancel and we obtain the very elegant expression
\be
R^{(4N_f)}_N(z_1,\ldots,z_k) \ = \ 
\frac{R_{N+N_f}^{(0)}(z_1,\ldots,z_k,im_1,\ldots,im_{N_f})}{
R_{N+N_f}^{(0)}(im_1,\ldots,im_{N_f})}\ .
\label{masterrho}
\ee
The same 
relation was obtained previously for real eigenvalues in the chSE \cite{AK} 
and for the non-chiral complex UE \cite{A01}. It is easy to see that it also
holds for the non-chiral complex SE, replacing squared by un-squared variables
everywhere. Together with eqs. (\ref{detformula}) 
and (\ref{kernel}) this gives all eigenvalue correlation functions 
for our massive 
model eq. (\ref{Zevm}). Due to the different numbers of eigenvalues occurring,
$N$ compared to $N+N_f$, these relations are only exact for an $N$-independent 
weight function $w(z,z^*)$. When we consider $N$-dependent 
weights as in the examples eqs. (\ref{Gweight}) and  (\ref{Kweight}) below
our results will be still valid for asymptotically 
large-$N$ as then $N\approx N+N_f$ in the exponent.

We have to add a word of caution to applying eqs. (\ref{masterrho}) and
(\ref{Zmrho}).
While these expressions are true for general 
complex masses $m_f$ the limit of real masses 
has to be taken with care. As we have explained after eq. (\ref{defR_k}) 
our model shows a depletion of eigenvalues both from the real and imaginary 
axis. Therefore both the numerator and the denominator in eq. (\ref{masterrho})
vanish when setting the masses $m_f$ to be real or purely imaginary. 
We thus have to take the limit of the imaginary part of 
$m_f$ going to zero using the rule of de l'H\^ospital.
This leads to a finite
expression for the massive correlation functions. We will 
illustrate this explicitly in examples in sect. \ref{weak} 
and \ref{strong} when taking the large-$N$ limit.

\subsection{Examples for skew orthogonal polynomials in the complex plane} 
\label{skewcomplexL}

In order to illustrate the finite-$N$ expressions for the eigenvalue 
correlation functions 
we give two explicit examples of weight functions together with their 
corresponding skew orthogonal polynomials 
eqs. (\ref{qevenvev}) and (\ref{qoddvev}) as they appear 
inside the pre-kernel. We will use some results for non-skew 
orthogonal polynomials with respect to the same weight functions which are
derived in appendix \ref{proofLOP}.

The first example is from the Gaussian two-matrix model representation of the
complex chSE (see sect. \ref{Dirac}),
\be
w_K^{(2\nu)}(z,z^\ast) \ =\  |z|^{4\nu+2} 
K_{2\nu}\left(\frac{N(1+\mu^2)|z|^2}{2\mu^2}\right)
\exp\left[\frac{N(1-\mu^2)}{4\mu^2}(z^2+z^{*\,2})\right]
\ ,\ \ \mu\in\,(0,1]\ .
\label{Kweight}
\ee
Here $\mu$ is a non-Hermiticity parameter. Compared to the QCD symmetry class
\cite{James} the parameter $\nu$ measuring the rectangularity of the
matrices is shifted, $\nu\to2\nu$, as for the real eigenvalue
model.  
The  real limit 
can be taken by letting $\mu\to0$, leading to a delta-function in the
imaginary part. This will be discussed in more detail in
sect. \ref{weak} below. We only mention here that in this Hermitian limit
the last term of the Jacobian
in eq. (\ref{Zev}) will provide two more powers of the real part, 
$(\re(z))^2$, arriving at the standard power of zero-modes 
$|\re(z)|^{4\nu+3}$ for the real chSE (see eq. (\ref{Zreal})).

Let us consider a special case in which the $K$-Bessel function simplifies.
For indices $\nu=\pm\frac14$ it exactly coincides  \cite{Grad}
with its asymptotic value, 
$K_{2\nu}(x)\sim\mbox{e}^{-x}\sqrt{\frac{\pi}{2x}}$,
and we obtain up to a constant
\be
w_K^{(2\nu=\pm\frac12)}(z,z^\ast) \ \sim\  \left\{
\begin{array}{c}
|z|^2\\
1\\
\end{array} 
\right\}
\exp\left[
\frac{-N(1+\mu^2)}{2\mu^2}\left(|z|^2+\frac12\frac{(1-\mu^2)}{(1+\mu^2)}
(z^2+z^{*\,2})
\right)\right]
\ .
\label{K1/4weight}
\ee
This is of the same form as the weight in the corresponding non-chiral ensemble
\cite{EK}. 
More generally speaking the $K$-Bessel function at half-integer 
index simplifies to an exponential times a finite sum over half-integer powers
of its argument. 
In the non-chiral complex SE \cite{EK}  complex
Hermite polynomials were used to construct the 
skew orthogonal polynomials eq. (\ref{skewdef}) for a Gaussian weight
function as 
eq. (\ref{K1/4weight}). This is consistent with the relation between
odd or even Hermite polynomials and Laguerre polynomials $L_N^{\pm\frac12}$,
respectively (see e.g. \cite{Grad}).

Following the same lines as in  \cite{EK}
we construct our skew orthogonal polynomials
here using complex Laguerre polynomials, which are show to be orthogonal with
respect to the weight eq. (\ref{Kweight}) in appendix \ref{proofLOP}. We
obtain 
\bea
q_{2k+1}^K(z)&=& -(2k+1)!\left(\frac{1-\mu^2}{N}\right)^{2k+1}
L_{2k+1}^{2\nu}\left(\frac{Nz^2}{1-\mu^2}\right)\ ,
\label{qodd}\\
q_{2k}^K(z)&=& k!\
\Gamma\left(k+\nu+1\right)\frac{2^{2k}(1+\mu^2)^{2k}}{N^{2k}} 
\sum_{j=0}^k \frac{(1-\mu^2)^{2j}}{(1+\mu^2)^{2j}}
\frac{(2j)!}{2^{2j}j!\,\Gamma(j+\nu+1)}
L_{2j}^{2\nu}\left(\frac{Nz^2}{1-\mu^2}\right),\nn\\
\label{qeven}
\eea
in monic normalisation $q_{2k+1}(z)=(z^2)^{2k+1}+\ldots$ and 
$q_{2k}(z)=(z^2)^{2k}+\ldots\ \ $ .

The fact that this choice satisfies  eq. 
(\ref{skewdef}) can be seen as follows. In the case where both indices in eq. 
(\ref{skewdef}) are equal, the second line for $k=l$, the scalar product 
vanishes trivially due to antisymmetry. If both indices
are odd we now show that 
the scalar product vanishes because of the orthogonality of
the Laguerre polynomials from appendix \ref{proofLOP}. To see that we use the
three-step recursion relation 
\be
Z^2 L_n^{2\nu}(Z^2)\ =\ -(n+1)L_{n+1}^{2\nu}(Z^2)\ +
\ (2n+2\nu+1)L_n^{2\nu}(Z^2)\ -\ (n+2\nu)L_{n-1}^{2\nu}(Z^2)\ ,
\label{3step}
\ee
in the squared variable 
$Z^2=\frac{Nz^2}{1-\mu^2}$. Consequently the
factor $(z^{\ast\,2}-z^2)$ in the scalar product eq. (\ref{scalar})
can only lower or raise the index of the Laguerre polynomials by one
unit. The terms where indices are raised or lowered vanish from
orthogonality, having different parity. The middle term in the recursion 
eq. (\ref{3step}) only contributes at $k=l$ and the corresponding terms cancel
(as we have seen already from antisymmetry above).
The same logic applies if both indices in eq. (\ref{skewdef}) are even. 
Only those terms in the recursion where the index is unchanged contribute, and
they also cancel after using orthogonality, for all values of $k$ and $l$. 

It remains to analyse the scalar product in  eq. (\ref{scalar}) with one index
even and one odd, $\langle q_{2k+1},q_{2l}\rangle_S$. For $k>l$ the expression
again vanishes trivially due to orthogonality, using eq. (\ref{3step}). 
So far the coefficients in the linear combination of even Laguerre polynomials
in the 
definition (\ref{qeven}) could have been arbitrary. 
It is the requirement $\langle q_{2k+1},q_{2l}\rangle_S=r_k\delta_{kl}$ 
for $k\leq l$ that fixes them uniquely in a given normalisation. 
One can easily check that the coefficients given in the definition
(\ref{qeven}) imply vanishing 
for $k<l$, upon using the recursion and the norms of
the Laguerre polynomials eq. (\ref{Knorm}). 
The overall constants in eqs. 
(\ref{qodd}) and (\ref{qeven}) are fixed by choosing the monic
normalisation. This choice allows us to take the limit $\mu\to1$ of
maximal non-Hermiticity on the polynomials.
The remaining coefficients $r_k^K$ in eq. (\ref{skewdef}) at $k=l$ follow to be
\be
r_k^K
\ =\ 8\pi\mu^4 (2k+1)!\ \Gamma(2k+2\nu+2)
\frac{(1+\mu^2)^{4k+2\nu}}{N^{4k+2\nu+4}}
\ .
\label{rkK}
\ee
This determines all quantities in the pre-kernel eq. (\ref{prekernel})
and thus all $k$-point correlation functions through
eq. (\ref{detformula}).

We now turn to the second example given by a Gaussian, chiral 
Ginibre type weight
\be
w_G^{(2\nu)}(z,z^\ast) \ =\  |z|^{4\nu+1}\! 
\exp\left[-N|z|^2\right]\ .
\label{Gweight}
\ee
In can be obtained from the above weight eq. (\ref{Kweight}) at maximal
non-Hermiticity $\mu=1$, when taking the asymptotic 
limit of large arguments (or at $\nu=\pm\frac14$). However, at
small arguments they differ, and so far no matrix representation is known for
the weight  eq. (\ref{Gweight}). As for all
rotational invariant measures the associated orthogonal polynomials are the
monomials $z^{k}$ and their complex conjugates, 
from which we can construct the skew orthogonal polynomials. 
We obtain the following result
\bea
q_{2k+1}^G(z)&=& (z^2)^{2k+1}\ ,
\label{qoddG}\\
q_{2k}^G(z)&=& \frac{2^{4k}}{N^{2k}}
\Gamma\!\left(k+\frac{\nu}{2}+\frac98\right)
\Gamma\!\left(k+\frac{\nu}{2}+\frac78\right) 
\sum_{j=0}^k \frac{N^{2j}}{2^{4j}}
\frac{(z^2)^{2j}}{\Gamma\left(j+\frac{\nu}{2}+\frac98\right)
  \Gamma\left(j+\frac{\nu}{2}+\frac78\right) }\ , \nn\\
\label{qevenG}
\eea
in monic normalisation. 
The skew orthogonality eq. (\ref{skewdef}) can be checked along the same lines
as described above for the Laguerre polynomials, 
with a much simpler recursion of course. We only give the
final result for the coefficients $r_k^G$, 
using the normalisation of the orthogonal
polynomials computed in the appendix \ref{proofLOP} eq. (\ref{Gnorm}),
\be
r_k^G
\ =\ 2\pi \frac{\Gamma(4k+2\nu+\frac72)}{N^{4k+2\nu+\frac72}}
\ .
\label{rkG}
\ee

Let us also give an example for the simplest 
correlation function at finite-$N$, the spectral density. 
Taking the simplest case without determinants and 
the weight eq. (\ref{Kweight}) of the
two-matrix model we can explicitly write
out eq. (\ref{rho}):
\bea
R_{N}(z) &=& (z^{\ast\,2}-z^2)|z|^{4\nu+2} 
K_{2\nu}\left(\frac{N(1+\mu^2)|z|^2}{2\mu^2}\right)
\exp\left[\frac{N(1-\mu^2)}{4\mu^2}(z^2+z^{*\,2})\right]
\label{rhoKN}\\
&&\times \frac{N^{2\nu+3}}{8\pi\mu^4}
\sum_{k=0}^{N-1}\sum_{j=0}^k
\frac{(1-\mu^2)^{2k+2j+1}}{(1+\mu^2)^{2k+2j+2\nu}}
\frac{2^{2k-2j}k!\ \Gamma(k+\nu+1)(2j)!}
{\Gamma(2k+2\nu+2)j!\ \Gamma(j+\nu+1)}
\nn\\
&& \ \ \ \ \ \ \ \ \ \ \ \ \ \ \ \ \ \ \ \ 
\times\left( 
L_{2k+1}^{2\nu}\left(\frac{Nz^{\ast\,2}}{1-\mu^2}\right)
L_{2j}^{2\nu}\left(\frac{Nz^{2}}{1-\mu^2}\right)
\ -\ 
L_{2k+1}^{2\nu}\left(\frac{Nz^{2}}{1-\mu^2}\right)
L_{2j}^{2\nu}\left(\frac{Nz^{\ast\,2}}{1-\mu^2}\right)\!
\right).
\nn
\eea
It is manifestly real and vanishes along the real and imaginary axis.


\sect{The large-$N$ limit at weak non-Hermiticity}\label{weak}

The limit of weak non-Hermiticity was first introduced in \cite{FKS} 
for unitary invariant complex non-Hermitian matrices. 
It is defined by simultaneously 
taking the Hermitian limit $\mu\to0$
and $N\to\infty$ such that the 
following combination
\be
\lim_{N\to\infty,\ \mu\to0}2N\mu^2\ \equiv \ \al^2  
\label{aldef} 
\ee
is kept constant. In this limit the macroscopic spectral density collapses 
onto the real axis, while the microscopic correlation functions still 
extend into the complex plane, on a stripe of the order of $\alpha^2$. 
This limit allows us to extrapolate between
the correlation functions of real eigenvalues, by taking $\al\to0$, and 
those of complex eigenvalues at strong non-Hermiticity to be introduced 
in the next section \ref{strong}, by taking $\al\to\infty$. 
We define the following rescaling of the complex eigenvalues at the origin
\be
\sqrt{2}\ N(\re\ z+i\im\ z)\ =\ \sqrt{2}\ Nz \ \equiv \xi \ \ \ ,
\label{microweak}
\ee
where we take over the convention from $\mu=0$. There eigenvalues are rescaled 
as $V\Sigma x=\xi$ where $2N=V$ is the volume, and the inverse variance or 
chiral condensate proportional to the macroscopic density at the origin 
is given by $\Sigma=1/\sqrt{2}$ in the matrix model
eq. (\ref{Z2MM}) at $\mu=0$. 
The pre-kernel and the resulting microscopic correlation functions 
have to be rescaled accordingly:
\bea
\kappa_{weak}(\xi_1,\xi_2^\ast) &\equiv&  
\lim_{N\to\infty,\ \mu\to0}\frac{1}{2N^2}\ 
\kappa_N\left(\frac{\xi_1}{\sqrt{2}\ N},\frac{\xi_2^\ast}{\sqrt{2}\
  N}\right)\ , \nn\\ 
\rho_{weak}(\xi_1,\ldots,\xi_k)&\equiv&  \lim_{N\to\infty,\ \mu\to0}
\frac{1}{2^kN^{2k}}\ 
R_N\left(\frac{\xi_1}{\sqrt{2}\ N},\ldots,\frac{\xi_k}{\sqrt{2}\ N}\right).
\label{microrhow}
\eea
In the following we derive these quantities explicitly for the weight eq. 
(\ref{Kweight}).

We start with the large-$N$ weak non-Hermiticity limit 
of the skew orthogonal 
Laguerre polynomials eqs. (\ref{qodd}) and (\ref{qeven}). Here, we can use 
some of the results from \cite{A02}, which we repeat for 
completeness. 
From the asymptotic of the Laguerre polynomials, 
\be
\lim_{k\to\infty} k^{-2\nu} L_k^{2\nu}(Z^2) \ =\ 
(kZ^2)^{-\nu} J_{2\nu}\left(2\sqrt{kZ^2}\right)\ ,
\label{Lasymp}
\ee
we obtain for the odd Laguerre polynomials of
eq. (\ref{qodd})
\be
\lim_{k,N\to\infty,\,\mu\to0} 
L^{2\nu}_{2k+1}\left(\frac{Nz^2}{1-\mu^2}\right) \ =\ 
N^{2\nu} 2^{2\nu} s^\nu \xi^{-2\nu} J_{2\nu}\left(2\sqrt{s}\ \xi\right) 
\ ,\ \ \frac{2k+1}{N}\equiv 2s\ .
\label{qoddasymp}
\ee
Here we have introduced the scaling variable $s\in\ [0,1]$.
For the even polynomials eq. (\ref{qeven}) we have to replace the sum by an 
integral, $\sum_{j=0}^k\to k\int_0^1 dt$, introducing a second 
scaling variable $t=j/k\in\ [0,1]$.  
In terms of the integration variable $t$ 
we obtain for the even Laguerre polynomial inside the integrand
\be
\lim_{j,k,N\to\infty,\ \mu\to0}L_{2j}^{2\nu}\left(\frac{Nz^2}{1-\mu^2}\right) 
\ =\ 
N^{2\nu}   2^{2\nu} (st)^\nu \xi^{-2\nu}
J_{2\nu}\left(2\sqrt{st}\ \xi\right)\ , \ \ \frac{j}{k}\equiv t\ ,
\label{Lasymp2}
\ee
using $2jNz^2\to ts\xi^2  $. 
The normalisation constant is mapped to 
\be
\lim_{j,k,N\to\infty}\ \frac{(2j)!}{2^{2j}j!\,\Gamma(j+\nu+1)}\ =\ 
\frac{(stN)^{-\nu-\frac12}}{\sqrt{\pi }}\ .
\ee
The resulting expression for the sum over Laguerre polynomials
eq. (\ref{qeven}) is then
\be
\lim_{k,N\to\infty,\ \mu\to0} \sum_{j=0}^k 
\left(\frac{1-\mu^2}{1+\mu^2}\right)^{2j}
\!\!\frac{(2j)!}{2^{2j}j!\,\Gamma(j+\nu+1)}
L_{2j}^{2\nu}\left(\frac{Nz^2}{1-\mu^2}\right)=
\frac{N^{\nu+\frac12}}{\sqrt{\pi}}\frac{2^{2\nu}}{\xi^{2\nu}}
\int_0^1\!dt \sqrt{\frac{s}{t}}\ \mbox{e}^{-2st\al^2}
J_{2\nu}(2\sqrt{st}\xi) .
\label{Lasymp3}
\ee
Here we have used the fact that the $\mu$-dependent pre-factors 
turn into an exponential, 
\be
\lim_{N\to\infty,\ \mu\to0}
\left((1\pm\mu^2)^{2j} \ =\ \mbox{e}^{2j\ln(1\pm\al^2/2N)}
\right)
\ =\ \mbox{e}^{\pm st\al^2} \ ,
\label{muasymp}
\ee
and similarly for powers of $k$. In order to obtain the pre-kernel
eq. (\ref{prekernel}) 
it remains to evaluate the $k$-dependent pre-factors of both skew orthogonal
polynomials eq. (\ref{qeven}) and (\ref{qodd}),
divided by the norm $r_k^K$, 
in the large-$N$ limit,
\be
\lim_{k,N\to\infty} -\frac{1}{r_k^K}(2k+1)! \frac{(1-\mu^2)^{2k+1}}{N^{2k+1}}
k!\ \Gamma(k+\nu+1) \frac{2^{2k}(1+\mu^2)^{2k}}{N^{2k}}
\ =\ -\frac{\sqrt{\pi}\ 
N^{\nu+4+\frac12}}{4\alpha^42^{2\nu} s^{\nu+\frac12}}
\mbox{e}^{-2s\al^2} .
\label{prefactor}
\ee
Having collected the asymptotic of all building blocks we can write down the 
limiting expression for the pre-kernel eq. (\ref{prekernel}) 
\bea
\kappa_{weak}(\xi_1,\xi_2^*)&=& \frac{1}{\alpha^4}
N^{4\nu+4}\frac{2^{2\nu-3}}{(\xi_1\xi_2^*)^{2\nu}}
\label{prekweak}\\
&\times&\int_0^1 ds \int_0^1 \frac{dt}{\sqrt{t}}\mbox{e}^{-2s(1+t)\al^2}
\left(J_{2\nu}(2\sqrt{st}\ \xi_1)  J_{2\nu}(2\sqrt{s}\ \xi_2^\ast)
\ -\  J_{2\nu}(2\sqrt{st}\ \xi_2^*)J_{2\nu}(2\sqrt{s}\ \xi_1)\right).
\nn
\eea
The asymptotic of the even and odd skew orthogonal polynomials
eqs. (\ref{qeven}) and  (\ref{qodd}) follow from
setting $s=1$ in eqs. (\ref{Lasymp3}) and (\ref{qoddasymp}) times
e$^{\mp\al^2}$, respectively. The former will
be used later in subsection \ref{ex:massdensity} as it enjoys a direct
interpretation as a characteristic polynomial or massive partition function.

In order to give the spectral density and all higher correlation functions 
we also need the limiting form of the weight eq. (\ref{Kweight}):  
\be
\lim_{N\to\infty,\ \mu\to0} w_K^{(2\nu)}(z,z^\ast)\ =\ 
2^{-2\nu-1}N^{-4\nu-2} |\xi|^{4\nu+2}
K_{2\nu}\left(\frac{|\xi|^2}{2\al^2}\right)
\exp\left[\frac{1}{4\al^2}(\xi^2+\xi^{*\,2})\right]\ .
\label{weakKweight}
\ee
Inserting this together with the limiting form of the pre-kernel eq. 
(\ref{prekweak}) 
into eqs. (\ref{kernel}) and (\ref{detformula}) rescaled according to 
eq. (\ref{microrhow}) we obtain 
all quenched microscopic correlation functions, and through eq. 
(\ref{masterrho}) also all massive microscopic correlations functions.
Since the massive correlation functions at real masses involve taking limits
we give explicit examples in subsection \ref{ex:massdensity} below, after
first discussing the simplest example, the spectral density without masses. 

\subsection{Example: the quenched spectral density} 
\label{ex:m=0density}

Here we give the quenched $(N_f=0)$ 
microscopic spectral density 
for an arbitrary topological charge $\nu$ as well as a comparison to its real
limit $\al\to0$. 
It is given by inserting eqs. (\ref{prekweak}) and (\ref{weakKweight})
into eq. (\ref{rho})
\bea
\rho_{weak}(\xi) &=& \frac{1}{32\al^4}
(\xi^{\ast\,2}-\xi^2)\ |\xi|^2\ 
K_{2\nu}\left(\frac{|\xi|^2}{2\al^2}\right)
\exp\left[+\frac{1}{4\al^2}(\xi^2+\xi^{*\,2})\right]
\label{rhoKweak}\\
&&\times\int_0^1 ds \int_0^1 \frac{dt}{\sqrt{t}}\mbox{e}^{-2s(1+t)\al^2}
\left(J_{2\nu}(2\sqrt{st}\ \xi)J_{2\nu}(2\sqrt{s}\ \xi^\ast)
\ -\ J_{2\nu}(2\sqrt{s}\ \xi)J_{2\nu}(2\sqrt{st}\ \xi^\ast)\right). 
\nn
\eea
We note that in addition to the $\nu$-dependence of the $J$-Bessel functions 
the weight is also explicitly topology dependent. This is in contrast to the 
spectral density of real eigenvalues, see eq. (\ref{rhoreal}) below.
It is only in the limit of small $\al^2$ or large arguments,
$|\xi|^2/(4\al^2)\gg1$, that the K-Bessel function becomes $\nu$-independent,
$K_{2\nu}(x)\sim\ $e$^{-x}\sqrt{\frac{\pi}{2x}}$, leading 
to a Gaussian weight asymptotically (see also eq. (\ref{K1/4weight})).

\begin{figure}[-h]
\centerline{
\epsfig{figure=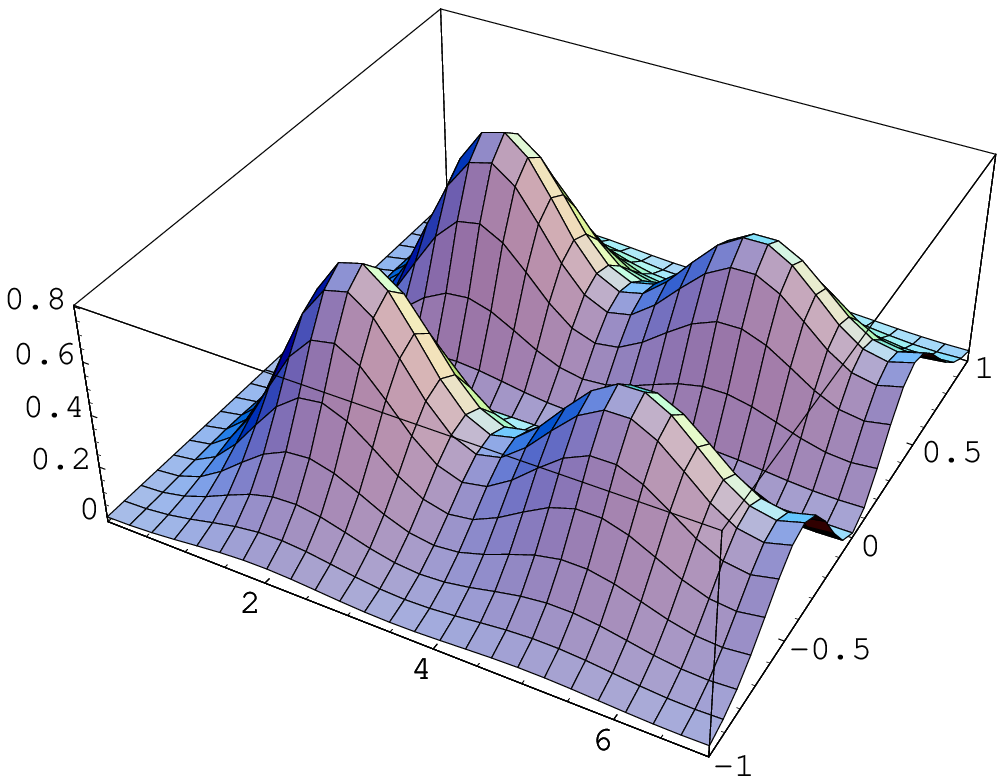,width=20pc}
\epsfig{figure=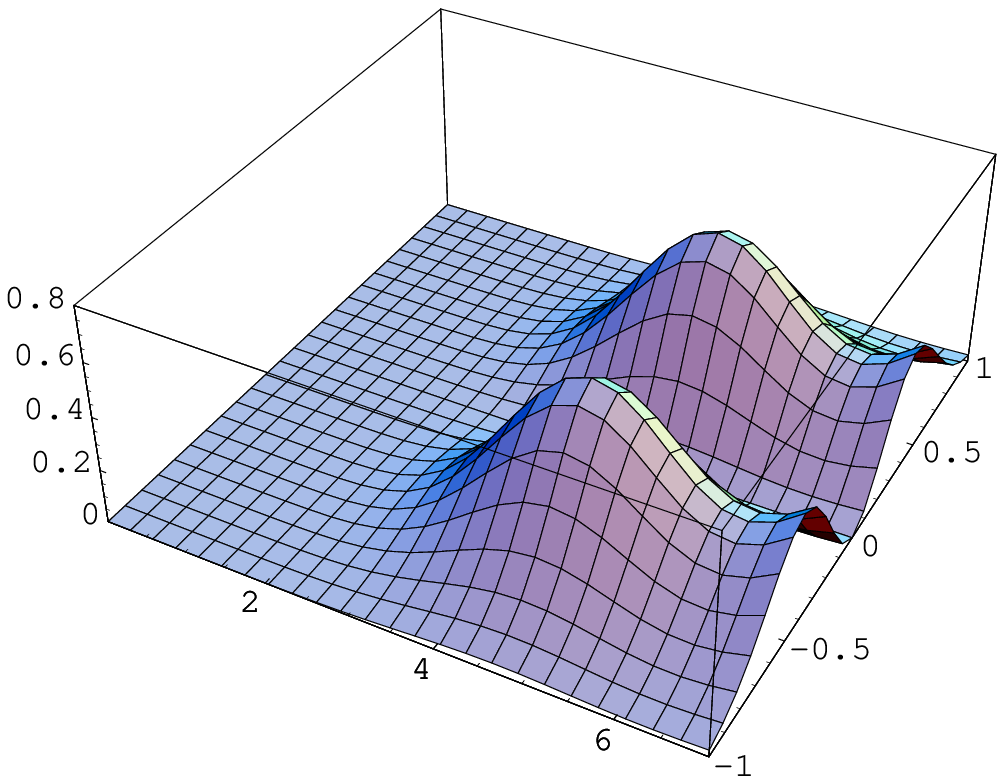,width=20pc}
\put(-460,10){$\re\ \xi$}
\put(-270,50){$\im\ \xi$}
\put(-480,180){$\rho_{weak}(\xi)$}
\put(-210,10){$\re\ \xi$}
\put(-20,50){$\im\ \xi$}
\put(-230,180){$\rho_{weak}(\xi)$}
}
\caption{The microscopic spectral density of complex eigenvalues 
at weak non-Hermiticity parameter {$\alpha=0.4$}
for $\nu=0$ (left) 
and $\nu=2$ (right).}
\label{rhoweakplot}
\end{figure}

In Fig. \ref{rhoweakplot} 
we have plotted the microscopic density for $\nu=0$ and 2
at the same value of weak non-Hermiticity parameter $\al=0.4$.
It is instructive to look at the density of real eigenvalues $x$ for the chSE
\cite{NF} 
\be
\rho_{real}(x)\ =\ 2x^2 \int_0^1 du u^2 \int_0^1 dv 
\left( J_{2\nu}(2uvx)J_{2\nu+1}(2ux)\ -\ v
J_{2\nu}(2ux)J_{2\nu+1}(2uvx)\right)
\ .
\label{rhoreal}
\ee
\begin{figure}[-h]
\centerline{
\epsfig{figure=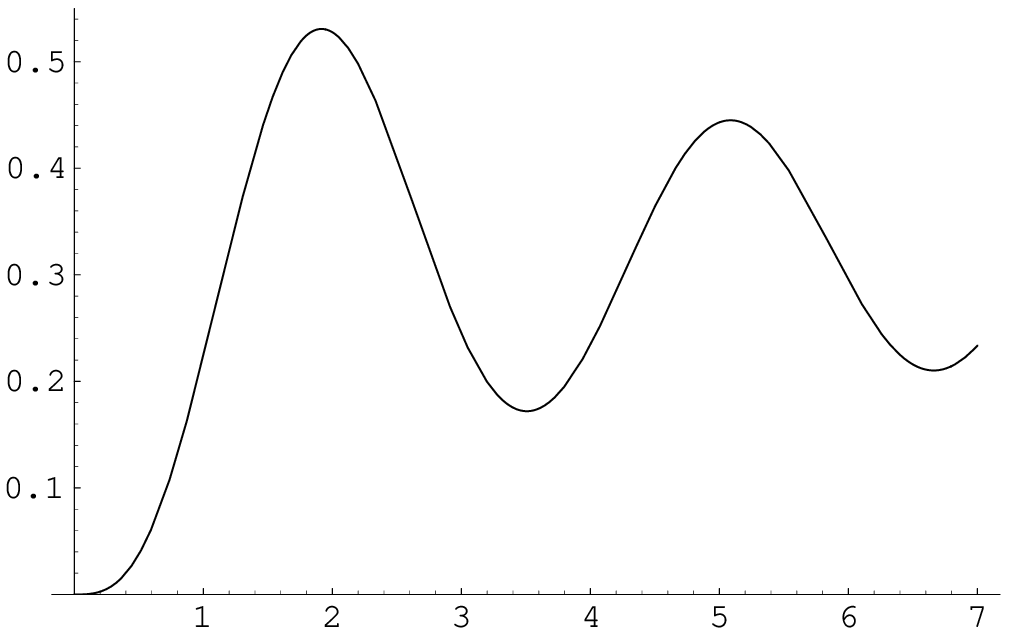,width=20pc}
\epsfig{figure=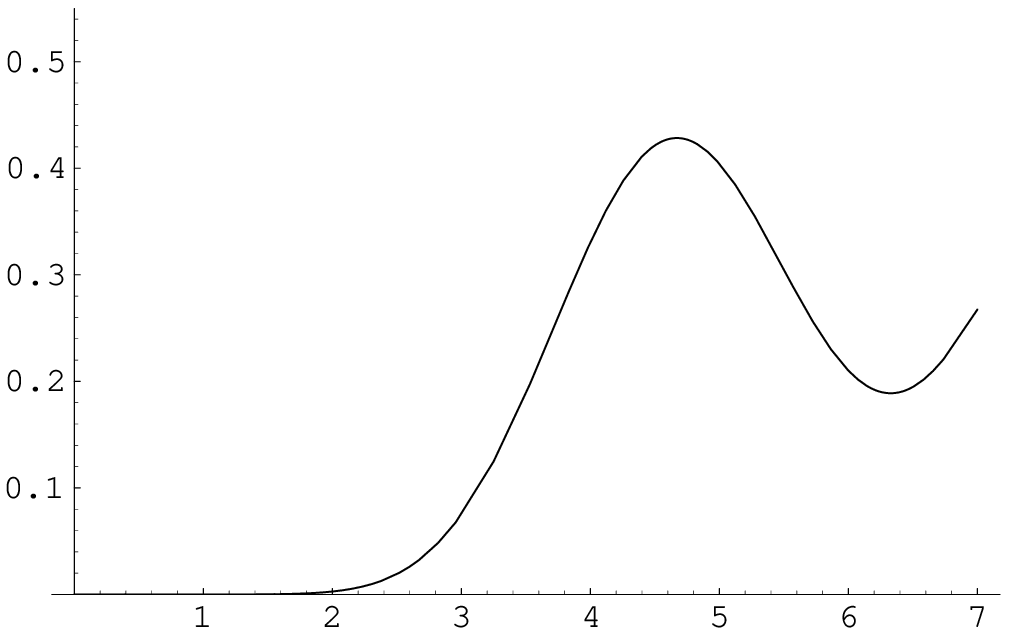,width=20pc}
\put(5,4){$x$}
\put(-240,4){$x$}
\put(-240,160){$\rho_{real}(x)$}
\put(-480,160){$\rho_{real}(x)$}
}
\caption{The microscopic spectral density of real eigenvalues 
for ${\nu=0}$ (left) and  ${\nu=2}$ (right).}
\label{rhorealplot}
\end{figure}
Compared to the real density the density in the complex plane keeps the 
oscillatory structure, with the maxima and minima at the same places. 
However, due to the repulsion of eigenvalues from 
the real axis the peak structure is doubled, with a valley along the real 
axis. On the right of Fig. \ref{rhorealplot}
the additional zero eigenvalues at $\nu=2$ repel the density further 
away from the origin, just as they do in the real case. 
The double peak structure is very different from the complex chUE
\cite{A02,SplitVerb2,James} and has been detected in
lattice simulations \cite{ABMLP}.

We can also perform an analytical check by taking the limit $\al\to0$ 
on the microscopic density eq. (\ref{rhoKweak}).
In this limit the pre-factor of the integral becomes  
\be
\lim_{\al\to0}  \frac{1}{32\al^4}
(\xi^{\ast\,2}-\xi^2)\ |\xi|^2\ 
K_{2\nu}\left(\frac{|\xi|^2}{2\al^2}\right)
\mbox{e}^{\frac{1}{4\al^2}(\xi^2+\xi^{*\,2})}\ =\ 
\frac{i\pi}{16}
x\sqrt{x^2+y^2}\ \delta(y)^\prime\ ,
\label{deltanorm}
\ee
where we denote 
$\xi=x+iy$.
Taylor expanding the integrand in the density with respect to $y=\im(\xi)$, 
we obtain to leading order
\bea
&&\lim_{y\to0}\left(J_{2\nu}(2\sqrt{st}\xi)J_{2\nu}(2\sqrt{s}\xi^\ast)
\ -\ (\xi\leftrightarrow \xi^\ast)\right) \nn\\ 
&&= \
4iy\sqrt{s}\left(
J_{2\nu}(2\sqrt{st}x)J_{2\nu+1}(2\sqrt{s}x)
-\sqrt{t} J_{2\nu+1}(2\sqrt{st}x)J_{2\nu}(2\sqrt{s}x)
\right)\ +\ {\cal O}(y^2) \ ,
\label{reallim}
\eea
after using a Bessel identity. For the limiting spectral density we thus obtain
\bea
\lim_{\al\to0}\rho_{weak}(\xi) &=&
-\,y \delta(y)^\prime \frac{\pi}{4}\nn\\
&\times& x^2 \int_0^1 ds \int_0^1dt \sqrt{\frac{s}{t}}
\left(J_{2\nu}(2\sqrt{st}\,x)J_{2\nu+1}(2\sqrt{s}\,x)
-\sqrt{t} J_{2\nu+1}(2\sqrt{st}\,x)J_{2\nu}(2\sqrt{s}\,x)\right)\nn\\
&=& -\, y \delta(y)^\prime \frac{\pi}{2} \rho_{real}(x)\ ,
\label{rhoa0check}
\eea
where in the last step we substituted $s=u^2$ and $t=v^2$. 
Compared to the Hermitian limit of the chiral complex ensemble 
\cite{A02} the density is not proportional to $\delta(y)$
but to $-y \delta(y)^\prime$. This reflects the repulsion from 
the real axis. Still it gives unity when integrated over the imaginary part
$y$, after an integration by parts.  
\begin{figure}[-h]
\centerline{
\epsfig{figure=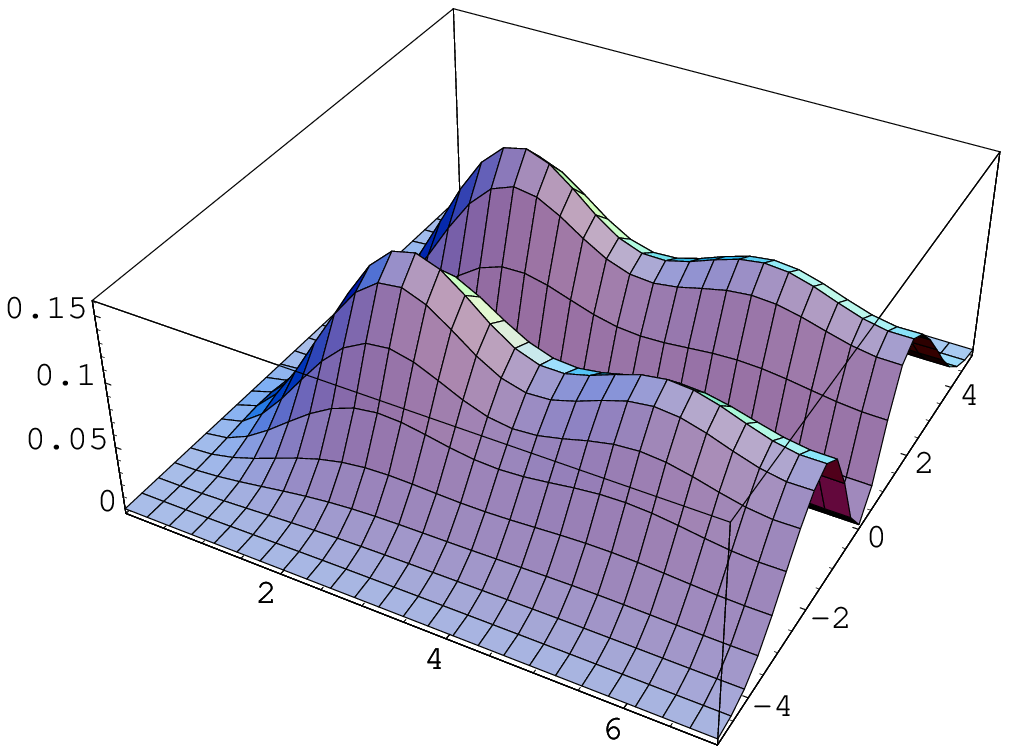,width=20pc}
\epsfig{figure=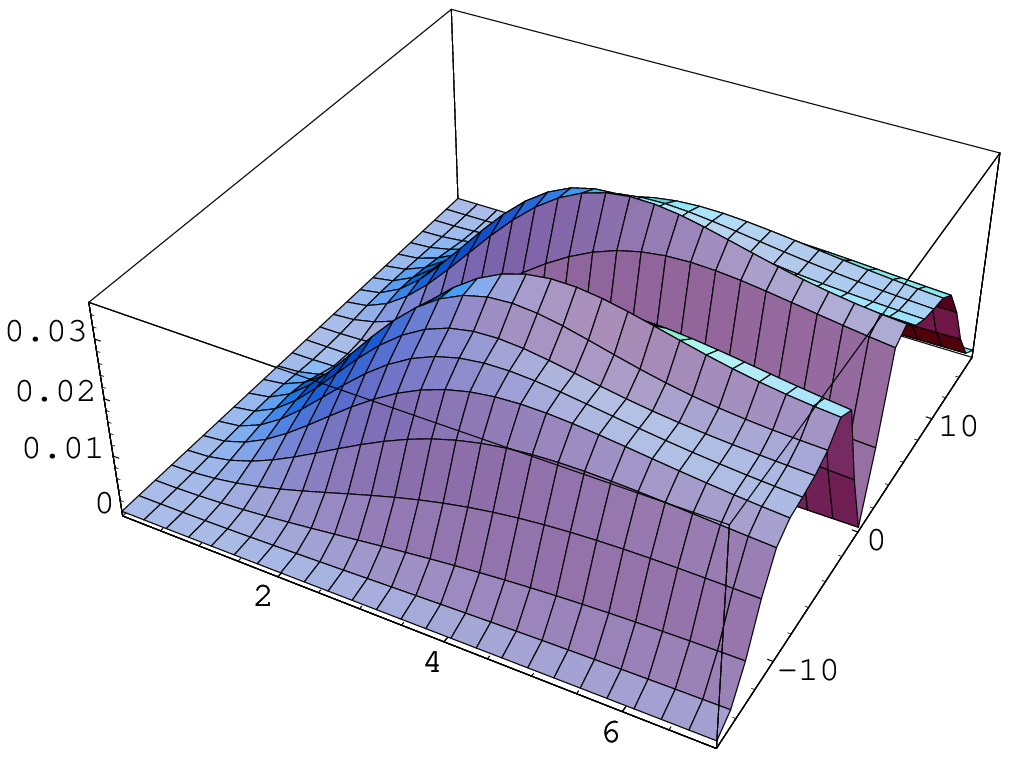,width=20pc}
\put(-460,10){$\re\ \xi$}
\put(-270,50){$\im\ \xi$}
\put(-480,180){$\rho_{weak}(\xi)$}
\put(-210,10){$\re\ \xi$}
\put(-20,50){$\im\ \xi$}
\put(-230,180){$\rho_{weak}(\xi)$}
}
\caption{The microscopic spectral density of complex eigenvalues 
at weak non-Hermiticity parameter $\alpha=1.2$
(left) and $\alpha=2.5$ (right),
both for $\nu=0$.}
\label{rhoweak2plot}
\end{figure}

To illustrate our results further we also give the complex density for larger
values of the non-Hermiticity parameter $\alpha$. As it is seen in
Fig. \ref{rhoweak2plot} the
oscillations are rapidly damped, with the density starting a to develop a 
plateau as 
at strong non-Hermiticity (see Fig. \ref{rhostrongplot}). 
For the latter as well as for an analytic check of this
relation by letting $\al\to\infty$ we refer to section \ref{strong}.
We also remind the reader that the overall height of the density varies with
$\alpha$. This is because of the normalisation to a delta-function in the
limit $\alpha\to0$ eq. (\ref{deltanorm}).

\subsection{Massive correlation functions at weak non-Hermiticity}
\label{ex:massdensity}

After discussing in detail the quenched spectral density 
we now turn to the presence of massive flavours. 
We present the spectral density for one pair of degenerate 
massive flavours $4N_f=4$ explicitly, as well as some characteristic
polynomials or massive partition functions as examples. Higher order
correlation functions and more flavours can then easily be obtained
from the given building blocks eq. (\ref{prekweak}), following the same
procedure. 

At finite-$N$ the simplest spectral density with non-zero flavour content has 
$4N_f=4$ masses. It is given by 
\be
R_N^{(4)}(z)\ =\ \frac{R_{N+1}^{(0)}(z,im)}{R_{N+1}^{(0)}(im)}\ ,
\label{rhomass}
\ee
using eq. (\ref{masterrho}).
Inserting the explicit expressions eqs. (\ref{2point}) and (\ref{rho}) 
we obtain 
\be
R_N^{(4)}(z)\ =\ (z^{*\,2}-z^2)w(z,z^*;\tau)\left( \kappa_{N+1}(z,z^*)
\ -\ \frac{|\kappa_{N+1}(z,im)|^2 
-|\kappa_{N+1}(z,-im^*)|^2}{\kappa_{N+1}(im,-im^*)}
\right)\ .
\label{rhomassexpl}
\ee 
While the first term gives exactly the quenched spectral density the second 
term is a correction term, just as for zero chemical potential \cite{AK}. 
When taking the large-$N$ limit we  
have to rescale the masses in the same way as the complex eigenvalues in
eq. (\ref{microweak}), 
\be
\sqrt{2}\ Nm_f\ \equiv\ \eta_f\ , \ \ f=1,\ldots,N_f \ \ .
\label{microweakm}
\ee
As we have mentioned at the end of subsection. \ref{Rmass}, 
eq. (\ref{rhomassexpl})
includes a limit when setting the masses to be real, $m_f=m_f^*$. 
This is because of the level repulsions both from the real and
imaginary axis: the denominator as well as the numerator of 
the second term in eq. (\ref{rhomassexpl}) vanishes.
The same feature also holds for more general correlation functions 
eq. (\ref{masterrho}).
We illustrate this 
by Taylor expanding the limiting pre-kernel $\kappa_{weak}(i\eta,-i\eta^*)$
in the denominator
in the imaginary part of the mass $\epsilon\to0$, setting 
$\eta=\eta_x+i\epsilon$,
\bea
&&\lim_{\epsilon\to0} \kappa_{weak}(i\eta_x-\epsilon,-i\eta_x-\epsilon) \ =\ 
\frac{1}{\alpha^4}
N^{4\nu+4}\frac{2^{2\nu-3}}{(\eta_x)^{4\nu}}\nn\\
&&\times \ 4i\epsilon \int_0^1dt\int_0^1ds \sqrt{\frac{s}{t}}
\mbox{e}^{-2s(1+t)\al^2}
\left( \sqrt{t} I_{2\nu+1}(2\sqrt{st}\eta_x)I_{2\nu}(2\sqrt{s}\eta_x)
- I_{2\nu}(2\sqrt{st}\eta_x)I_{2\nu+1}(2\sqrt{s}\eta_x)\right) \nn\\
&&\ +\ {\cal O}(\epsilon^2) \ .
\label{prekercont}
\eea
The imaginary argument rotates $J$- to $I$-Bessel functions, where we could
have obtained the same result from eq. (\ref{reallim}), setting $x=i\eta$ and
$y=i\epsilon$ there.
Making the same expansion for the other terms, 
$\kappa_{weak}(z,\pm i\eta^{(*)})$ and its conjugates
in the numerator, we arrive at the following expression for the microscopic
limit of the density eq. (\ref{rhomassexpl}):
\be
\rho_{weak}^{(4)}(\xi) \ =\  \rho_{weak}^{(0)}(\xi)\ -\
\Delta\rho_{weak}^{(4)}(\xi) \ \ ,
\label{rhoKweakmass}
\ee
with the first term from eq. (\ref{rhoKweak}) and 
the correction term to the quenched density being given by 
\bea
&&\Delta\rho_{weak}^{(4)}(\xi) \ =\  
\frac{1}{32\al^4}
(\xi^{\ast\,2}-\xi^2)\ |\xi|^2\ 
K_{2\nu}\left(\frac{|\xi|^2}{2\al^2}\right)
\exp\left[+\frac{1}{4\al^2}(\xi^2+\xi^{*\,2})\right]
\label{deltarhoweak}\\
&&\times\left\{ 
\left( 
\int_0^1\!ds\int_0^1dt \sqrt{\frac{s}{t}}\ \mbox{e}^{-2s(1+t)\al^2}
( J_{2\nu}(2\sqrt{st}\ \xi)I_{2\nu+1}(2\sqrt{s}\ \eta)
- \sqrt{t}I_{2\nu+1}(2\sqrt{st}\ \eta) J_{2\nu}(2\sqrt{s}\ \xi))
\right)\right.\nn\\
&&\ \ \left.\times\left( 
\int_0^1\!ds\int_0^1\frac{dt}{\sqrt{t}}\ \mbox{e}^{-2s(1+t)\al^2}
( J_{2\nu}(2\sqrt{st}\, \xi^*)I_{2\nu}(2\sqrt{s}\, \eta)
- I_{2\nu}(2\sqrt{st}\, \eta) J_{2\nu}(2\sqrt{s}\, \xi^*))\!
\right)
-(\xi\leftrightarrow\xi^*)\right\}\nn\\
&&\times
\left[\int_0^1\!ds\int_0^1dt \sqrt{\frac{s}{t}}\ \mbox{e}^{-2s(1+t)\al^2}
\left( \sqrt{t} I_{2\nu+1}(2\sqrt{st}\ \eta)I_{2\nu}(2\sqrt{s}\ \eta)
- I_{2\nu}(2\sqrt{st}\ \eta)I_{2\nu+1}(2\sqrt{s}\ \eta)\right) \right]^{-1}\!.
\nn
\eea
Here we have again called $\eta_x=\eta$ which is now real. Before discussing
this main result of the subsection we also briefly give 2 examples for massive
partition functions. The simplest case with $2N_f=2$ flavours follows from the
even skew-orthogonal polynomial at imaginary mass $z=im$, comparing
eq. (\ref{qevenvev}) and (\ref{Z2evfinal}):
\be
{\cal Z}_N^{(2N_f=2)}(m)/{\cal Z}_N^{(0)}=|m|^{2\nu}q_{2N}(im)\ .
\label{Z2N}
\ee 
We can set the mass to be real here without taking 
limits, as  $q_{2N}(im)$ itself does not display any level-repulsion from
the axis.
We obtain the remarkably simple expression in the weak limit,
\be
{\cal Z}_{weak}^{(2)}(\eta)\ \sim\ 
\int_0^1\frac{dt}{\sqrt{t}}\ \mbox{e}^{-2t\al^2} I_{2\nu}(2\sqrt{t}\,\eta) \ , 
\label{Z2weak}
\ee
where we have rotated the result eq. (\ref{Lasymp3}) from the previous
subsection to an imaginary argument, skipping all constants (including $N$).
As a second example we give the partition function for $4N_f=4$ using our
formula eq. (\ref{Zmrho}),
\be
{\cal Z}_N^{(4)}(m)/{\cal Z}_{N+1}^{(0)}\ =\ \frac{|m|^{4\nu}}{(N+1)}
\frac{\kappa_N(im,-im^*)}{(m^{2}-m^{*\,2})} \ .
\label{Z4N}
\ee
Upon using the same expansion as for the density eq. (\ref{prekercont}) we
immediately obtain 
\be
{\cal Z}_{weak}^{(4)}(\eta)\sim\frac{1}{\eta\alpha^4}
\int_0^1\!dt\int_0^1\!ds \sqrt{\frac{s}{t}}\,
\mbox{e}^{-2s(1+t)\al^2}\!
\left( 
\sqrt{t} I_{2\nu+1}(2\sqrt{st}\eta)I_{2\nu}(2\sqrt{s}\eta)
- 
I_{2\nu}(2\sqrt{st}\eta)I_{2\nu+1}(2\sqrt{s}\eta)
\right),
\label{Z4weak}
\ee
where we have again suppressed all constants. Due to the presence of conjugate
pairs of complex eigenvalues the dependence on $\al$ cannot be factored out,
similar to the same situation in the unitary ensembles \cite{AV03,SplitVerb2,
AOSV}. 
Such a simplification only occurs in the absence of conjugate flavours, in the
QCD matrix model 
with a sign problem \cite{AFV,James,AOSV}, 
where the partition function can be
normalised to be $\mu$-independent. 
As a check both expressions eqs. (\ref{Z2weak}) and (\ref{Z4weak})
reduce to the know results \cite{NN,AK} 
in the limit
$\al\to0$.

\begin{figure}[-h]
\centerline{
\epsfig{figure=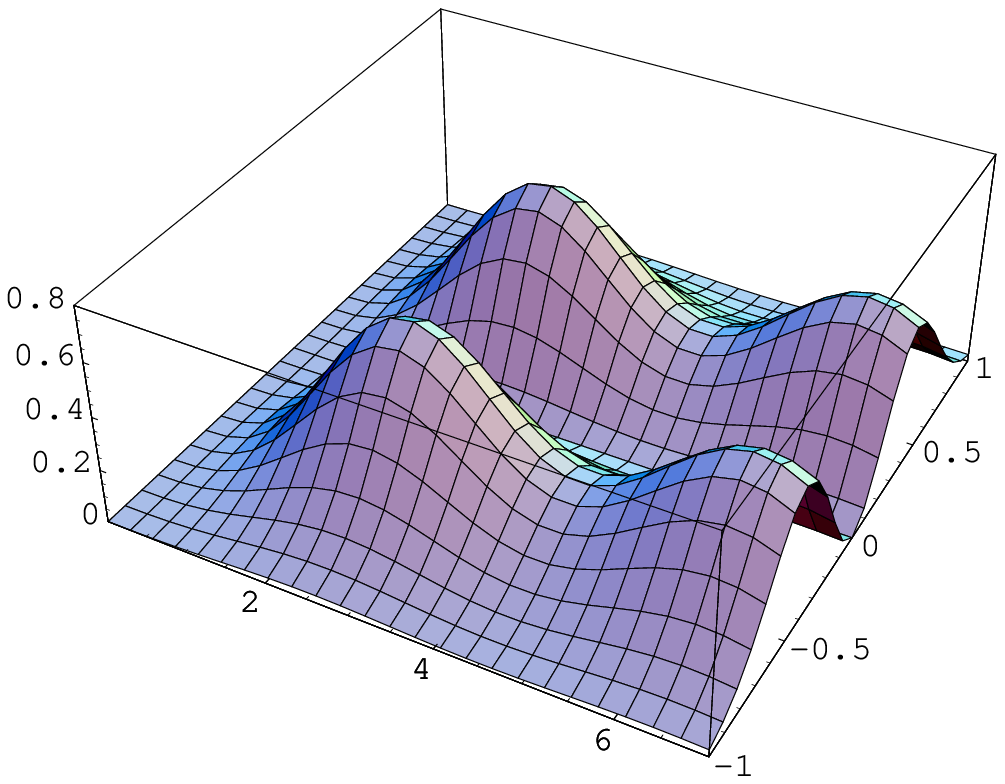,width=20pc}
\epsfig{figure=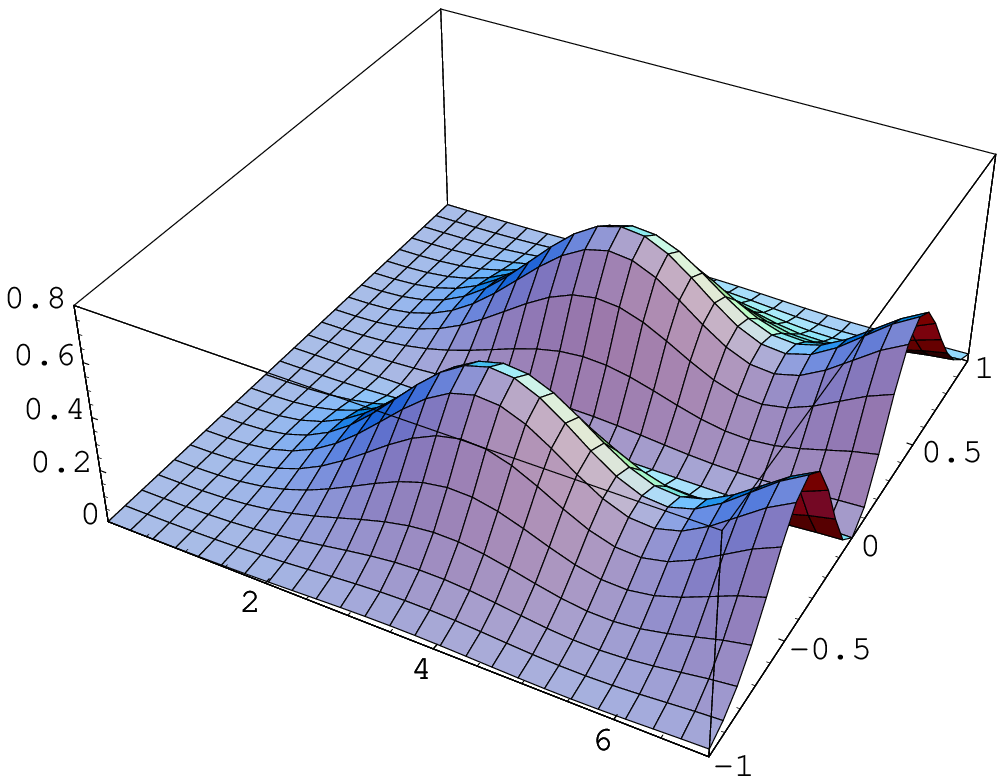,width=20pc}
\put(-460,10){$\re\ \xi$}
\put(-270,50){$\im\ \xi$}
\put(-480,180){$\rho_{weak}^{(4)}(\xi)$}
\put(-210,10){$\re\ \xi$}
\put(-20,50){$\im\ \xi$}
\put(-230,180){$\rho_{weak}^{(4)}(\xi)$}
}
\caption{The massive microscopic spectral density 
at weak non-Hermiticity parameter {$\alpha=0.4$}
for masses $\eta=8.74$ (left) 
and $\eta=4.26$ (right), both at $\nu=0$.}
\label{rhoweakmassplot}
\centerline{
\epsfig{figure=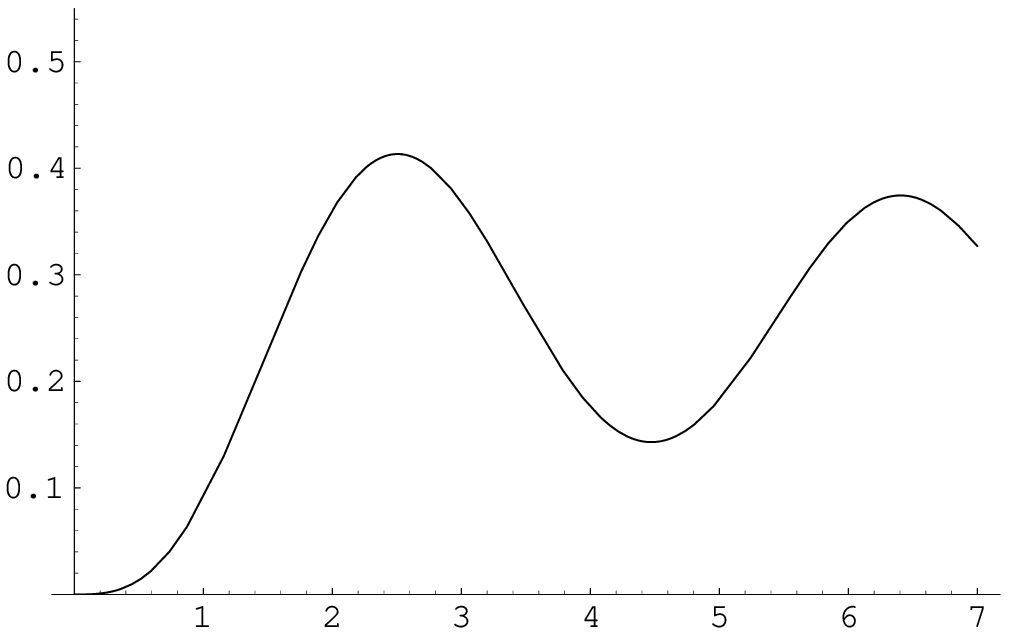,width=20pc}
\epsfig{figure=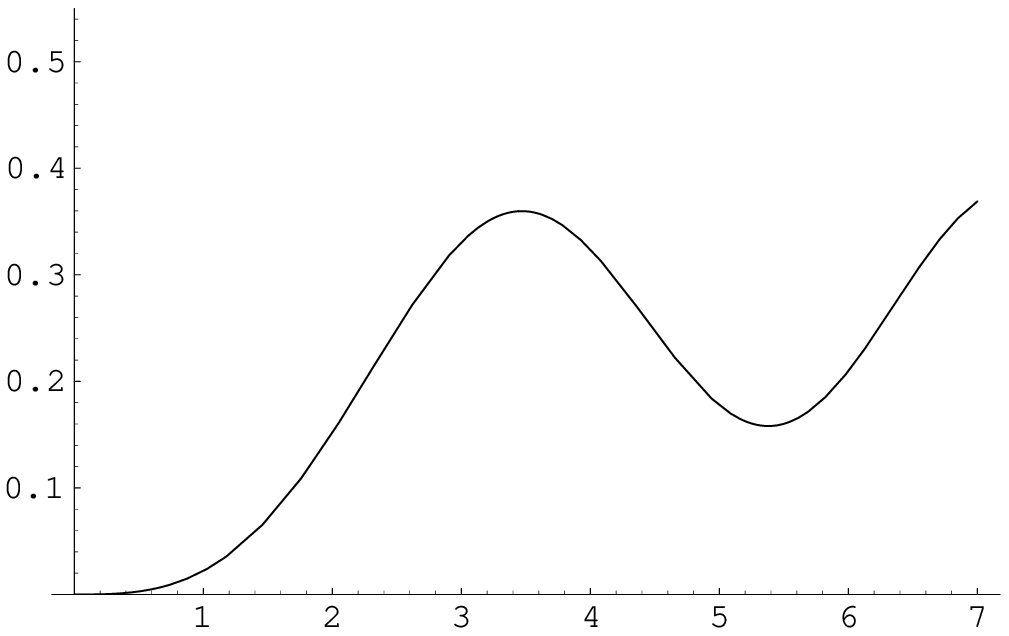,width=20pc}
\put(5,4){$x$}
\put(-240,4){$x$}
\put(-240,160){$\rho_{real}^{(4)}(x)$}
\put(-480,160){$\rho_{real}^{(4)}(x)$}
}
\caption{The massive microscopic spectral density of real eigenvalues 
for masses $\eta=8.74$ (left) and $\eta=4.26$ (right), both at ${\nu=0}$.}
\label{rhorealmassplot}
\end{figure}
Let us now discuss the massive density eq. (\ref{rhoKweakmass}).
In Fig. \ref{rhoweakmassplot} 
we plot the microscopic, massive spectral density 
for two different values of masses $\eta=8.74$ and $\eta=4.26$,
both at weak non-Hermiticity parameter $\al=0.4$ and $\nu=0$, 
and compare it the 
corresponding expression for real eigenvalues \cite{AK}. We have deliberately
chosen the same values for the mass parameter as there, where numerical data
for lattice simulations with dynamical fermions at $\mu=0$ \cite{BMW}
were compared. 
Similar to having $\nu=2$ exact zero eigenvalues 
in Fig. \ref{rhoweakplot} the 
eigenvalues are pushed further away from the origin through the masses.
Below we compare the corresponding expression for real eigenvalues 
at the same values of the mass parameters \cite{AK}. As for the quenched case
the correlation functions keep their oscillatory structure, with the maxima
located at the same positions as for real eigenvalues, at the given value of
$\alpha$. 
\begin{figure}[-h]
\centerline{
\epsfig{figure=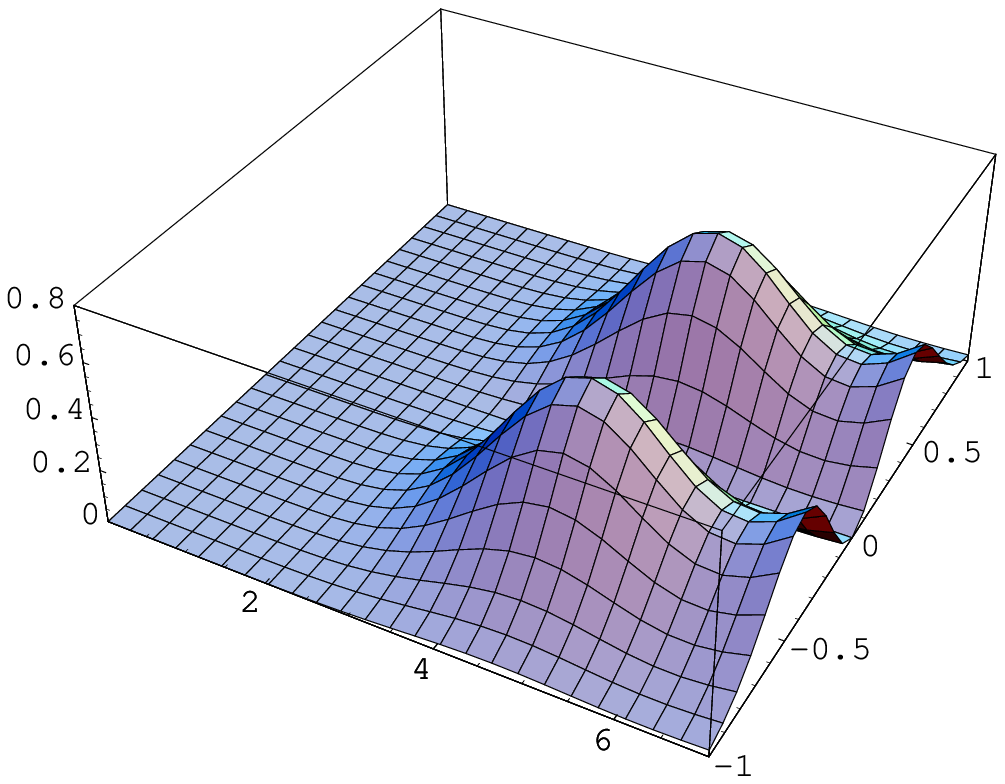,width=20pc}
\epsfig{figure=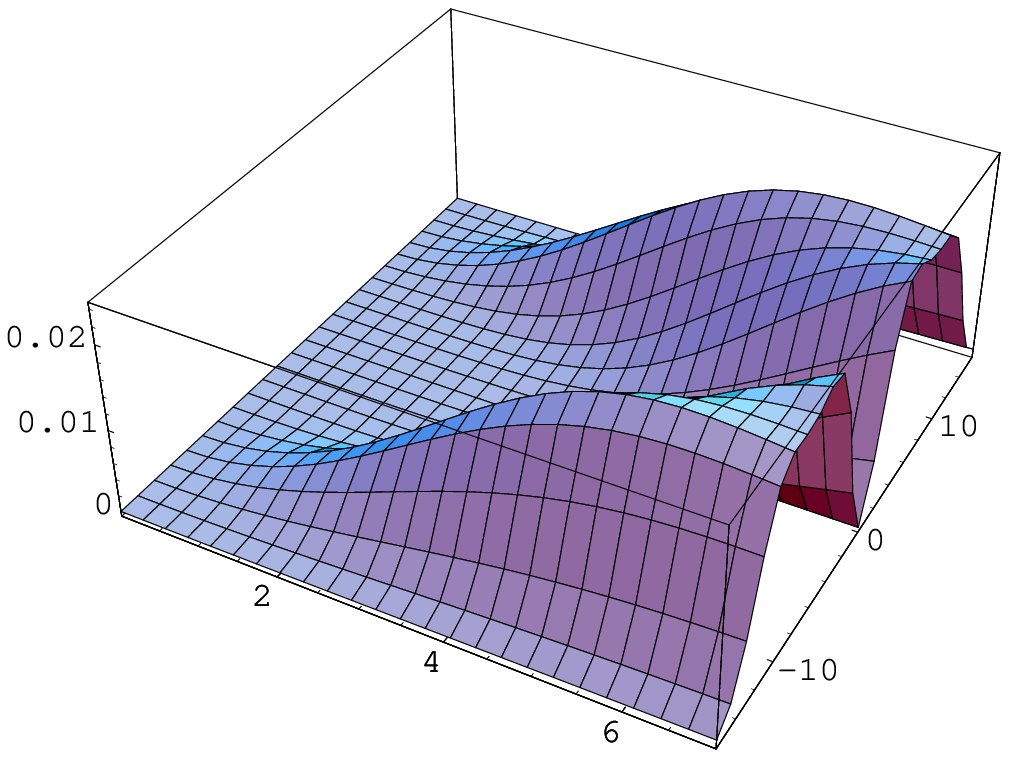,width=20pc}
\put(-460,10){$\re\ \xi$}
\put(-270,50){$\im\ \xi$}
\put(-480,180){$\rho_{weak}^{(4)}(\xi)$}
\put(-210,10){$\re\ \xi$}
\put(-20,50){$\im\ \xi$}
\put(-230,180){$\rho_{weak}^{(4)}(\xi)$}
}
\caption{The massive microscopic spectral density 
at weak non-Hermiticity parameter at {$\alpha=0.4$}
with mass $\eta=1$ (left) and at {$\alpha=2.5$} with 
$\eta=4.26$ (right), both at $\nu=0$.}
\label{rhoweakmass2plot}
\end{figure}
Furthermore we observe that for larger masses the density moves toward the
quenched density in Fig. \ref{rhoweakplot}, 
as the infinite mass limit quenches the fermion
determinant. On the other hand smaller masses moving 
towards the origin push
the eigenvalues further out. In the zero-mass limit the density should
approach the quenched density at topological charge $\nu=2$ approximately, as
can be seen in Fig. \ref{rhorealmassplot} left. Although for
non-zero chemical potential there is no more flavour-topology duality
\cite{Jacftdual} because
of the explicit $\nu$-dependence in the $K$-Bessel functions in the weight,
this duality holds approximately, when the $K$-Bessel function reaches its
asymptotic, $\nu$-independent form. For those values of
$|\xi^2|/(2\alpha^2)$ in Fig. \ref{rhoweakmass2plot} left, where the density is
non-vanishing, this is indeed the case.

As a final example we display the massive density at a larger value of
$\alpha=2.5$ in Fig. \ref{rhoweakmass2plot} right. As seen already for the
quenched density the oscillations are washed out (compared to 
Fig. \ref{rhoweakmassplot} right). 
A further feature can be seen here. For the chosen value of $\alpha$ and $\eta$
the $\xi$-value where the mass factor $|\xi^2+\eta^2|^2$ vanishes,
$\pm\im(\xi)=\eta$, now lies on the imaginary axis inside the shown picture. 
It can be seen very clearly that the
eigenvalues are pushed away from the very location of the mass, in addition to
the level repulsion from the $y$-axis.

In the previous subsection we performed an analytical check by taking 
the Hermitian limit $\al\to0$ of the spectral density 
and comparing it to the known expression for the 
density of real eigenvalues. 
We have performed the same check here 
where we refer to \cite{AK} for the corresponding expression 
for the massive spectral density of real eigenvalues. Without going through 
the details we just comment of the structure to be obtained.

Since our example eq. (\ref{rhomass}) involves the two-point function 
it contains all four elements of the matrix kernel eq. (\ref{kernel}), the 
pre-kernel $\kappa_N$ at different (complex conjugated arguments).
In contrast to that for real eigenvalues the two-point or higher correlation
functions are given by a quaternion determinant of a two by two matrix kernel
with {\it different} matrix elements: a pre-kernel which is basically given by
eq. (\ref{rhoreal}), its derivative and its integral (see
\cite{NF}). How can we obtain these three different matrix
elements from one pre-kernel? 
Taking the limit $\al\to0$ invokes a further 
derivative with respect to the imaginary part $y=\im(\xi)$, in addition to the
expansion in eq. (\ref{deltarhoweak}) due to real masses.
As a result we will get the real limit of the pre-kernel eq. (\ref{prekweak}),
its first and second derivative, correctly reproducing 
all terms in \cite{AK} for the real massive density.


\sect{The strong non-Hermiticity limit}\label{strong}

The regime of strong non-Hermiticity in defined by taking the limit 
$N\to\infty$ while keeping $\mu\in(0,1]$ {\it fixed}. The rescaling of the 
eigenvalues is modified compared to eq. (\ref{microweak}), keeping
\be
\sqrt{N}\ (\re\ z+i\im\ z)\ =\ \sqrt{N}\ z \ \equiv \xi \ \ \ 
\label{microstrong}
\ee
finite\footnote{We comment on a $\mu$-dependent rescaling 
  later.}. Hence the definition of the large-$N$ correlation functions is 
also altered, reading
\bea
\kappa_{strong}(\xi_1,\xi_2^\ast) &\equiv&  
\lim_{N\to\infty}\frac{1}{N}\ 
\kappa_N\left(\frac{\xi_1}{\sqrt{N}},\frac{\xi_2^\ast}{\sqrt{N}}\right), \nn\\
\rho_{strong}(\xi_1,\ldots,\xi_k)
&\equiv&  \lim_{N\to\infty}
\frac{1}{N^{k}}\ 
R_N\left(\frac{\xi_1}{\sqrt{N}},\ldots,
\frac{\xi_k}{\sqrt{N}}\right).
\label{microrhostr}
\eea
The difference in scaling can be easily understood 
as follows. At weak non-Hermiticity the eigenvalues remain close to the 
real axis, with their imaginary part being small compared to the real part. 
In order to obtain a macroscopic density at the origin 
of the order $\rho_{macro}(0)\sim N$
their average distance has to be ${\cal O}(1/N)$. 
In contrast to that at strong non-Hermiticity the eigenvalues spread out 
in the complex plane, having a comparable real and imaginary part. 
To form again a constant macroscopic density $\rho_{macro}(0)\sim N$
the average distance between nearest neighbours has to be ${\cal O}(1/N^2)$.
Here we measure distance with respect to the two-dimensional metric. 

A clear separation of scales between the two different limits can be defined 
as follows. At weak non-Hermiticity the microscopic density describes 
the universal fall off to zero 
of the eigenvalues away from the imaginary axis. 
In direction of the real axis along the maxima we obtain as usual 
$\lim_{\Re e z\to\infty}\rho_{weak}(z)=\rho_{macro}(0)$ \footnote{ 
Because of the repulsion
of eigenvalues from the real axis for $\beta=4$
we of course have to take this limit 
away from the axis, along the maxima of the density, see Fig.
\ref{rhoweakplot}.}. 
In contrast to that at strong non-Hermiticity the microscopic density attains 
the constant value of the macroscopic one in all directions,
$\lim_{|z|\to\infty}\rho_{strong}(z)=\rho_{macro}(0)$, as can be seen below
in Figs. \ref{rhostrongplot} and \ref{rhostrsym}. 
This is of course except along 
the real and imaginary axis, due to the repulsion from the 
Jacobian in eq. (\ref{Zev}). The fall-off of the macroscopic density to zero 
in any direction (which is not seen in the microscopic regime here) 
can then be expected to be given 
by another microscopic function at the edge, in analogy to the Airy-kernel for
real ensembles at the edge of the spectrum.
When numerically solving the matrix model at finite but large-$N$ care has to 
be taken to ensure a clear separation between micro- and macroscopic density
when comparing to the analytic formulas.

The computation of correlation functions at strong non-Hermiticity 
is more involved, as is already the case in the non-chiral ensemble
\cite{EK}. In fact we will follow the same strategy as there, by 
first deriving 
a differential equation for the large-$N$ pre-kernel at maximal non-Hermiticity
$\mu=1$. A solution of this equation can be obtained either directly or 
by taking the 
limit $\al\to\infty$ of the pre-kernel at weak non-Hermiticity, 
keeping $\xi_{weak}/\al$ fixed.
Since the two methods match it is suggestive to recover the full range of 
strong non-Hermiticity $\mu\in(0,1]$ by inserting a $\mu$-dependent parameter
through $\xi\to\xi \Sigma(\mu)$. However, we can also argue that such a
rescaling can always be absorbed by redefining the scaling limit
eq. (\ref{microstrong}).
Details of the derivation of the differential equation and its solution 
are given in the appendix \ref{Appdiff}.

We begin by taking the limit of maximal non-Hermiticity $\mu\to1$ 
on the skew orthogonal polynomials eqs. (\ref{qodd}) and (\ref{qeven}).
In this limit only the highest powers survive and 
they become monic or a sum over monomials, respectively. 
We thus obtain the following expression for the 
pre-kernel eq. (\ref{prekernel}) in the microscopic large-$N$ limit
eq. (\ref{microrhostr}):
\be
\left.\kappa_{strong}(\xi,\zeta^*)\right|_{\mu=1}\ =\ 
\frac{N^{2\nu+2}}{\pi2^{2\nu+3}}
\sum_{k=0}^\infty\sum_{j=0}^k
\frac{k!\Gamma(k+\nu+1)}{\Gamma(2k+2\nu+2)(2k+1)!}
\frac{(\xi^{4k+2}\zeta^{*\,4j}
\ -\ \xi^{4j}\zeta^{*\,4k+2})
}{2^{4j}j!\Gamma(j+\nu+1)}
\ .
\label{kstrdef}
\ee
It can be shown that the function $\kappa_{strong}(\xi,\zeta^*)$ obeys the 
following set of inhomogeneous second order differential equations
\bea
\frac12\left[\partial_\xi^2+\frac{(4\nu+1)}{\xi}\partial_\xi-\xi^2\right]
\kappa_{strong}(\xi,\zeta^*)|_{\mu=1}&=&
+\ \frac{N^{2\nu+2}}{4\pi}
\frac{I_{2\nu}\left(\xi\zeta^*\right)}{(\xi\zeta^*)^{2\nu}}\ ,
\nn\\
\frac12
\left[\partial_{\zeta^*}^2+
\frac{(4\nu+1)}{\zeta^*}\partial_{\zeta^*}-\zeta^{*\,2}\right]
\kappa_{strong}(\xi,\zeta^*)|_{\mu=1}&=&
-\ \frac{N^{2\nu+2}}{4\pi}
\frac{I_{2\nu}\left(\xi\zeta^*\right)}{(\xi\zeta^*)^{2\nu}}\ ,
\label{diff}
\eea
where we refer to the appendix \ref{Appdiff} for details.
Due to the antisymmetry of the pre-kernel the right hand side changes sign.
The solution of this equation can be constructed 
as explained in the appendix \ref{Appdiff}.
The result is given by
\bea
\kappa_{strong}(\xi,\zeta^*)|_{\mu=1}&=& 
\frac{N^{2\nu+2}}{2\pi}
\frac{1}{(\xi\zeta^*)^{2\nu}}\mbox{e}^{\frac12(\xi^2+\zeta^{*\,2})}\nn\\
&&\times\int_0^\infty dq \int_0^q dp\,\mbox{e}^{-q^2-p^2}
\left[
J_{2\nu}(2q\zeta^*)J_{2\nu}(2p\xi)   
\ -\ 
J_{2\nu}(2q\xi) J_{2\nu}(2p\zeta^*) 
\right]\ .
\label{prekstrong}
\eea
In order to obtain all correlation functions from 
eq. (\ref{prekstrong}) we next give the weight function eq. (\ref{Kweight}) 
in the limit of strong non-Hermiticity eq. (\ref{microstrong}):
\be
w_K^{(2\nu)}(\xi,\xi^*) \ =\  N^{-2\nu-1}|\xi|^{4\nu+2} 
K_{2\nu}\left(\frac{(1+\mu^2)|\xi|^2}{2\mu^2}\right)
\exp\left[\frac{1-\mu^2}{4\mu^2}(\xi^2+\xi^{*\,2})\right]
\ ,\ \ \mu\in\,(0,1]\ .
\label{strongweights}
\ee
Here we display the weight for any value of the strong non-Hermiticity
$\mu$. At $\mu=1$ the expression simplifies, the exponential factor
cancels with the kernel 
and the argument inside the $K$-Bessel function reduces to unity times
$|\xi|^2$. Inserting this into eq. (\ref{rho}) we obtain the following result
for the microscopic spectral density at
maximal non-Hermiticity
\bea
\rho_{strong}(\xi)|_{\mu=1} &=& 
\frac{1}{2\pi} (\xi^{\ast\,2}-\xi^2)|\xi|^2
K_{2\nu}\left({|\xi|^2}\right)\mbox{e}^{\frac{1}{2}(\xi^2+\xi^{*\,2})}
\label{rhoKstr}\\
&\times&\int_0^\infty dq 
\int_0^q dp\ \mbox{e}^{-q^2-p^2}
\left[
J_{2\nu}\left({2q\xi^\ast}\right)J_{2\nu}\left({2p\xi}\right)
\ -\ 
J_{2\nu}\left({2q\xi}\right)J_{2\nu}\left({2p\xi^\ast}\right)
\right]\ .
\nn
\eea
As an analytic check we have performed the limit $\al\to\infty$ of the 
microscopic spectral density at weak non-Hermiticity eq. (\ref{rhoKweak}),
where we refer to the end of appendix \ref{Appdiff} for details. 
We obtain exactly the same functional form of the spectral density,
eq. (\ref{rhoweakstr}) (up to a different constant normalisation), in terms of
the variable $\xi_S=\sqrt{N}\,z/\mu\sqrt{2}$.

\begin{figure}[-h]
\centerline{
\epsfig{figure=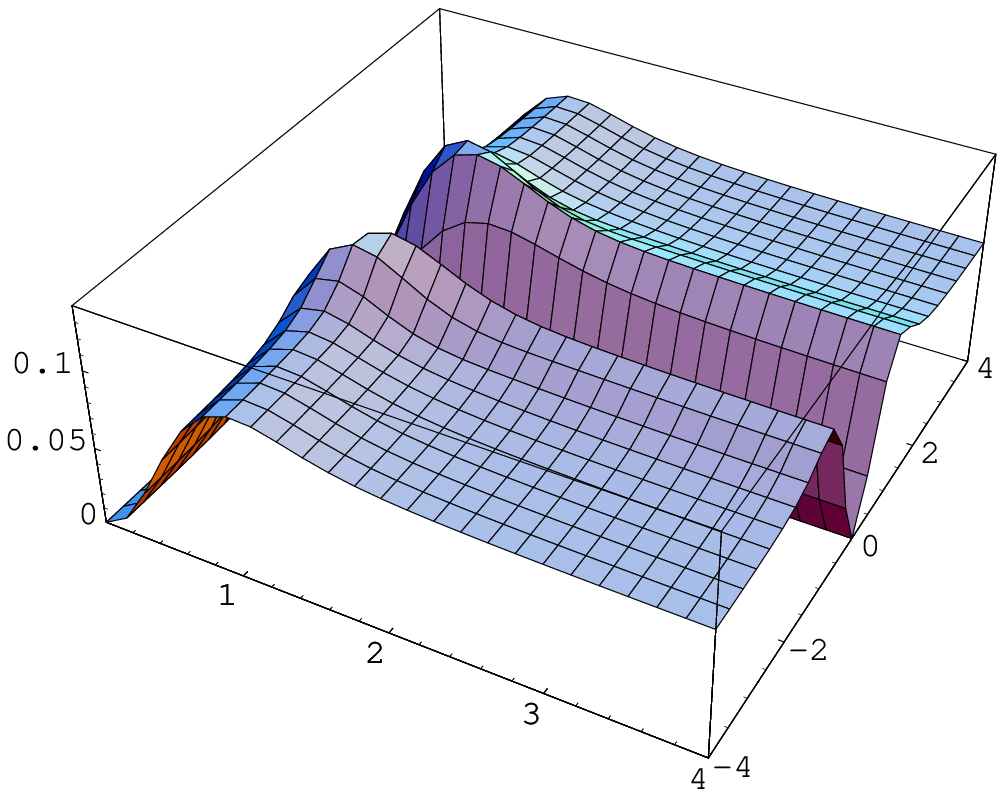,width=20pc}
\epsfig{figure=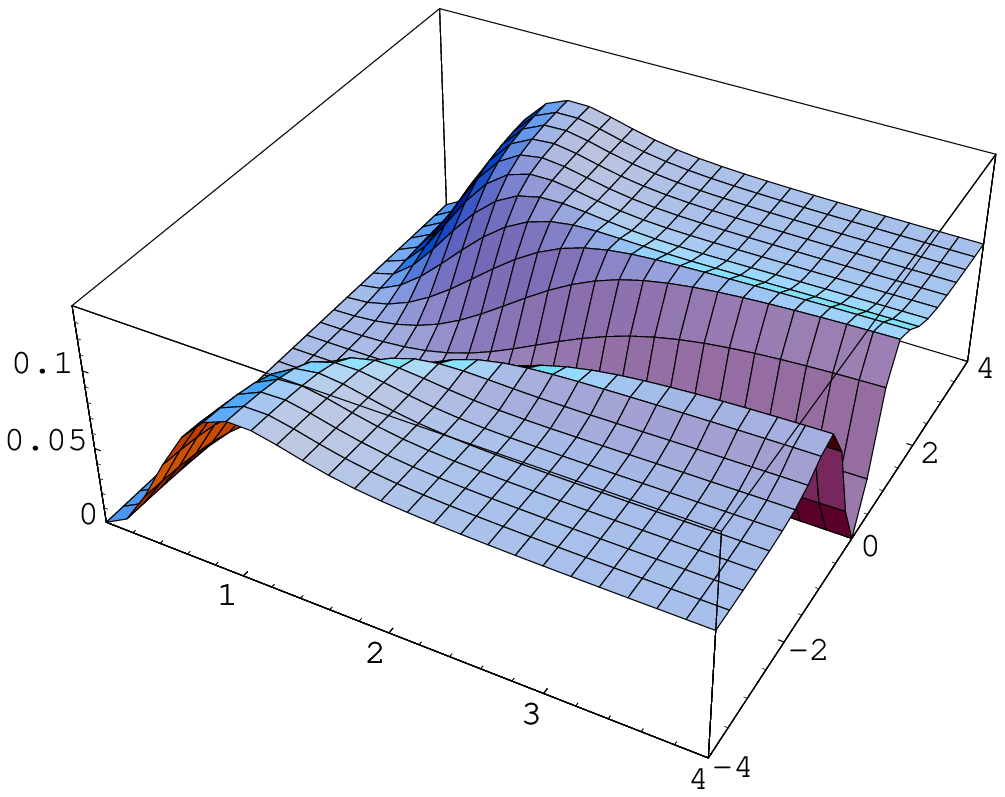,width=20pc}
\put(-460,10){$\re\ \xi$}
\put(-270,50){$\im\ \xi$}
\put(-480,180){$\rho_{strong}(\xi)$}
\put(-210,10){$\re\ \xi$}
\put(-20,50){$\im\ \xi$}
\put(-230,180){$\rho_{strong}(\xi)$}
}
\caption{The microscopic spectral density at maximal non-Hermiticity $\mu=1$ 
with $\nu=0$ (left) and $\nu=2$ (right)
.}
\label{rhostrongplot}
\end{figure}
Together with eq. (\ref{strongweights}) 
this leads us to conjecture that the $\mu$-dependence at strong
non-Hermiticity with $\mu<1$ can be restored as follows, replacing 
\be
\xi^2\ \longrightarrow\ \frac{(1+\mu^2)}{2\mu^2} \xi^2
\label{microstrmu} 
\ee
everywhere in eq. (\ref{rhoKstr}). We have shown it to be true for $\mu=1$ and 
for small $\mu\ll1$. However, if this is true we could then redefine our
strong non-Hermiticity limit eq. (\ref{microstrong}) to completely reabsorb
this $\mu$-dependence, defining 
\bea
\frac{\sqrt{1+\mu^2}}{\mu\sqrt{2}}\sqrt{N}\ z \ &\equiv& \xi \\
\rho_{strong}(\xi_1,\ldots,\xi_k)
&\equiv&  \lim_{N\to\infty}
\frac{2^k\mu^{2k}}{(1+\mu^2)^kN^{k}}\ 
R_N\left(\frac{\xi_1\mu\sqrt{2}}{\sqrt{(1+\mu^2)N}},\ldots,
\frac{\xi_k\mu\sqrt{2}}{\sqrt{(1+\mu^2)N}}
\right). \nn
\eea

As an illustration of our results 
we show the microscopic spectral density eq. (\ref{rhoKstr}) 
and its dependence on $\nu$ at maximal non-Hermiticity $\mu=1$ in
Fig. \ref{rhostrongplot}. 
One can clearly see that for $\nu=2$ the additional zero eigenvalues push the
density away from the origin.
For the numerically evaluation of the integral in eq. (\ref{rhoKstr}) necessary
to plot it an equivalent representation of the integral is very
convenient.
Using the identity derived in the appendix eq. (\ref{intid}) we have
\be
\rho_{strong}(\xi)
\ =\  
\frac{1}{4\pi} (\xi^{\ast\,2}-\xi^2)|\xi|^2
K_{2\nu}\left({|\xi|^2}\right)
\int_0^1\frac{dr}{\sqrt{1-r^2}}\ I_{2\nu}(r|\xi|^2)
\sinh\left(\frac12\sqrt{1-r^2}\,(\xi^2-\xi^{\ast\,2})\right) ,
\label{rhoKstr2}
\ee
where the exponential pre-factor has cancelled out. This form of the density
allows to dicuss its rotation and reflection symmetries most clearly.
\begin{figure}[-h]
\centerline{
\epsfig{figure=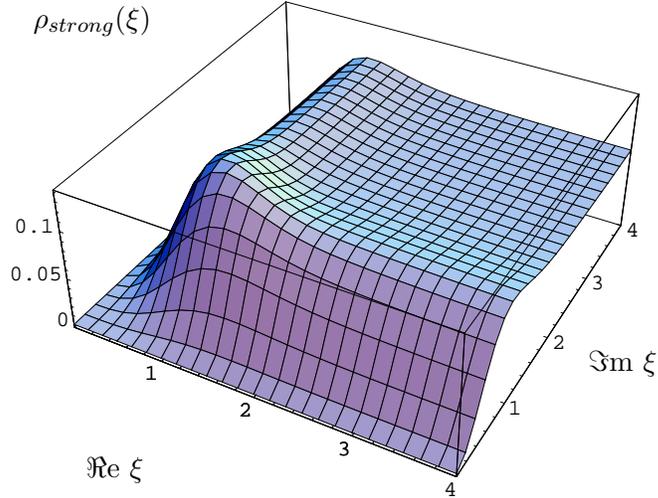,width=20pc}
\put(-210,10){$\re\ \xi$}
\put(-20,50){$\im\ \xi$}
\put(-230,180){$\rho_{strong}(\xi)$}
}
\caption{Part of Fig. \ref{rhostrongplot} (left) showing the first quadrant
  only. The symmetry with respect to the diagonal $\re\xi=\im\xi$ is clearly
  visible 
.}
\label{rhostrsym}
\end{figure}
The fact that it is an even function of $(\xi^2-\xi^{\ast\,2})$ and otherwise
only depends on the modulus $|\xi|^2$ leads to the following
properties. First, one can obviously rotate the density by $\pi/2$, $\pi$, or 
$3\pi/2$, multiplying $\xi$ by $i$ , $(-1)$, or $-i$ without changing it.
Second, the density is invariant under reflections along the $x$-, $y$-,
$(x=y)$- and
$(x=-y)$-axis where $\xi=x+iy$.  
For example for the $(x=y)$-axis in the first quadrant 
we write the pair of $\xi$-values related by reflection as 
\be
\xi_\pm \equiv r \mbox{e}^{i\frac{\pi}{4}\,\pm i\delta}
\ \ \Rightarrow \ \ \xi_\pm^*= -i\xi_\mp. 
\ee
The modulus remains unchanged, $|\xi_+|=|\xi_-|$, as well as the difference 
$(\xi_+^2-\xi_+^{\ast\,2})=-(\xi_-^{*\,2}-\xi_-^{2})$.
This symmetry can be observed in Fig. 
\ref{rhostrsym}. The remaining reflection symmetries easily follow in a
similar way.


\subsection{Massive correlation functions at strong non-Hermiticity}
\label{ex:massdensitystr}

In this subsection we give the simplest nontrivial result including one pair
of degenerate massive flavours, $4N_f=4$, where we will follow closely 
the corresponding subsection \ref{ex:massdensity} at weak non-Hermiticity.
Higher order correlations as well as more flavours follow along the same lines.
We start by defining how to rescale the masses 
\be
\sqrt{N}\, m_f\ \equiv\ \eta_f\ ,\ \ f=1,\ldots,N_f\ ,
\label{massstr}
\ee
in analogy to the eigenvalues in eq. (\ref{microstrong}). In order to derive
the massive spectral density we could simply take the limit $\al\to\infty$ of
the expressions in eqs. (\ref{rhoKweak}) and (\ref{deltarhoweak}) at weak
non-Hermiticity, together with the results from appendix \ref{Appdiff}. 

In order to give more compact expressions we derive results using
the alternative form of the strong density eq. (\ref{rhoKstr2}) and the
corresponding kernel. The massive spectral density follows again from
eq. (\ref{rhomassexpl}) at finite-$N$. The kernel at strong non-Hermiticity we
use is obtained from eq. (\ref{prekstrong}) by inserting eq. (\ref{intid}):
\be
\kappa_{strong}(\xi,\zeta^*) \ =\ 
\frac{N^{2\nu+2}}{4\pi}
\frac{1}{(\xi\zeta^*)^{2\nu}}
\int_0^1dr \frac{1}{\sqrt{1-r^2}}\,I_{2\nu}(r\xi_S\zeta_S^*)
\sinh\left(\frac12\sqrt{1-r^2}\,(\xi_S^2-\zeta_S^{*\,2})\right).
\label{prekstr2}
\ee
Expanding again in the imaginary part of the mass $\eps\to0$, setting
$\eta=\eta_x+i\eps$ we easily obtain for the denominator of
eq. (\ref{rhomassexpl}) 
\be
\lim_{\epsilon\to0} \kappa_{strong}(i\eta_x-\epsilon,-i\eta_x-\epsilon) \ =\ 
\frac{N^{2\nu+2}}{4\pi}\frac{1}{(\eta_x)^{4\nu}}
(-2i)\,\eps\,\eta_x \int_0^1dr I_{2\nu}(r\eta_x^2) 
\ +\ {\cal O}(\epsilon^2) \ .
\label{prekstrden}
\ee
The same expansion can be done for $ \kappa_{strong}(\xi,\pm\eta^{(*)})$ and
its conjugates, leading to the following expression for the massive spectral
density at strong non-Hermiticity:
\be
\rho_{strong}^{(4)}(\xi) \ =\  \rho_{strong}^{(0)}(\xi)\ -\
\Delta\rho_{strong}^{(4)}(\xi)\ ,
\ee
\label{rhostrmass} 
with the quenched density given by eq. (\ref{rhoKstr2}). The correction term
reads  
\bea
&&\Delta\rho_{strong}^{(4)}(\xi) \ =\  -
\frac{1}{4\pi} (\xi^{\ast\,2}-\xi^2)|\xi|^2
K_{2\nu}\left({|\xi|^2}\right)
\left\{
\int_0^1\frac{dr}{\sqrt{1-r^2}}J_{2\nu}(r\xi\eta)
\sinh[\frac12\sqrt{1-r^2}\,(\xi^2+\eta^2)] 
\right.
\nn\\
&&\times\left( 
\int_0^1\!\!\!\!\frac{dr\,r\xi^*}{\sqrt{1-r^2}}J_{2\nu+1}(r\xi^*\eta)
\sinh[\frac12\sqrt{1-r^2}(\xi^{*\,2}+\eta^2)]
-\!\!
\int_0^1\!\!\!\!
dr\eta J_{2\nu}(r\xi^*\eta)\cosh[\frac12\sqrt{1-r^2}(\xi^{*\,2}+\eta^2)]\!
\right)\nn\\
&&\ \ \ \ -\ (\xi\leftrightarrow\xi^*)\left\}\times
\left[
\eta\int_0^1dr\ I_{2\nu}(r\eta^2)\right.
\right]^{-1}\!.
\label{rhoKstrmass}
\eea
We have again called $\eta=\eta_x$ which is now real. Before discussing the 
massive spectral density let us also give some examples for
partition functions. Taking the strong non-Hermiticity 
limit of eq. (\ref{Z4N}) at $4N_f=4$ 
we simply have to insert the expansion eq. (\ref{prekstrden}) to obtain
\be
{\cal Z}_{strong}^{(4)}(\eta)\ \sim\ \int_0^1dr I_{2\nu}(r\eta^2) 
\ ,
\label{Z4str}
\ee
up to suitable normalisation constants. Since we did not compute the
asymptotic even skew-orthogonal polynomials at generic strong non-Hermiticity 
$0<\mu\leq1$ we cannot directly use eq. (\ref{Z2N}) for $2N_f=2$. 
We can circumvent this
problem by taking the infinite $\al$ limit on eq. (\ref{Z2weak}), leading to
the following
\be
{\cal Z}_{strong}^{(2)}(\eta)\ =\ 
\lim_{\al\to\infty}{\cal Z}_{weak}^{(2)}(\eta=\eta_w/\al\sqrt{2})\ \sim\ 
\mbox{e}^{\frac{\eta^2}{2}} I_\nu\left(\frac{\eta^2}{2}\right)\ .
\label{Z2str}
\ee

Let us come back to the spectral density eq. (\ref{rhoKstrmass}). 
Below we show a few
examples for the influence of the mass parameter at maximally strong
non-Hermiticity. 
\begin{figure}[-h]
\centerline{
\epsfig{figure=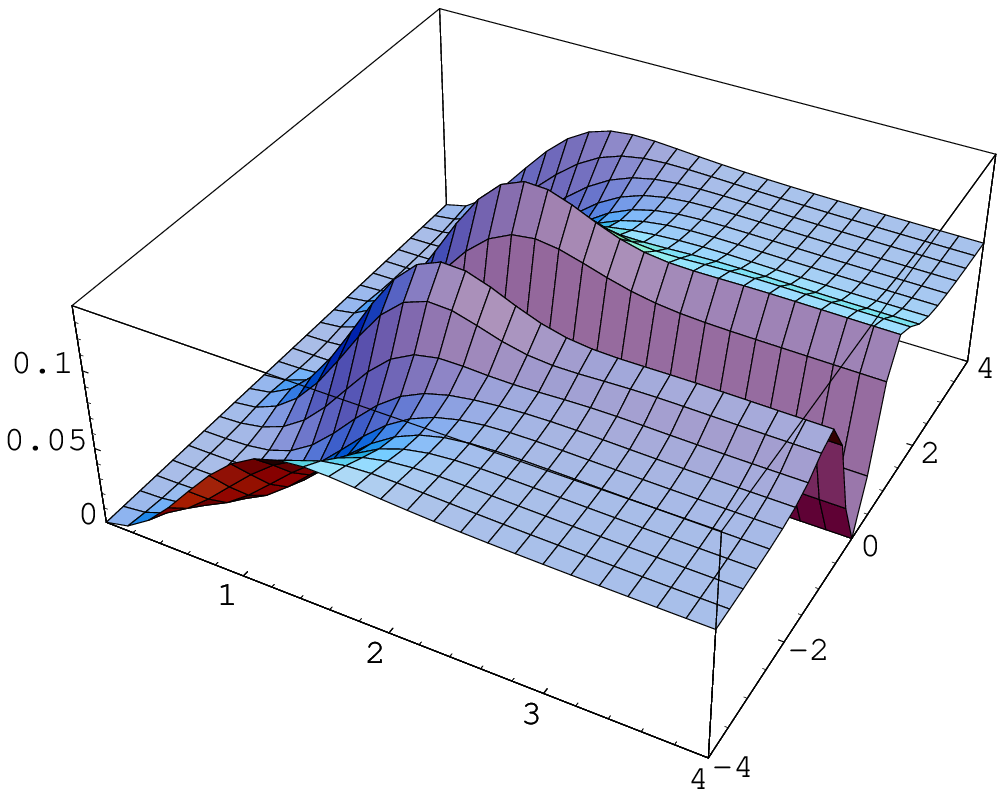,width=20pc}
\epsfig{figure=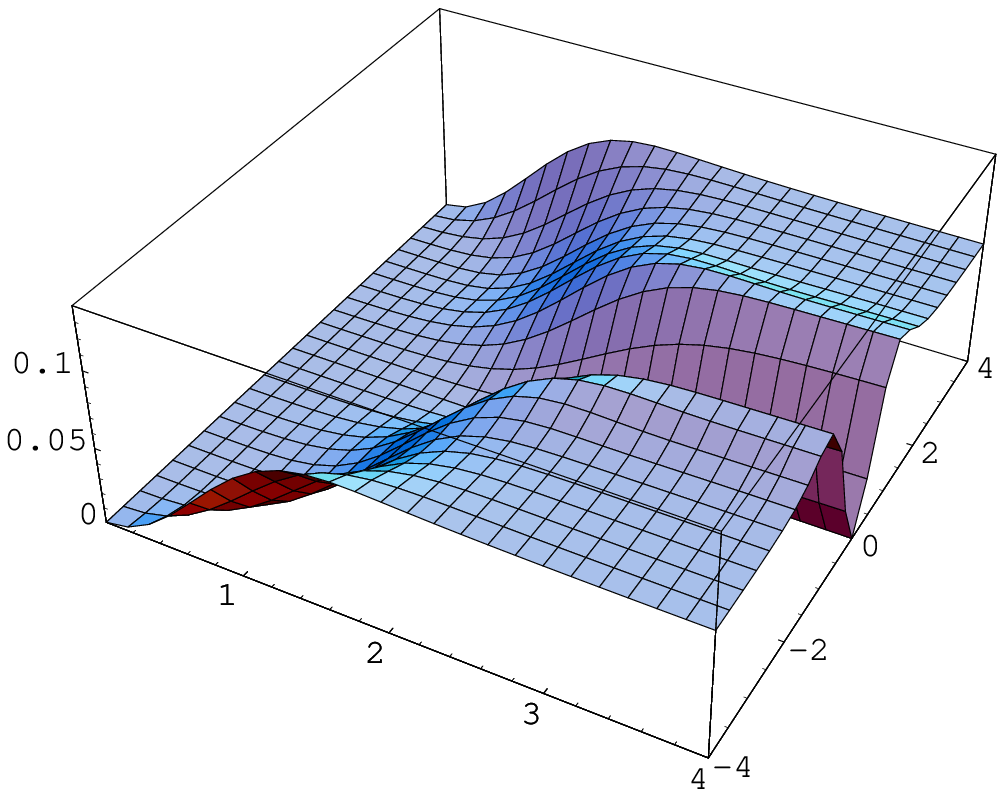,width=20pc}
\put(-460,10){$\re\ \xi$}
\put(-270,50){$\im\ \xi$}
\put(-480,180){$\rho_{strong}^{(4)}(\xi)$}
\put(-210,10){$\re\ \xi$}
\put(-20,50){$\im\ \xi$}
\put(-230,180){$\rho_{strong}^{(4)}(\xi)$}
}
\caption{The massive spectral density at at mass $\eta=2.5$
with $\nu=0$ (left) and $\nu=2$ (right)
.}
\label{rhostrmassplot}
\end{figure}
Compared to the quenched density in Fig. \ref{rhostrongplot} left there is an
additional level repulsion from the mass, when $\pm\im(\xi)=\eta$. Non-zero
topology leads to an additional level repulsion from the origin as can be seen
in Fig. \ref{rhostrmassplot} right, 
and both the effects of mass and $\nu$ can be clearly separated
(see also Fig. \ref{rhostrongplot} right).

\begin{figure}[-h]
\centerline{
\epsfig{figure=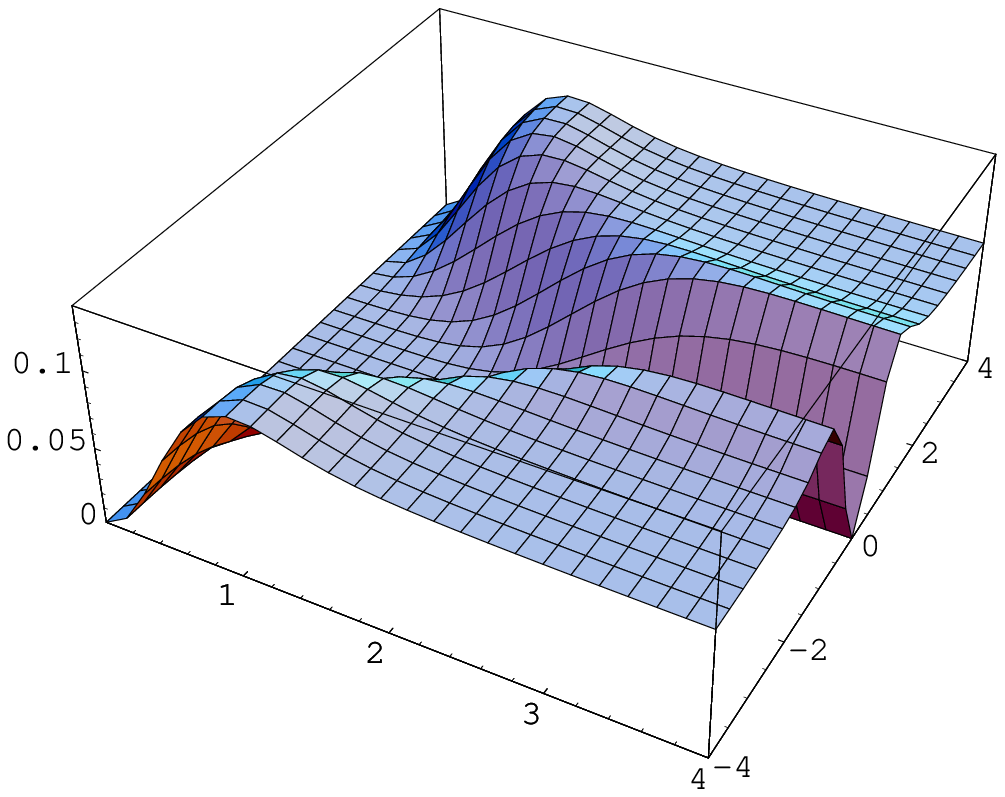,width=20pc}
\epsfig{figure=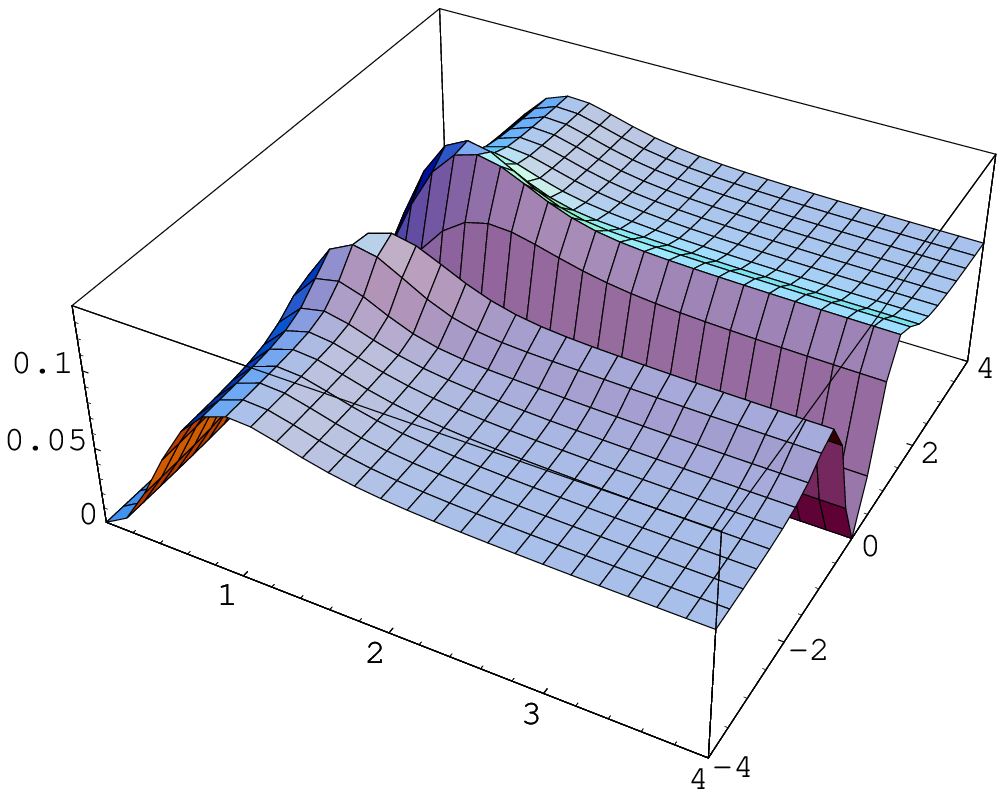,width=20pc}
\put(-460,10){$\re\ \xi$}
\put(-270,50){$\im\ \xi$}
\put(-480,180){$\rho_{strong}^{(4)}(\xi)$}
\put(-210,10){$\re\ \xi$}
\put(-20,50){$\im\ \xi$}
\put(-230,180){$\rho_{strong}^{(4)}(\xi)$}
}
\caption{The massive spectral density at at mass $\eta=0.5$ (left) and $\eta=8$
  (right), both with $\nu=0$ 
.}
\label{rhostrmassplot2}
\end{figure}

For smaller values of the mass parameter the density approaches
approximatively the quenched
density at $\nu=2$, as the mass acts as additional zero eigenvalues (comparing
Fig. \ref{rhostrmassplot2} left and 
Fig. \ref{rhostrongplot} right). On the other hand going to larger masses we
get back the quenched density as can be seen in Fig. \ref{rhostrmassplot2}
right. 

\begin{figure}[-h]
\centerline{
\epsfig{figure=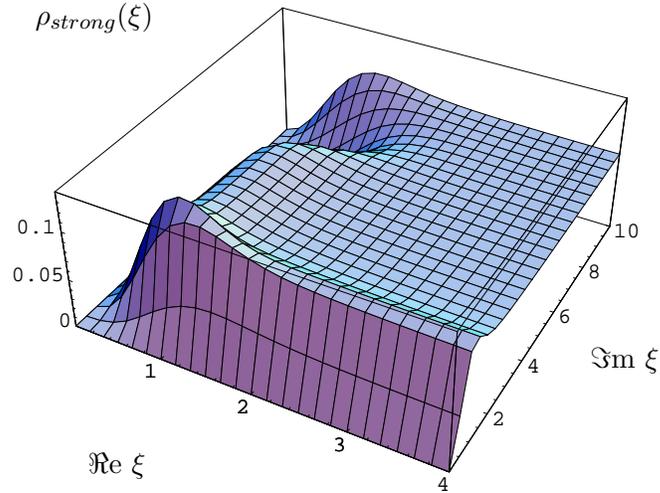,width=20pc}
\put(-210,10){$\re\ \xi$}
\put(-20,50){$\im\ \xi$}
\put(-230,180){$\rho_{strong}(\xi)$}
}
\caption{Part of Fig. \ref{rhostrmassplot2} (right) showing the first
  quadrant for a larger argument $\im(\xi)$. 
The level repulsion at the mass value $\xi=8i$ is clearly
  visible.}
\label{rhostrasym}
\end{figure}

Of course there is still level repulsion at the large mass value, in this case
$\eta=8$, as displayed in Fig. \ref{rhostrasym}. However, this does no longer
affect the density at the origin. Furthermore, there are no strong
oscillations observed here at the location of the mass, 
as seen in unquenched QCD \cite{AOSV}. Here, the situation is very
similar to phase quenched QCD also reported in \cite{AOSV} on the
matrix model in the $\beta=2$ symmetry class. 
These findings underline once more the importance of the presence or absence
of the sign problem on the level of the microscopic complex eigenvalue
correlations of the Dirac operator. 
We note in passing that because of the analytic relationship between the
correlations at strong non-Hermiticity and weak non-Hermiticity at large $\al$
similar figures could have obtained in this section by plotting 
the corresponding pictures at large enough $\al$. We have checked this
numerically in several cases for both the quenched and massive density, with
$\al=5$ (see also Fig. \ref{rhoweakmass2plot}). However, such a density at
weak non-Hermiticity will always vanish for large enough $\pm\im(\xi)$, at any
given $\alpha$. In contrast to that the density at strong non-Hermiticity
remains constant.

\sect{Relation to the Dirac operator spectrum with chemical 
potential}\label{Dirac}   

The purpose of this section is twofold. First, we 
give a more detailed discussion of the 
matrix representation of our complex eigenvalue model eq. (\ref{Zev})
in terms of two non-Hermitian quaternion real matrices, including mass terms.
This justifies to speak of a matrix model ensemble. 
It also gives an explicit realization of the joint probability distribution 
in eq. (\ref{Zev}) by computing the Jacobian that leads to complex eigenvalues.
In particular it yields the special weight function 
eq. (\ref{Kweight}) containing a $K$-Bessel function.
The second purpose is to relate the model to the Dirac operator 
spectrum of QCD-like gauge theories in the presence of a chemical potential.
To do this matching we need to compare the global symmetries of such Dirac 
operators to a random matrix representation.

The Dirac operator as it appears in four dimensional 
gauge theories can be represented 
by an anti-Hermitian block matrix, $\Dirac_{gauge}=\left(\begin{array}{cc} 
0&iW\\ iW^\dag&0\\ 
\end{array}\right)$, with non vanishing entries only on the off diagonal due 
to chiral symmetry. For different gauge groups and representations 
the off-diagonal blocks $W$ 
have either a orthogonal, unitary, or symplectic 
symmetry, as was pointed out in \cite{Jac3fold}. 
Adding a chemical potential $\mu$ for the quarks amounts to shifting 
the Dirac operator by $\Dirac_{gauge}\to \Dirac_{gauge}+\mu\gamma_0$
where $\gamma_0$ is a Dirac-$\gamma$ matrix. In a given basis it is 
represented as a block matrix as well, with unit elements on the off diagonals 
only \cite{Steph}. 
Being Hermitian this addition destroys the anti-Hermiticity property,
where the sum $\Dirac_{gauge}+\mu\gamma_0$ now has complex eigenvalues. 
The generalisation of the classification \cite{Jac3fold} to $\mu\neq0$ we use
here was made in \cite{HOV}.

If we replace the Dirac operator by a random matrix we have to satisfy
two criteria. First, $\Dirac_{gauge}$ may have exact zero eigenvalues
corresponding to different sectors of topological charge through an
index theorem. For a fixed number $\nu$ of zero modes, which we shall 
always consider here, we thus have to replace the blocks $W$ by 
rectangular matrices, of size $(N+\nu)\times N$. Second and more importantly, 
these rectangular matrices have to be in the same symmetry class 
as the corresponding gauge theory, consisting of 
real, complex or quaternion real elements \cite{Jac3fold,HOV}. The latter case 
which we consider here, also labelled by the Dyson index $\beta=4$, 
describes $SU(N_c)$ gauge theories in the adjoint representation. 
When representing the gauge theory on a lattice using staggered fermions
the corresponding symmetry is an  $SU(N_c=2)$ gauge theory in the 
fundamental representation.

The addition of a chemical potential in the matrix model can be done in 
different ways. It was first suggested \cite{Steph,HOV}
to add it as a constant times $\gamma_0$, the off diagonal unity matrix,
assuming this term to be diagonal in the random matrix space.
While this model successfully explained the failure of the quenched
approximation \cite{Steph} 
an analytical treatment of all complex 
eigenvalue correlations from orthogonal polynomials 
similar to the $\mu=0$ case was not achieved to 
date, apart from the quenched spectral density obtained 
using the replica method \cite{SplitVerb2}.
Therefore a different matrix model representation of the chemical 
potential term was suggested very recently in \cite{James} 
for the QCD symmetry class $\beta=2$. It assumes that the chemical 
potential term is non-diagonal in matrix space and is represented by 
a second, uncorrelated matrix which has the same symmetries 
as the one modelling the $\mu=0$ part. Making the model more complicated 
at first sight it has the remarkable feature of leading 
to a complex eigenvalue 
representation, which then allows for a treatment using (bi-)orthogonal 
polynomials \cite{James,AOSV}. 
A different model given only in terms of complex eigenvalues 
was first suggested in \cite{A02} and tested in quenched QCD lattice
simulations \cite{AW}. Although it is lacking the matrix 
symmetries of $\Dirac_{gauge}$ it asymptotically agrees with the results
of \cite{SplitVerb2,James} (see the weight eq. (\ref{K1/4weight}) vs. eq.
(\ref{Kweight})).

In the following we will adopt the strategy of \cite{James} 
and apply it to the complex $\beta=4$ symmetry class. Our corresponding 
matrix model is thus given in terms of two rectangular $(N+\nu)\times N$
matrices, $\Phi$ and $\Psi$, with quaternion real elements  
without further symmetry properties: 
\be
{\cal Z}_N^{(2N_f)}(\{m_f\}) \equiv
 \int d\Phi  d\Psi 
\exp\left[-N\Tr (\Phi^\dag \Phi\ +\ \Psi^\dag\Psi)\right]
\prod_{f=1}^{N_f} 
\det\left( \begin{array}{cc}
m_f\mbox{\bf 1} & i \Phi + \mu \Psi \\
i \Phi^{\dagger} + \mu \Psi^{\dagger} & m_f\mbox{\bf 1}
\end{array} \right) ,
\label{Z2MM}
\ee
where \mbox{\bf 1} is the quaternion unity element. Compared to
eq. (\ref{Zmatrix}) we have now included the quark masses. 
As a first step we define the linear combinations
\bea
C &\equiv& i \Phi\ + \mu \Psi \nn\\
D &\equiv& i \Phi^{\dagger} + \mu \Psi^{\dagger} \ ,
\label{CDdef}
\eea
with a trivial constant Jacobian.
In terms of these new matrices the partition function reads
\bea
{\cal Z}_N^{(2N_f)}(\{m_f\}) &\sim& 
 \int dC  dD 
\exp\left[-N\frac{(1+\mu^2)}{4\mu^2}\Tr (C^\dag C + DD^\dag )
-N\frac{(1-\mu^2)}{4\mu^2}\Tr (DC + C^\dag D^\dag)
\right]\nn\\
&&\times \prod_{f=1}^{N_f} 
\det\left( \begin{array}{cc}
m_f\mbox{\bf 1} & C\\
D               & m_f\mbox{\bf 1}
\end{array} \right) .
\label{Z2MMCD}
\eea
In this form we have only convergent integrals for $\mu\in\ [0,1]$ at
finite-$N$, as can be seen from eq. (\ref{Z2evfinal}) below. 
We will not discuss the possibility of phase transitions here, where for 
lattice simulations of this situation we refer to \cite{BMLP}.

The Dirac matrix inside the determinant has $\nu$ zero eigenvalues 
and $N$ complex eigenvalues which come in pairs of opposite sign 
as well as complex conjugate pairs.
The former is due to chiral symmetry and the latter is  due to the 
symplectic structure as the eigenvalues of a quaternion real matrix come
in complex conjugate pairs.
It is this symplectic structure that is responsible for the 
fact that the determinant will remain positive definite, without having 
a sign problem as unquenched QCD. It 
makes this symmetry class an ideal testing ground for lattice simulations 
with dynamical fermions in the presence of a chemical potential. 

In the next step we parametrise the matrices $C$ and $D$ as follows,
\bea
C&=& U(X+R)V \nn\\
D&=& V^\dag(Y+S)U^\dag \ ,
\label{CDdiag}
\eea
which is equivalent to making an independent Schur decomposition of $CD$ and
$DC$. 
Here $U$ and $V$ are symplectic matrices. The matrices 
$R$ and $S$ are upper triangular with real quaternion elements, and we have 
split off their diagonal parts $X$ and $Y$
which are also quaternion real, without 
containing the second and third quaternion unit (see e.g. \cite{Mehta} 
for explicit representations). They contain the complex 
eigenvalues of $C$ and $D$ which come in $N$ pairs $(x_k,x_k^*)$ for the matrix
$X$ and $(y_k,y_k^*)$ for the matrix $Y$ respectively. 
The parametrisation is not unique as under an additional 
diagonal unitary transformation the matrices remain triangular. It
only changes the upper triangular matrices $R$ and $S$ which we will 
integrate out afterwards. This symmetry 
can be used to restrict the matrices to be 
$V\in\ Sp(N)/U(1)^N$ and $U\in\
Sp(N+\nu)/(Sp(\nu)\times U(1)^N)$. The parametrisation eq. (\ref{CDdiag}) 
as well as the counting of degrees of freedom is discussed most explicitly in
appendix \ref{Jacobian}.
The relation to the original eigenvalues of the Dirac matrix is
\be
x_k y_k \ = \ -z_k^2\ ,
\label{xyz}
\ee
which explains that the eigenvalues of $CD$ come in complex conjugate 
pairs $(z_k^2,z_k^{*\,2})$, $k=1,\ldots N$.
The final Jacobian of the transformation computed in appendix \ref{Jacobian}
reads 
\be
\det\left(\frac{\partial(C,D)}{\partial(X,Y,U,V,R,S)}\right)\ =\ 
\prod_j^N |x_j|^{4\nu}
\prod_{k>l}^N |z_k^2-z_l^2|^2\ |z_k^2-z_l^{\ast\,2}|^2
\prod_{h=1}^N |z_h^2-z_h^{\ast\,2}|^2 \ ,
\label{Jacobi}
\ee
after a suitable ordering of the independent variables.
We can now integrate out the symplectic matrices $U$ and $V$, as well 
as the Gaussian integrals over $R$ and $S$ to obtain
\bea
{\cal Z}^{(2N_f)}_N(\{m_f\}) &\sim&  
\int \prod_{j=1}^N 
d^2x_j\, d^2y_j 
\exp\left[-N\frac{(1+\mu^2)}{4\mu^2}( |x_j|^2 +|y_j|^2)
+N\frac{(1-\mu^2)}{4\mu^2}( z_j^2 +z_j^{*\,2})\right] \nn\\
&&\times
|x_j|^{4\nu} \prod_{f=1}^{N_f} |m_f|^{2\nu}|z_j^2 + m_f^2|^2 
\prod_{k>l}^N |z_k^2-z_l^2|^2\ |z_k^2-z_l^{\ast\,2}|^2
\prod_{h=1}^N |z_h^2-z_h^{\ast\,2}|^2\ .
\label{Z2ev}
\eea
Here we have included the mass dependent factor $\sim|m_f|^{2\nu}$ 
in the normalisation, to the power of mass degeneracy 2 in $2N_f$. 
Although we have now achieved an eigenvalues 
representation in terms of $x_k$ and $y_k$ we still have twice as many 
degrees of freedom that what we aimed at, a single set of $N$ complex 
eigenvalues $z_k$. If we substitute $y_k$ by $z_k$: $y_k=-z_k^2/x_k$
we can integrate out the variables $x_k$ in analogy to \cite{James}, 
after decomposing them into 
polar coordinates,
\bea
{\cal Z}^{(2N_f)}_N(\{m_f\}) &\sim&  
\int  \prod_{j=1}^N d^2z_j \ 
K_{2\nu}\!\left(\frac{N(1+\mu^2)}{2\mu^2}|z_j|^2\right)
\exp\left[\frac{N(1-\mu^2)}{4\mu^2}(z_j^2+z_j^{*\,2})\right]
\label{Z2evfinal}
\\
&&\times 
|z_j|^{4\nu+2} \prod_{f=1}^{N_f} |m_f|^{2\nu}|z_j^2 + m_f^2|^2 
\prod_{k>l}^N |z_k^2-z_l^2|^2\ |z_k^2-z_l^{\ast\,2}|^2
\prod_{h=1}^N |z_h^2-z_h^{\ast\,2}|^2
\nn\\
&\sim&\left\langle \prod_{f=1}^{N_f} 
\det\left( \begin{array}{cc}
m_f & C\\
D & m_f
\end{array} \right)\right\rangle_{{\cal Z}^{(0)}_N}.\nn
\eea
This gives us a complex 
eigenvalue model as in eq. (\ref{Zev}) with the specific weight 
function eq. (\ref{Kweight}) plus mass terms, 
as well as a matrix representation
of the $q_{2N}(z)$ eq. (\ref{qevenvev}) for $N_f=1$.
We note that compared to the QCD symmetry class \cite{James} the index of the 
$K$-Bessel function is 
shifted: $\nu\to2\nu$.
The massive partition functions eq.(\ref{Zevm}) computed 
in section \ref{Rmass} are obtained 
by taking $4N_f$ flavours of twofold degenerate real masses $m_f$ in eq. 
(\ref{Z2MM}), or $2N_f$ flavour pairs of complex conjugate 
masses $(m_f,m_f^*)$ for complex masses: 
\be
{\cal Z}^{(4N_f)}_N(\{m_f\})\ \sim\ \left\langle \prod_{f=1}^{N_f} 
\det\left( \begin{array}{cc}
m_f & C\\
D & m_f
\end{array} \right)
\det\left( \begin{array}{cc}
m_f^* & C\\
D & m_f^*
\end{array} \right)
\right\rangle_{{\cal Z}^{(0)}_N}.
\label{charpol}
\ee
Let us stress that it is not this imposed 
degeneracy that is responsible for the
absence of a phase in the weight. The non-degenerate partition function
eq. (\ref{Z2evfinal}) has already a positive definite weight.
Eq. (\ref{charpol}) provides a determinant 
representation for the mass terms in eq. (\ref{Zevm}) for which we have 
been able to compute all complex correlation functions. The computation of
partition functions without 2-fold degenerate masses eq. (\ref{Z2evfinal})
remains an open problem for $2N_f>2$, where the special case 
$2N_f=2$ is given by
eq. (\ref{Z2N}). 
In the limit $\mu\to0$ the partition function eq. (\ref{charpol}) reduces to 
\be
\lim_{\mu\to0}{\cal Z}^{(4N_f)}_N(\{m_f\}) \sim  
\int_{-\infty}^{\infty}\prod_{j=1}^N   dx_j
\exp\left[-Nx_j^2\right]
|x_j|^{4\nu+3} \prod_{f=1}^{N_f} |m_f|^{4\nu}|x_j^2 + m_f^2|^4 
\prod_{k>l}^N |x_k^2-x_l^2|^4\ ,
\label{Zreal}
\ee
using eq. (\ref{deltanorm}).
This is the partition function of
the chGSE with real eigenvalues computed in \cite{AK,NN} and
successfully compared to lattice simulations with dynamical fermions at 
$\mu=0$ in \cite{BMW,AK}. 
It matches the corresponding chiral Lagrangian in the $\epsilon$-regime 
calculated for equal masses in \cite{SV}.

\sect{Conclusions}\label{disc}   

In this paper we have solved the complex extension 
of the chiral or Laguerre 
Symplectic Ensemble to non-Hermitian matrices. The solution 
was shown to be expressible in terms skew orthogonal 
polynomials for general weight functions 
in the complex plane. We gave explicit expression 
for finite-$N$ for all correlations functions of eigenvalues 
in the absence and presence 
of twofold degenerate mass terms, as well as for
characteristic polynomials. Two examples of a Gaussian Ginibre type 
weight and a
weight with a $K$-Bessel function were given. We proved that the
(skew-)orthogonal polynomials corresponding to the latter are 
Laguerre polynomials in the complex plane.

We investigated the large-$N$ limit of complex skew orthogonal 
Laguerre polynomials, 
first treating weak non-Hermiticity, 
and derived all correlation functions in terms of the limiting 
kernel and weight function. These finding were illustrated by 
examples of the microscopic spectral density without and with mass terms as
well as by partition functions, where we discussed 
the influence of both exact zero eigenvalues and the masses. 
The same analysis was then repeated for strong non-Hermiticity
where we had to solve two inhomogeneous differential equations to determine 
the limiting kernel.
The virtue of the weak non-Hermiticity limit is that it allows to extrapolate 
in terms of the weak non-Hermiticity parameter $\al$ 
between real eigenvalue correlations at $\al\to0$, and strong non-Hermiticity
in the limit $\al\to\infty$. This provided several analytical checks.

These mathematical results may have applications in various
physical systems with complex operators, and we have focused on the 
application to the Dirac operator 
spectrum in QCD-like gauge theories. We have identified 
the correct global symmetries of our ensemble by finding a two-matrix 
representation that matches with the symmetries 
of the Dirac operator with chemical 
potential in $SU(N_c)$ gauge theories in the adjoint representation, or 
$SU(N_c=2)$ gauge theories in the fundamental representation using 
staggered fermions in a lattice discretisation. Our two-matrix model 
representation led to the particular form of a $K$-Bessel weight function of 
complex eigenvalues, which is identical to that of the corresponding 
model in the QCD symmetry class. 
However, the solution we presented in terms of complex eigenvalues
holds for more general classes of weight functions as well.

It would be very interesting to verify
our predictions in lattice simulations with dynamical fermions in 
the corresponding theories with chemical potential.
In contrast to QCD these
do not suffer from a sign problem and thus can be solved using standard
techniques. This project is work in progress.

\indent

\noindent
\underline{Acknowledgements}: F. Basile, 
P. Damgaard, Y. Fyodorov, E. Kanzieper, B. Khoruzhenko, J. Osborn, 
K. Splittorff, R. Szabo, J. Verbaarschot and T. Wettig 
are thanked for useful conversations as well as 
E. Bittner, M.-P. Lombardo, H. Markum and R. Pullirsch
for getting involved into lattice simulations of this symmetry class.
The CERN Theory Division and the CIC Cuernavaca are thanked for hospitality 
where part of this work was being done. 
Partial support by a Heisenberg fellowship of the 
Deutsche Forschungsgemeinschaft and by a BRIEF award of Brunel
University is gratefully acknowledged.

\begin{appendix}

\sect{Orthogonality of complex Laguerre polynomials}
\label{proofLOP}

In this appendix we prove that Laguerre polynomials in the complex plane
are orthogonal with respect to the weight function eq. (\ref{Kweight}). 
This justifies our construction of skew orthogonal Laguerre polynomials 
described in subsection \ref{skewcomplexL}. 
We wish to show that
\be
\int d^2z\  |z|^{2\nu+2} K_\nu(a|z|^2) \exp\left[ \frac{b}{2}(z^2+z^{*\,2})
\right]\ L^\nu_k\left(\frac{a^2-b^2}{2b}z^2\right) 
L^\nu_l\left(\frac{a^2-b^2}{2b}z^{*\,2}\right) \ \sim\ \delta_{kl}\ ,
\label{delta}
\ee
holds for $a>b$ when integrating over the full complex plane , 
where we have introduced
\be
a\ =\ \frac{N(1+\mu^2)}{2\mu^2} \ \ \mbox{and}\ \ 
b\ =\ \frac{N(1-\mu^2)}{2\mu^2}\ .
\label{abdef}
\ee
Note that throughout this
appendix we use the shifted topological index $2\nu\to\nu$ as it 
appears in the complex chUE \cite{James},
compared to the weight eq. (\ref{Kweight}).
In the following we can restrict ourselves to the case $k\geq l$ in eq. 
(\ref{delta}), as then $k\leq l$ follows from complex conjugation. 
To show eq. (\ref{delta}) it is thus sufficient to prove 
\be
\int d^2z\ |z|^{2\nu+2} K_\nu(a|z|^2) \exp\left[ \frac{b}{2}(z^2+z^{*\,2})
\right]\ L^\nu_k\left(\frac{a^2-b^2}{2b}z^2\right) z^{*\,2l}\ =\ 0\ ,
\ \ \mbox{for all}\  k>l\ .
\label{deltamod}
\ee
The remaining case $k=l$ in eq. (\ref{delta}) is trivially non-vanishing, 
as we integrate over a positive quantity.
The corresponding norm of the $k$-th polynomial is computed below as well.
To set up our orthogonality proof \`a la Gram-Schmidt we first
compute the following general integral,
\bea
c_{kl} &=& \int d^2z\  |z|^{2\nu+2} K_\nu(a|z|^2) 
\exp\left[ \frac{b}{2}(z^2+z^{*\,2}) \right]\ z^{2k} z^{*\,2l} \nn\\
&=& \int_0^\infty dr\, r \int_0^{2\pi}d\phi\ r^{2\nu+2}  K_\nu(ar^2) 
\ \mbox{e}^{br^2\cos(2\phi)}\ r^{2k+2l}\ \mbox{e}^{2i\phi(k-l)}\nn\\
&=& \pi \int_0^\infty ds\ s^{k+l+\nu+1}  K_\nu(as) I_{k-l}(bs)\nn\\
&=& \pi\ 2^{k+l+\nu} \frac{(k+\nu)!k!}{(k-l)!} \frac{b^{k-l}}{a^{2k+2+\nu}}
F\left(1+k+\nu, 1+k;1+k-l;\frac{b^2}{a^2}\right) \ , \ \ k\geq l\ .
\label{ckl}
\eea
After introducing polar coordinates we can integrate out the angle to obtain
the modified $I$-Bessel function. Changing variables $r^2=s$ the
remaining integral can be found in terms of the hypergeometric function
\cite{Grad} where convergence is guaranteed due to $a>b$. We note that all the
moments 
$c_{kl}$ are real. The case $k\leq l$ can thus simply be obtained from  
interchanging $k$ and $l$: 
\be
c_{lk} \ =\ c_{kl}\ \ \mbox{for}\  l\geq k\ .
\ee
We first consider
the orthogonality of the zero-th polynomial
to $L_k^\nu$. 
For $l=0$ the hypergeometric function in eq. (\ref{ckl}) reduces to a
simple rational function, $F(n,\beta;\beta;z)=(1-z)^{-n}$, and we obtain
\be
c_{k0}\ = \  \pi\ 2^{k+\nu} (k+\nu)!\frac{a^\nu b^k}{(a^2-b^2)^{k+1+\nu}}\ .
\label{ck0}
\ee
Using the fact that the Laguerre polynomials have an explicit representation,
\be
L_k^\nu(x) \ = \ \sum_{m=0}^k (-)^m 
\binom{k+\nu}{k-m}
\frac{x^m}{m!}\ ,
\label{L}
\ee
we can simply write down the orthogonality relation between the $k$-th and 0-th
polynomial as 
\bea
\int d^2z\  w_K^{(\nu)}(z,z^*)\   
L^\nu_k\left(\frac{Nz^2}{1-\mu^2}\right)  1
&=&\sum_{m=0}^k (-)^m 
\binom{k+\nu}{k-m}
\frac{1}{m!} \left(\frac{a^2-b^2}{2b}\right)^m 
c_{m0} \nn\\
&=& \pi N^{-2-\nu} (1+\mu^2)^\nu \frac{(k+\nu)!}{k!} \sum_{m=0}^k (-)^m
\frac{k!}{(k-m)!\, m!} 
\nn\\
&\sim& 
\ \delta_{k0}\ .
\label{k0OP}
\eea
Here we have used the definitions (\ref{abdef}) as well as 
$\frac{N}{(1-\mu^2)}= \frac{a^2-b^2}{2b}$ 
and $a^2-b^2=N^2/\mu^2$. The insertion of the coefficient $c_{m0}$
eq. (\ref{ck0})  
leads to the binomial sum $(1-1)^k$ that vanishes for all $k>0$. For $k=0$ 
we obtain the norm of the 0-th polynomial reading
\be
|| L^\nu_0||^2\ = \ \pi N^{-2-\nu} (1+\mu^2)^\nu \ .
\label{0norm}
\ee
In the general case eq. (\ref{deltamod}) the strategy of our the proof will be
similar. We shall try to reduce the expressions to vanishing binomial sums. 
However, there will be remaining terms, which we have to show to cancel
as well. As a preliminary step we can use a transformation formula 
\cite{Grad} to show
that the hypergeometric functions in the coefficients $c_{kl}$ eq. (\ref{ckl})
terminate,
\be
F\left(1+k+\nu, 1+k;1+k-l;\frac{b^2}{a^2}\right)\ =\ 
F\left(-l-\nu, -l;1+k-l;\frac{b^2}{a^2}\right)\left(1- \frac{b^2}{a^2}
\right)^{-1-l-k-\nu} \ .
\label{hyper}
\ee
Because of $\nu\geq0$, and $k\geq l\geq 0$ 
being integers the hypergeometric function on
the right hand side terminates, containing $l+1$ terms.
Writing down the definitions we obtain for eq. (\ref{deltamod})
\bea
&&\int d^2z\  w_K^{(\nu)}(z,z^*)\   
L^\nu_k\left(\frac{Nz^2}{(1-\mu^2)}\right)  z^{*\,2l}\ =\nn\\
&&=\sum_{m=0}^k (-)^m 
\binom{k+\nu}{k-m}
\frac{1}{m!} \left(\frac{a^2-b^2}{2b}\right)^m 
\int d^2z\  w_K^{(\nu)}(z,z^*)  z^{2m}  z^{*\,2l}
\nn\\
&&=\sum_{m=0}^{l-1} (-)^m 
\binom{k+\nu}{k-m}
\frac{1}{m!} \left(\frac{a^2-b^2}{2b}\right)^m c_{lm}
\ +\ 
\sum_{m=l}^k (-)^m \binom{k+\nu}{k-m}
\frac{1}{m!} \left(\frac{a^2-b^2}{2b}\right)^m c_{ml}
\nn\\
&&
\equiv\  \Sigma_1\ +\ \Sigma_2\ ,
\label{sigdef}
\eea
where we have split the sum into two contributions for 
$m<l$ and $m\geq l$ in $c_{ml}$. In the sequel we manipulate the $\Sigma_1$ 
and $\Sigma_2$ separately by rearranging the hypergeometric functions, 
showing that they indeed cancel for $k>l$.
We start with the simpler second sum, using eq. (\ref{hyper}):
\bea 
\Sigma_2 &\equiv& \sum_{m=l}^k (-)^m  \binom{k+\nu}{k-m}
\frac{1}{m!} \left(\frac{a^2-b^2}{2b}\right)^m c_{ml}
\label{presig2}\\
&=& \sum_{m=l}^k (-)^m 
\frac{(k+\nu)!}{(k-m)!(\nu+m)!\,m!}
\left(\frac{a^2-b^2}{2b}\right)^m \pi\ 
2^{m+l+\nu} \frac{(m+\nu)!\,m!}{(m-l)!} \frac{b^{m-l}}{a^{2m+2+\nu}}
\nn\\
&&\times \left(1- \frac{b^2}{a^2}\right)^{-1-l-m-\nu}
F\left(-l-\nu, -l;1+m-l;\frac{b^2}{a^2}\right)
\nn\\
&=&
\frac{\pi\,(k+\nu)!}{N^{\nu+l+2}}\mu^2 \frac{(1+\mu^2)^{\nu+2l}}{(1-\mu^2)^l}
\sum_{m=l}^k \frac{(-)^m}{(k-m)!(m-l)!} \nn\\
&&\times\left(1+\frac{l(l+\nu)}{(1+m-l)}
\frac{b^2}{a^2}+\frac{l(l-1)(l+\nu)(l+\nu-1)}{(1+m-l)(2+m-l)\,2}\frac{b^4}{a^4}
+\ldots + \frac{(l+\nu)!(m-l)!}{\nu!\,m!}\frac{b^{2l}}{a^{2l}}
\right) .\nn 
\eea
In the next step we reduce the individuals sums sorted by powers of $b^2/a^2=
(1-\mu^2)^2/(1+\mu^2)^2$ to binomial sums by shifting the summation indices:
\bea 
\Sigma_2 &=& 
\frac{\pi\,(k+\nu)!}{N^{\nu+l+2}}\mu^2 \frac{(1+\mu^2)^{\nu+2l}}{(1-\mu^2)^l}
(-)^l\left\{ \sum_{m=0}^{k-l} (-)^m  \binom{k-l}{m}
\frac{1}{(k-l)!} 
\right.\nn\\
&-&\left. 
\sum_{m=1}^{k-l+1} (-)^m  \binom{k-l+1}{m}
\frac{l(l+\nu)}{(k-l+1)!}\frac{(1-\mu^2)^2}{(1+\mu^2)^2}
\pm\ldots +
(-)^l\sum_{m=l}^{k} (-)^m  \binom{k}{m}
\frac{l!(l+\nu)!}{k!\,l!\,\nu!}\frac{(1-\mu^2)^{2l}}{(1+\mu^2)^{2l}}
\right\}\nn\\
&=&\frac{\pi\,(k+\nu)!}{N^{\nu+l+2}}\mu^2 (1+\mu^2)^{\nu}
(-)^l\left\{ 0\cdot\frac{1}{(k-l)!} \frac{(1+\mu^2)^{2l}}{(1-\mu^2)^l}
+1\cdot\frac{l(l+\nu)}{(k-l+1)!}\frac{(1+\mu^2)^{2l-2}}{(1-\mu^2)^{l-2}}\ \pm\
\ldots  
\right.\nn\\
&-& (-)^n\sum_{m=0}^{n} (-)^m  \binom{k-l+1+n}{m}
\frac{l\cdots(l-n)(l+\nu)\cdots(l+\nu-n)}{(k-l+1+n)!(n+1)!}
\frac{(1+\mu^2)^{2l-2n-2}}{(1-\mu^2)^{l-2n-2}}\nn\\
&\pm&\left.\ \ldots + (-)^{l-1}\sum_{m=0}^{l-1} (-)^m  \binom{k}{m}
\frac{l!(l+\nu)!}{k!\,l!\,\nu!}
\frac{1}{(1-\mu^2)^{-l}}\right\}\ \ \mbox{for}\ k>l\ .
\label{sig2}
\eea
In the last step we have used 
$\sum_{m=n+1}^{k-l+1+n} (-)^m  \binom{k-l+1+n}{m}=
-\sum_{m=0}^{n} (-)^m  \binom{k-l+1+n}{m}$,
replacing incomplete binomial sums. In particular we need 
$k>l$ for the first term. Also we have multiplied in the
factor $\frac{(1+\mu^2)^{2l}}{(1-\mu^2)^l}$. The generic $n$-th 
term in the sum is explicitly displayed. In this form we can 
compare to
the first sum $\Sigma_1$, again after doing some manipulations:
\bea
\Sigma_1 &\equiv&\sum_{m=0}^{l-1} (-)^m 
\binom{k+\nu}{k-m}
\frac{1}{m!} \left(\frac{a^2-b^2}{2b}\right)^m c_{lm}
\label{presig1}\\
&=& \sum_{m=0}^{l-1} (-)^m
\frac{(k+\nu)!}{(k-m)!(m+\nu)!\,m!}\left(\frac{a^2-b^2}{2b}\right)^m  
\pi\ 2^{l+m+\nu} \frac{(l+\nu)!\,l!}{(l-m)!} \frac{b^{l-m}}{a^{2l+2+\nu}}\nn\\
&&\times\left(1- \frac{b^2}{a^2}\right)^{-1-m-l-\nu}
F\left(-m-\nu, -m;1+l-m;\frac{b^2}{a^2}\right) \nn\\
&=& \frac{\pi\,(k+\nu)!}{N^{\nu+l+2}}\mu^2 (1+\mu^2)^\nu (1-\mu^2)^l
(l+\nu)!\,l! \sum_{m=0}^{l-1}\frac{(-)^{m}}{(k-m)!(m+\nu)!\,m!(l-m)!}
\frac{(1+\mu^2)^{2m}}{(1-\mu^2)^{2m}}\nn\\
&&\times\left( 1+\frac{m(m+\nu)}{(1+l-m)}\frac{b^2}{a^2}
+\frac{m(m-1)(m+\nu)(m+\nu-1)}{(1+l-m)(2+l-m)\,2}\frac{b^4}{a^4}
+\ldots+
\frac{m!(m+\nu)!(l-m)!}{l!\,\nu!\,m!}\frac{b^{2m}}{a^{2m}}
\right).\nn
\eea
In contrast to eq. (\ref{presig2}) the length of each hypergeometric sum now
increases from term to term. Let us therefore write them out most
explicitly: 
\bea
&\Sigma_1&\ =\ \frac{\pi\,(k+\nu)!}{N^{\nu+l+2}}\mu^2(1+\mu^2)^\nu (1-\mu^2)^l
(l+\nu)!\,l! \label{sig1}\\
&\times& \!\!\!
\left\{ \frac{1}{k!\,\nu!\,l!}\cdot 1 - \frac{1}{(k-1)!(\nu+1)!(l-1)!}
\frac{a^2}{b^2}\!\left(1+ \frac{1(1+\nu)}{(1+l-1)} \frac{b^2}{a^2}
\right) \pm\ldots+
\frac{(-)^q}{(k-q)!(\nu+q)!\,q!(l-q)!}\frac{a^{2q}}{b^{2q}} 
\right.\nn\\
&\times& \!\!\!
\left(1+ \frac{q(q+\nu)}{(1+l-q)} \frac{b^2}{a^2}+\ldots +
\frac{q\cdots(q-p+1)(q+\nu)\cdots(q+\nu-p+1)}{(1+l-q)\ldots(p+l-q)p!} 
\frac{b^{2p}}{a^{2p}}
+\ldots +
\frac{
(q+\nu)!(l-q)!}{\nu!\,l!
} \frac{b^{2q}}{a^{2q}}
\right)\nn\\
&\pm& \!\!\!\ldots+\left.
\frac{(-)^{l-1}}{(k-l+1)!(\nu+l-1)!(l-1)!}\frac{a^{2(l-1)}}{b^{2(l-1)}}
\left(1+ \frac{(l-1)(l-1+\nu)}{2} \frac{b^2}{a^2}+\ldots +
\frac{
(l-1+\nu)!}{\nu!\,l!
} \frac{b^{2(l-1)}}{a^{2(l-1)}}
\right)\!
\right\}
\nn
\eea
In order to see that $\Sigma_1$ and $\Sigma_2$ cancel we now have to compare 
individual powers $\frac{a^2}{b^2}=\frac{(1+\mu^2)^2}{(1-\mu^2)^2}$. 
For the for highest power,
$\frac{a^{2(l-1)}}{b^{2(l-1)}}=\frac{(1+\mu^2)^{2(l-1)}}{(1-\mu^2)^{2(l-1)}}$,
it is easy to see that this is indeed the case, with only the first term in
the last line of $\Sigma_1$ in eq. (\ref{sig1}) contributing.
In contrast to that for the lowest power  $\frac{a^0}{b^0}$ the last term 
in each hypergeometric sum in  
eq. (\ref{sig1}) does contribute, leading to 
\be
\Sigma_1:\ {\cal O}\left(\frac{a^0}{b^0}\right)\ =
\frac{\pi\,(k+\nu)!}{N^{\nu+l+2}}\mu^2 (1+\mu^2)^\nu (1-\mu^2)^l
(l+\nu)!\,l! (-)^l\frac{1}{l!\,\nu!}\sum_{q=0}^{l-1}\frac{(-)^q}{(k-q)!\,q!}\ .
\ee
This clearly cancels the last term in eq. (\ref{sig2}). To see that all other 
terms
cancel too we now write down all contributions to the generic power 
$\frac{(1+\mu^2)^{2l-2n-2}}{(1-\mu^2)^{2l-2n-2}}$ to be compared with that of 
$\Sigma_2$ in eq. (\ref{sig2}). In $\Sigma_1$ we thus have to pick all terms 
$\frac{a^{2q-2p}}{b^{2q-2p}}$ with $2q-2p=2l-2n-2$. Because of $n\leq l-1$ 
and $p\leq q$ only those hypergeometric sums with $2q\geq 2l-2n-2$ start
contributing. Eliminating $p=q-l+n+1$
we obtain the following contribution
\bea
\Sigma_1:\ {\cal O}\left(\frac{a^{2l-2n-2}}{b^{2l-2n-2}}\right)
&=&\frac{\pi\,(k+\nu)!}{N^{\nu+l+2}}\mu^2(1+\mu^2)^\nu (1-\mu^2)^l
(l+\nu)!\,l!\frac{(1+\mu^2)^{2l-2n-2}}{(1-\mu^2)^{2l-2n-2}}\nn\\
&&\times\sum_{q=l-n-1}^{l-1}
\frac{(-)^{q}}{(k-q)!(l-n-1)!(l-n-1+\nu)!(n+1)!(q-l+n+1)!}\nn\\
&=& \frac{\pi\,(k+\nu)!}{N^{\nu+l+2}}\mu^2
\frac{(1+\mu^2)^{\nu+2l-2n-2}}{(1-\mu^2)^{l-2n-2}}
\frac{(-)^{n-l+1}(l+\nu)!l!}{(l-n-1)!(l-n-1+\nu)!(n+1)!}\nn\\
&&\times\sum_{q=0}^{n}\frac{(-)^q}{(k-q-l+n+1)!\,q!}\ ,
\eea
where we have shifted summation variables in the last step. This term clearly
cancels the corresponding term in $\Sigma_2$ in the last but one line of 
eq. (\ref{sig2}). This ends our proof of eq. (\ref{deltamod}) and therefore of
the orthogonality of the Laguerre polynomials in the complex plane. 

What remains to be computed 
are the norms of the $k$-th Laguerre polynomial, which
can be obtained as follows. The only step where we used $k>l$ was when
producing a zero in the first term of $\Sigma_2$ in eq. (\ref{sig2}), deducing
$(1-1)^{k-l}=0$. For $k=l$ this is no longer true and this term does
contribute, leading to the only non-vanishing contribution from
$\Sigma_1+\Sigma_2$. Because of the orthogonality only 
$L_k^\nu(z^2)z^{*\,2k}$ will contribute to $||L^\nu_k||^2$. To obtain
the norm for the polynomials in monic normalisation,
\be
P_k(z)\ \equiv\ (-)^k\frac{k!}{N^k}(1-\mu^2)^kL^\nu_k\left(
\frac{Nz^2}{1-\mu^2}\right)\ ,
\ee
we still have to normalise 
eq. (\ref{sigdef}) appropriately,
leading to 
\bea
||P_k||^2 &=& (-)^k\frac{k!}{N^k}(1-\mu^2)^k
\int d^2z\  w_K^{(\nu)}(z,z^*)\   
L^\nu_k\left(\frac{Nz^2}{1-\mu^2}\right)  z^{*\,2k}\nn\\
&=&
\frac{\pi\,k!(k+\nu)!}{N^{\nu+2k+2}}\mu^2 (1+\mu^2)^{\nu+2k}\ ,
\label{Knorm}
\eea
which is our final result on orthogonal Laguerre polynomials in the complex
plane. 

In order to clarify the issue of orthogonality of complex Laguerre
polynomials with respect to different weight functions 
we have explicitly checked that $L_0^\nu$ and $L_1^\nu$ are {\it not} 
orthogonal with respect to the exponential Gaussian weight in 
eq. (\ref{K1/4weight}) times $|z|^{2\nu+1}$ for a
general value of $\nu$, as was claimed in \cite{A02}. The integral is given by
a linear combination of Legendre functions of index $\nu+\frac12$ and
$\nu+\frac52$ which does not vanish in general. Only in the special case of
$\nu=\pm\frac12$ where the Laguerre polynomials are related to Hermite the
orthogonality holds. This does not come as a surprise now in view of the 
simplification of the $K$-Bessel weight eq. (\ref{Kweight}) to
eq. (\ref{K1/4weight}) in that case.

For comparison we shall also give the norms for the Gaussian weight
eq. (\ref{Gweight}). It trivially holds that for all rotational invariant 
weight functions $w(z,z^*)=w(|z|)$ 
the monic powers $z^k$ form a set of orthogonal polynomials 
\be
\int d^2z\  w(|z|) z^k\,z^{*\,l}\ \sim\ \delta_{kl}\ .
\ee
For different weight functions 
only the corresponding norms differ, which then enter the kernel, or the 
construction of
the skew orthogonal polynomials as in section \ref{skewcomplexL}.
For the Gaussian weight eq. (\ref{Gweight}), 
\be
w_G^{(\nu)}(z,z^*)\ =\ |z|^{2\nu+1} \exp[-N|z|^2] z^{2k}\ ,
\ee
with shifted $2\nu\to\nu$
one easily obtains the following result, after changing
to polar coordinates 
\be
||P_k^G||^2\ =\ \int d^2z\  |z|^{2\nu+1} \exp[-N|z|^2] z^{2k}\,z^{*\,2k}\ =\ 
\pi\frac{\Gamma\left(2k+\nu+\frac32\right)}{N^{2k+\nu+\frac32}} \ .
\label{Gnorm}
\ee


\sect{Differential equation for the kernel at strong non-Hermiticity}
\label{Appdiff}

\subsection{Construction of the differential equation}

In order to derive the differential equation we first rewrite the 
coefficients inside the pre-kernel eq. (\ref{kstrdef}) in a more symmetric 
way. Using the following identity for the $\Gamma$-function called doubling
formula \cite{Grad},
\be
\frac{\Gamma(z)}{\Gamma(2z)}\ =\ 
\frac{(2\pi)^{\frac12}}{2^{2z-\frac12}\Gamma(z+\frac12)}\ ,
\label{gammarel}
\ee
we obtain
\be
\left.\kappa_{strong}(\xi,\zeta^*)\right|_{\mu=1}\ =\ 
\frac{N^{2\nu+2}}{2^{4\nu+4}\sqrt{\pi}}
\sum_{k=0}^\infty\sum_{j=0}^k
\frac{(\xi^{4k+2}\zeta^{*\,4j}
\ -\ \xi^{4j}\zeta^{*\,4k+2})
}{2^{3(k+j)}(2k+1)!!(2j)!!\Gamma(k+\frac12+\nu+1)\Gamma(j+\nu+1)}
\ .
\label{kstrsym}
\ee
It is convenient to change variables,
\bea
u&=& \frac12 \xi^2\ ,\nn\\
v&=& \frac12 \zeta^{*\,2}\ ,
\label{UVdef}
\eea
and to define the following functional which is antisymmetric 
in the variables $u,v$,  
\be
\sigma(u,v)\ \equiv\ \sum_{k=0}^\infty\sum_{j=0}^k\left[
\frac{u^{2k+1}v^{2j}\ -\ v^{2k+1}u^{2j}}
{2^k\Gamma(k+\nu+\frac32)(2k+1)!! 2^j\Gamma(j+\nu+1)(2j)!!}\right].
\label{sigmadef}
\ee
This leads to 
\be
\kappa_{strong}(\xi,\zeta^*)|_{\mu=1}\ =\ 
\frac{N^{2\nu+2}}{2^{4\nu+3}\sqrt{\pi}}
\  \sigma(u,v)\ .
\label{sigkrel}                 
\ee
Applying the operator $v^{-2\nu}\partial_v v^{2\nu+1}\partial_v$ to $\sig(u,v)$
relates to $v\,\sig(u,v)$. In detail we have
\bea
\left[v^{-2\nu}\partial_v v^{2\nu+1}\partial_v\right]\sig(u,v) &=&
\sum_{k=0}^\infty\sum_{j=0}^{k-1}
\frac{u^{2k+1}}{2^k\Gamma(k+\nu+\frac32)(2k+1)!!}
\frac{v^{2j+1}}{2^j\Gamma(j+\nu+1)(2j)!!} \label{dsig}\\
&&-\sum_{k=0}^\infty\sum_{j=0}^k
\frac{v^{2k}}{2^{k-1}\Gamma(k+\nu+\frac12)(2k-1)!!}
\frac{u^{2j}}{2^j\Gamma(j+\nu+1)(2j)!!}\ ,
\nn
\eea
where we have shifted the sum over $j$ in the first term and denote $(-1)!!=1$.
This compares to multiplication by $v$ reading as follows
\bea
v\ \sig(u,v) &=&
\sum_{k=0}^\infty\sum_{j=0}^k\frac{u^{2k+1}}{2^k\Gamma(k+\nu+\frac32)(2k+1)!!}
\frac{v^{2j+1}}{2^j\Gamma(j+\nu+1)(2j)!!}\nn\\
&&-\sum_{k=1}^\infty\sum_{j=0}^{k-1}
\frac{v^{2k}}{2^{k-1}\Gamma(k+\nu+\frac12)(2k-1)!!}
\frac{u^{2j}}{2^j\Gamma(j+\nu+1)(2j)!!}\ ,
\label{Vsig}
\eea
where we have shifted the sum over $k$ in the second term. 
Subtracting this from eq. (\ref{dsig}) we obtain
\bea
\left[v^{-2\nu}\partial_v v^{2\nu+1}\partial_v-v\right]\sig(u,v) &=&
-\sum_{k=0}^\infty
\frac{(uv)^{2k+1}}{2^{2k}\Gamma(k+\nu+\frac32)\Gamma(k+\nu+1)(2k+1)!}
\nn\\
&&-\frac{2(uv)^0}{\Gamma(\nu+\frac12)\Gamma(\nu+1)}-
\sum_{k=1}^\infty 
\frac{(uv)^{2k}}{2^{2k-1}\Gamma(k+\nu+\frac12)\Gamma(k+\nu+1)(2k)!}\nn\\
&=& -\frac{2^{2\nu+1}}{\sqrt{\pi}} \sum_{k=0}^\infty 
\frac{(uv)^{k}}{\Gamma(k+2\nu+1)k!}\nn\\
&=&  -\frac{2^{2\nu+1}}{\sqrt{\pi}} 
\frac{I_{2\nu}(2\sqrt{uv})}{(uv)^\nu}\ .
\label{prediff}
\eea
In the first step we have again used the relation (\ref{gammarel})
and the last step follows upon the series representation of the $I$-Bessel
function. Going back to the kernel with eq. (\ref{sigkrel}) in the changed 
variables eq. (\ref{UVdef}) the differential equation (\ref{diff}) follows,
\be
\frac12\left[\partial_{\zeta^*}^2+\frac{(4\nu+1)}{\zeta^*}
\partial_{\zeta^*}-\zeta^{*\,2}\right]
\kappa_{strong}(\xi,\zeta^*)|_{\mu=1}\ =\ 
-\ \frac{N^{2\nu+2}}{4\pi}
\frac{I_{2\nu}\left(\xi\zeta^*\right)}{(\xi\zeta^*)^{2\nu}}\ .
\label{diffA}
\ee
The corresponding equation in the variable $u=\frac{\xi^2}{2}$ 
is obtained in the same way, 
or by simply using the antisymmetry of the function $\sigma(u,v)$
\be
\frac12\left[\partial_{\xi}^2+\frac{(4\nu+1)}{\xi}
\partial_{\xi}-\xi^{2}\right]
\kappa_{strong}(\xi,\zeta^*)|_{\mu=1}\ =\ 
+\ \frac{N^{2\nu+2}}{4\pi}
\frac{I_{2\nu}\left(\xi\zeta^*\right)}{(\xi\zeta^*)^{2\nu}}\ .
\label{diffB}
\ee


\subsection{Solution of the differential equation}

Next we seek for a solution of these two inhomogeneous differential equations.
The homogeneous equation can be easily solved, as it holds
\be
\left[v^{-2\nu}\partial_v v^{2\nu+1}\partial_v-v\right] \frac{I_\nu(v)}{v^\nu}
\ =\ 0\ \ ,
\ee
in both variables $u$ and $v$. The second independent solution 
in terms of $K$-Bessel functions is excluded because of the regularity of the
pre-kernel at the origin. We therefore conclude that the following product 
\be
\left[v^{-2\nu}\partial_v v^{2\nu+1}\partial_v-v\right] 
\left(\frac{I_\nu(u)I_\nu(v)}{(uv)^\nu}\right)
\ =\ 0\ = \ 
\left[u^{-2\nu}\partial_u u^{2\nu+1}\partial_u-u\right] 
\left(\frac{I_\nu(u)I_\nu(v)}{(uv)^\nu}\right) \ ,
\label{hom}
\ee
solves both homogeneous equations. 
However, this solution is symmetric in both arguments, in contrast to the 
antisymmetric pre-kernel. 
Each factor of the homogeneous solution has the 
following integral representation that will be useful below. 
In terms of the original variables it is given by
eq. (\ref{UVdef}) 
\be
I_{\nu}\left(\frac{\xi^2}{2}\right)\ = \ 
\frac{2}{\sqrt{\pi}}\ \mbox{e}^{\frac12\xi^2} 
\int_0^\infty dq \,\mbox{e}^{-q^2} J_{2\nu}(2q\xi) \ ,
\label{homint}
\ee
and similarly for $\zeta^*$.

Instead of making an Ansatz that solves the full set of inhomogeneous 
differential equations 
(\ref{diffA}) and (\ref{diffB})
we construct a solution by rewriting the right hand side 
as an integral and manipulating it until we obtain the correct differential 
operator acting on it. Omitting the constant pre-factor we start 
from the following integral identity,
\bea
\frac{I_{2\nu}\left(\xi\zeta^*\right)}{(\xi\zeta^*)^{2\nu}} &=&
\frac{4\mbox{e}^{\frac12(\xi^2+\zeta^{*\,2})}}{(\xi\zeta^*)^{2\nu}}
\int_0^\infty dq \,\mbox{e}^{-2q^2} q J_{2\nu}(2q\xi) J_{2\nu}(2q\zeta^*)
\label{inhomint1}\\
&=& \frac{-4\mbox{e}^{\frac12(\xi^2+\zeta^{*\,2})}}{(\xi\zeta^*)^{2\nu}}
\int_0^\infty dq \,\mbox{e}^{-q^2}\left[(1-2q^2+q\partial_q)J_{2\nu}(2q\xi)
\right]\int_0^q dp \,\mbox{e}^{-p^2}J_{2\nu}(2p\zeta^*)\nn\\
&=&\frac{-4\mbox{e}^{\frac12(\xi^2+\zeta^{*\,2})}}{(\xi\zeta^*)^{2\nu}}
\frac12\left[\partial_\xi^2+\frac{1}{\xi}\partial_\xi -\frac{4\nu^2}{\xi^2}
+2\xi\partial_\xi +2\right]\int_0^\infty dq \,\mbox{e}^{-q^2}J_{2\nu}(2q\xi)
\int_0^q dp \,\mbox{e}^{-p^2}J_{2\nu}(2p\zeta^*)
\nn\\
&=& \frac12\left[\partial_\xi^2+\frac{(4\nu+1)}{\xi}\partial_\xi-\xi^2\right]
\left( \frac{-4\mbox{e}^{\frac12(\xi^2+\zeta^{*\,2})}}{(\xi\zeta^*)^{2\nu}}
\int_0^\infty dq \,\mbox{e}^{-q^2}J_{2\nu}(2q\xi)
\int_0^q dp \,\mbox{e}^{-p^2}J_{2\nu}(2p\zeta^*)
\right).\nn
\eea
Here we first integrated by parts $\mbox{e}^{-q^2}J_{2\nu}(2q\zeta^*)$, with a
vanishing surface term. Next we replace 
$q\partial_q J_{2\nu}(2q\xi)=\xi\partial_\xi J_{2\nu}(2q\xi)$, 
and then use that $J_{2\nu}(2q\xi)$ satisfies a Bessel differential equation:
\be
\frac12\left[\partial_\xi^2+\frac{1}{\xi}\partial_\xi -\frac{4\nu^2}{\xi^2}
\right]J_{2\nu}(2q\xi)\ =\ -2q^2J_{2\nu}(2q\xi)\ .
\ee
We can now take the differential operator out of the integral
and interchange it with the $\xi$-dependent pre-factor, giving us precisely 
the 
desired differential operator. This provides a solution of the 
second differential equation (\ref{diffB}), after multiplying with 
the appropriate constant. A solution of the corresponding equation
(\ref{diffA}) for $\zeta^*$ follows along the same lines 
\be
-\frac{I_{2\nu}\left(\xi\zeta^*\right)}{(\xi\zeta^*)^{2\nu}} \ =\ 
\frac12\left[\partial_{\zeta^*}^2+\frac{(4\nu+1)}{\zeta^*}\partial_{\zeta^*}
-\zeta^{*\,2}\right]
\left( \frac{4\mbox{e}^{\frac12(\xi^2+\zeta^{*\,2})}}{(\xi\zeta^*)^{2\nu}}
\int_0^\infty dq \,\mbox{e}^{-q^2}J_{2\nu}(2q\zeta^*)
\int_0^p dp \,\mbox{e}^{-p^2}J_{2\nu}(2p\xi)
\right).
\label{inhomint2}
\ee
However, none of two solutions is antisymmetric in the variables $\xi$ and 
$\zeta^*$ to be proportional to the pre-kernel, 
nor do they obviously solve both differential equations with the 
different signs simultaneously. 
We will therefore add and subtract a multiple of 
the homogeneous equation eq. (\ref{hom}) in its integral representation
eq. (\ref{homint}) to eq.  (\ref{inhomint1}) and  from (\ref{inhomint2}),
respectively, in order to obtain an antisymmetric solution of both
differential equations. The requirement of antisymmetry of the pre-kernel
fixes the solution uniquely.
It can be seen that the solution is given as follows
\bea
&&\frac{2\mbox{e}^{\frac12(\xi^2+\zeta^{*\,2})}}{(\xi\zeta^*)^{2\nu}}
\int_0^\infty dq \int_0^q dp\,\mbox{e}^{-q^2-p^2}
\left[ J_{2\nu}(2q\zeta^*)  J_{2\nu}(2p\xi)\ -\ 
J_{2\nu}(2q\xi)  J_{2\nu}(2p\zeta^*)\right]
\label{diffsol}\\
&=& 2\left(\frac{\mbox{e}^{\frac12\zeta^{*\,2}}}{\zeta^{*\,2\nu}}
\int_0^\infty dq \,\mbox{e}^{-q^2} J_{2\nu}(2q\zeta^*)\right)
\left(\frac{\mbox{e}^{\frac12\xi^{2}}}{\xi^{2\nu}}
\int_0^\infty dp \,\mbox{e}^{-p^2} J_{2\nu}(2p\xi)\right)\nn\\
&&-\ \frac{4\mbox{e}^{\frac12(\xi^2+\zeta^{*\,2})}}{(\xi\zeta^*)^{2\nu}}
\int_0^\infty dq \int_0^q dp\,\mbox{e}^{-q^2-p^2}
J_{2\nu}(2q\xi)  J_{2\nu}(2p\zeta^*)\nn\\
&=&  \frac{4\mbox{e}^{\frac12(\xi^2+\zeta^{*\,2})}}{(\xi\zeta^*)^{2\nu}}
\int_0^\infty dq \int_0^q dp\,\mbox{e}^{-q^2-p^2}
J_{2\nu}(2q\zeta^*)  J_{2\nu}(2p\xi)\nn\\
&&-\ 2\left(\frac{\mbox{e}^{\frac12\xi^2}}{\xi^{2\nu}}
\int_0^\infty dq \,\mbox{e}^{-q^2} J_{2\nu}(2q\xi)\right)
\left(\frac{\mbox{e}^{\frac12\zeta^{*\,2}}}{\zeta^{*\,2\nu}}
\int_0^\infty dp \,\mbox{e}^{-p^2} J_{2\nu}(2p\zeta^*)\right) .
\nn
\eea
The first line is antisymmetric in its arguments $\xi$ and $\zeta^*$. 
In the subsequent lines we have made an integration by parts of 
either $\mbox{e}^{-q^2}J_{2\nu}(2q\zeta^*)$ or 
$-\mbox{e}^{-q^2}J_{2\nu}(2q\xi)$ in order to show, that this expression 
solves both differential equations, (\ref{inhomint1}) and  (\ref{inhomint2})
plus or minus a solution of the homogeneous equation.
The full solution for the pre-kernel including all factors thus reads
\bea
\kappa_{strong}(\xi,\zeta^*)|_{\mu=1}&=& 
\frac{N^{2\nu+2}}{2\pi}
\frac{1}{(\xi\zeta^*)^{2\nu}}\ \mbox{e}^{\frac12(\xi^2+\zeta^{*\,2})}
\label{fullkstr}\\
&&\times\int_0^\infty dq \int_0^q dp\ \mbox{e}^{-q^2-p^2}
\left[
J_{2\nu}(2q\zeta^*) J_{2\nu}(2p\xi) \ -\  
J_{2\nu}(2q\xi)J_{2\nu}(2p\zeta^*)
\right] \ .
\nn
\eea

As a check and alternative way to derive this result we
take the limit $\al\to\infty$ of the pre-kernel at
weak non-Hermiticity eq. (\ref{prekweak}). Substituting 
first $t$ by $p^2=2st\al^2$ and then $s$ by  
$q^2=2s\al^2$  
we obtain from eq. (\ref{prekweak})
\bea
\lim_{\al\to\infty}
\kappa_{weak}(\xi,\zeta^*) 
&=& \lim_{\al\to\infty}
\frac{4}{\al^6}N^{4\nu+4}\frac{2^{2\nu-3}}{(\xi\zeta^*)^{2\nu}}
\label{alooint}\\
\times&&
\int_0^{\al\sqrt{2}}\!\!\!dq \int_0^q dp\ \mbox{e}^{-q^2-p^2}
\left[  J_{2\nu}\left(\frac{2q\zeta^*}{\al\sqrt{2}}\right)
J_{2\nu}\left(\frac{2p\xi}{\al\sqrt{2}}\right)- 
J_{2\nu}\left(\frac{2q\xi}{\al\sqrt{2}}\right) 
J_{2\nu}\left(\frac{2p\zeta^*}{\al\sqrt{2}}\right) \right].\nn
\eea
If we then replace the first integral 
$\int_0^{\al\sqrt{2}}dq\to \int_0^\infty dq$, 
keeping $\frac{\zeta^*}{\al\sqrt{2}}$
and  $\frac{\xi}{\al\sqrt{2}}$ fixed, we reproduce the strong kernel
eq. (\ref{fullkstr}) up to constants and the exponential pre-factor. Therefore
we cannot directly identify the kernels, but have to take into account the
weight function as well when mapping correlation functions from the weak to
the strong limit. From eq. (\ref{weakKweight}) we thus obtain
\bea
\lim_{\al\to\infty}2N
\rho_{weak}(\xi) &=& \frac{1}{4\mu^2}
 (\xi_S^{\ast\,2}-\xi_S^2)|\xi_S|^2
K_{2\nu}\left({|\xi_S|^2}\right)\mbox{e}^{\frac{1}{2}(\xi_S^2+\xi_S^{*\,2})}
\label{rhoweakstr}\\
&\times&\int_0^\infty dq 
\int_0^q dp\ \mbox{e}^{-q^2-p^2}
\left[J_{2\nu}\left({2q\xi_S^\ast}\right)J_{2\nu}\left({2p\xi_S}\right)
\ -\ J_{2\nu}\left({2p\xi_S^\ast}\right)J_{2\nu}\left({2q\xi_S}\right)\right]\
, \nn
\eea
exactly reproducing the strong density eq. (\ref{rhoKstr}) obtained from
eq. (\ref{fullkstr}). 
Here we have introduced the scaling variable
\be
\xi_S\ \equiv\ \frac{\sqrt{N}\, z}{\sqrt{2}\,\mu}\ =\
\frac{\xi}{\al\sqrt{2}}
\label{xis}
\ee
with $\xi$ being the scaling variable in the weak limit eq. (\ref{microweak}).
$\xi_S$ scales as the  scaling variable in the strong 
limit eq. (\ref{microstrong}).
The explicit factor $2N$ in front of the density is added since the weak and
strong density are rescaled with different powers of $N$, see
eqs. (\ref{microrhow}) and (\ref{microrhostr}).
The explicit $\mu$-dependent factor in front can be absorbed by defining a 
$\mu$-dependent strong scaling limit by $\xi_S$, and rescaling the density
accordingly. 
It can be easily seen that all higher eigenvalue correlation functions
at weak and strong non-Hermiticity can be matched in the same way 
when taking $\al\to\infty$.

We end this appendix by noting that another integral representation can be 
obtained from the $\al\to\infty$ limit of the weak pre-kernel 
eq. (\ref{prekweak}). First we substitute $s=u^2$ and $t=v^2$. Next we
interchange the integrals with respect 
to $u$ and $v$ and substitute $u$ by $y^2=u^2\al^2(1+v^2)$ to arrive at
\bea
&&\lim_{\al\to\infty}2\al^2 
\int_0^1 ds \int_0^1 \frac{dt}{\sqrt{t}}\ \mbox{e}^{-2s(1+t)\al^2}
\left(J_{2\nu}(2\sqrt{st}\ \xi)  J_{2\nu}(2\sqrt{s}\ \zeta^\ast)
\ -\  J_{2\nu}(2\sqrt{st}\ \zeta^*)J_{2\nu}(2\sqrt{s}\ \xi)\right)\nn\\
&&=
\lim_{\al\to\infty}4
\int_0^1\!dv \frac{1}{1+v^2}\int_0^{\al\sqrt{2(1+v^2)}}\!\!\!\!\!dy\,y
\,\mbox{e}^{-y}\left[  
J_{2\nu}\left(\frac{2yv\xi}{\al\sqrt{2(1+v^2)}}\right)
J_{2\nu}\left(\frac{2y\zeta^*}{\al\sqrt{2(1+v^2)}}\right)-
(\xi\leftrightarrow\zeta^*)
\right]
\nn\\
&&=2
\int_0^1dv \frac{1}{1+v^2}\,
I_{2\nu}\left(\frac{v\xi\zeta^*}{\al^2(1+v^2)}\right)
\left[
\mbox{e}^{-\frac{v^2\xi^2+\zeta^{*\,2}}{2\al^2(1+v^2)}}
\ -\ 
\mbox{e}^{-\frac{\xi^2+v^2\zeta^{*\,2}}{2\al^2(1+v^2)}}
\right].
\label{altstrint}
\eea
In the last step 
we have replaced $\int_0^{\al\sqrt{2(1+v^2)}}dy\to\int_0^\infty dy$
in which case the integral can be evaluated. Keeping the same ratios
$\frac{\xi}{\al\sqrt{2}}$ and 
$\frac{\zeta^*}{\al\sqrt{2}}$ fixed, we obtain the following useful integral
identity by comparing eqs. (\ref{alooint}) and (\ref{altstrint}): 
\bea
&&4\int_0^\infty dq 
\int_0^q dp\ \mbox{e}^{-q^2-p^2}
\left[
J_{2\nu}\left({2q\zeta_S^\ast}\right)J_{2\nu}\left({2p\xi_S}\right)
\ -\ 
J_{2\nu}\left({2q\xi_S}\right)J_{2\nu}\left({2p\zeta_S^\ast}\right)
\right]
\nn\\
&&=2\int_0^1dv \frac{1}{1+v^2}\,
I_{2\nu}\left(\frac{2v\xi_S\zeta_S^*}{(1+v^2)}\right)
\left[
\mbox{e}^{-\frac{v^2\xi_S^2+\zeta_S^{*\,2}}{(1+v^2)}}
\ -\ 
\mbox{e}^{-\frac{\xi_S^2+v^2\zeta_S^{*\,2}}{(1+v^2)}}
\right]\nn\\
&&=2\ \mbox{e}^{-\frac12(\xi_S^2+\zeta_S^{*\,2})}
\int_0^1dr \frac{1}{\sqrt{1-r^2}}\,I_{2\nu}(r\xi_S\zeta_S^*)
\sinh\left(\frac12\sqrt{1-r^2}\,(\xi_S^2-\zeta_S^{*\,2})\right).
\label{intid}
\eea
In the last step we have substituted $r=2v/(1+v^2)$. It is this last form
which is very convenient for a numerical evaluation in order to plot the
spectral density at strong non-Hermiticity.

\sect{Jacobian of the two-matrix model}
\label{Jacobian}

In this appendix we compute the Jacobian for the parametrisation
eq. (\ref{CDdiag})  
\bea
C&=& UTV \ \ \ ,\ \ T=X+R\nn\\
D&=& V^\dag QU^\dag \ ,\ \ Q=Y+S\ ,
\label{CDpara}
\eea
resulting from Schur decompositions of $CD$ and $DC$.
To this aim we proceed in two steps. First, we provide a labelling of the
independent degrees of freedom to obtain a block triangular form of the
Jacobi matrix in terms of quaternion matrix elements. Here we follow some
analogy to \cite{Mehta} appendix A.35 where a similar decomposition was given
for a single quadratic complex non-Hermitian matrix.
Second, we calculate the Jacobi
determinant, evaluating the diagonal blocks. 

To start let us match the independent degrees of freedom present in the
matrices $C,D$ on the one hand and $T,Q$ and $V,U$ on the other hand.
The matrices $C$ and $D$ are of size $(N+\nu)\times N$ and $N\times (N+\nu)$,
respectively. Both are quaternion real, with each matrix element containing 4
real parameters. This leads to the following total number of 
degrees of freedom (d.o.f.)
\be
\mbox{d.o.f.}\{C,D\}\ =\ 8N^2+8N\nu\ .
\label{CDdof}
\ee
On the other hand the matrix $T$ in the parametrisation eq. (\ref{CDpara}) is
triangular of size $(N+\nu)\times N$: only $T_{n\leq l}\neq0$. The diagonal
elements are of the following form \cite{Mehta}\footnote{Throughout the
  following we will {\it not} use summation conventions.}
\be
T_{l,l}\ \equiv \ t_{l,l}=
\left( \begin{array}{cc}
x_l & 0\\
0 & x^*_l
\end{array} \right),\ l=1,\ldots,N\ ,
\ee
and carry only 2 real d.o.f., the eigenvalues of $C$.
All other $T_{n< l}\neq0$ carry 4  real d.o.f.:
\be
\mbox{d.o.f.}\{T\}= 2N+4\frac{N(N-1)}{2}=2N^2\ .
\label{Tdof}
\ee
The same analysis holds for the  $N\times (N+\nu)$-matrix $Q$, 
$Q_{n\leq l}\neq0$, having $N\times\nu$ more non-vanishing elements:
\be
\mbox{d.o.f.}\{Q\}= 2N+4\frac{N(N-1)}{2}+ 4N\nu=2N^2+4N\nu\ .
\label{Qdof}
\ee
In the same way as above we define for the diagonal elements 
\be
Q_{l,l}\ \equiv \ q_{l,l}=
\left( \begin{array}{cc}
y_l & 0\\
0 & y^*_l
\end{array} \right),\ l=1,\ldots,N \ ,
\ee
containing the eigenvalues of $D$. The symplectic matrices $U$ and $V$ of size
$(N+\nu)^2$ and $N^2$ a priori carry $2(N+\nu)^2+(N+\nu)$ and $2N^2+N$
d.o.f., respectively. However, the parametrisation eq. (\ref{CDpara}) is not
unique. Replacing the quaternion elements in all matrices by $2\times2$ blocks
one can see that a diagonal unitary matrix $U'$ of size $2N$ leaves the
matrices $T$ and $Q$ triangular, keeping the eigenvalues unchanged:
\be
U\ \to\ U\tilde{U}'\ \ \mbox{and} \ \ V\ \to \ U^{'\,\dag}V
\label{U'sym}
\ee
where $\tilde{U}'$ is $U'$ extended to size $2(N+\nu)$ by the unity matrix.
Therefore $2N$ real parameters are redundant in
eq. (\ref{CDpara}). Furthermore, the whole lower right $\nu\times\nu$ sub-block
in $U$ can be projected out, subtracting a symplectic matrix with 
$2\nu^2+\nu$ d.o.f.. We are left with $V\in\ Sp(N)/U(1)^N$ and $U\in\
Sp(N+\nu)/(Sp(\nu)\times U(1)^N)$: 
\be
\mbox{d.o.f.}\{U\}+\mbox{d.o.f.}\{V\}
= 2(N+\nu)^2+(N+\nu) -N-(2\nu^2+\nu)+2N^2+N-N=4N^2+4N\nu
\label{UVdof}
\ee
which together with eqs. (\ref{Tdof}) and (\ref{Qdof}) 
adds up to eq. (\ref{CDdof}).

In oder to
label independent d.o.f. for the computation of the Jacobian it is
most convenient to define the following differentials:
\bea
dC&=& U\,\dc\, V \ \ ,\ \ \ \dc\equiv dT+idH\,T -iTdJ\nn\\
dD&=& V^\dag \dd U^\dag \ ,\ \dd\equiv dQ+idJ\,Q -iQdH\ \ ,
\label{CDdiff}
\eea
where
\bea
dH&\equiv&+idU^\dag U\ = \ -iU^\dag dU \nn\\
dJ&\equiv&-iVdV^\dag \ =\ +idV\,V^\dag
\label{dHdJ}
\eea
Both $dH$ and $dJ$ are anti-self dual \cite{Mehta}. Their (upper N) 
diagonal elements 
are given by 
\bea
dH_{l,l}&\equiv& dh_{l,l}=
\left( \begin{array}{cc}
0 & ds_l\\
ds_l^* & 0
\end{array} \right),\ l=1,\ldots,N\nn\\
dJ_{l,l}&\equiv& dj_{l,l}=
\left( \begin{array}{cc}
0 & dp_l\\
dp_l^* & 0
\end{array} \right),\ l=1,\ldots,N\ .
\eea
where we have used the $2N$ parameters of the symmetry eq. (\ref{U'sym}) 
to set the real parameters on the diagonal to zero.
The remaining parameters on the upper and lower triangle are related due to
the anti-self duality, and we only keep the latter in the following set of
independent variables: 
\bea
&dH:& \ \ \{ dh_{l,l}, dH_{n,l}|n>l;\ n=1,\ldots N+\nu,\ l=1,\ldots N\}
\nn\\
&dJ:& \ \ \{ dj_{l,l}, dJ_{n,l}\ |n>l;\ n,l=1,\ldots N\} \ ,
\label{dJdHdof}
\eea
where in $dH$ we have also projected out the $\nu^2$-block. They label 
$4N(N-1)/2+4N\nu++2N=2N^2+4N\nu$ and
$4N(N-1)/2+2N=2N^2$ 
real variables
respectively, matching those of $U,V$ in eq. (\ref{UVdof}).

We can now arrange the independent variables $\{\dc,\dd\}$ and
$\{dT,dQ,dH,dJ\}$ in such a way that the Jacobian is block triangular. 
It is most efficient to use quaternion real matrix elements for this
ordering. In a second step we then have to compute the determinant of the
Jacobi matrix by taking the product of the determinants of the diagonal blocks
of quaternion elements. 
The variables $\{\dc,\dd\}$ are ordered row by row, 
starting with the lowest row of $\dc$, until we reach its quadratic part. We
then alternate  $\dc$ and $\dd$ element by element, 
completing the non-quadratic part of the row
with  $\dd$ only. In detail the ordering looks as follows:
\bea
&&\dc_{N+\nu,1}\ldots\dc_{N+\nu,N}\dc_{N+\nu-1,1} \ldots\dc_{N+\nu-1,N}
\ldots\dc_{N+1,1}\ldots\dc_{N+1,N}\nn\\
&&
\dc_{N,1}\dd_{N,1}\ldots\dc_{N,N}\dd_{N,N}\dd_{N,N+1}\ldots\dd_{N,N+\nu}\nn\\
&&
\dc_{N-1,1}\dd_{N-1,1}\ldots\dc_{N-1,N}\dd_{N-1,N}\dd_{N-1,N+1}\ldots
\dd_{N-1,N+\nu}\ldots\nn\\
&&\dc_{1,1}\dd_{1,1}\ldots\dc_{1,N}\dd_{1,N}\dd_{1,N+1}\ldots\dd_{1,N+\nu}\ .
\label{CDlabel}
\eea
The independent variables $\{dT,dQ,dH,dJ\}$ can also be put into two full
matrices without zeros, $dA$ and $dB$ of size  $(N+\nu)\times N$ and $N\times
(N+\nu)$ respectively, and then ordered in the same fashion as $\dc$ and
$\dd$. The matrix $dA$ consists in its strictly lower triangular part of $dH$. 
The diagonal part is composed of $dh_{l,l}$ and $dt_{l,l}$ giving full 
quaternion real matrix elements:
\be
d\hat{T}_{l,l}\equiv dh_{l,l}+dt_{l,l} = \left( \begin{array}{cc}
dx_l & ds_l\\
ds_l^* & dx_l^*
\end{array} \right), 
\label{dhTdef}
\ee
and its strictly upper triangular elements are completed by $dT$. 
Similarly we define a matrix $dB$ by putting $dJ$ in its strictly lower
triangular part, the sum of  $dj_{l,l}$ and $dq_{l,l}$ on its diagonal:
\be
d\hat{Q}_{l,l}\equiv dj_{l,l}+dq_{l,l} = \left( \begin{array}{cc}
dy_l & dp_l\\
dp_l^* & dy_l^*
\end{array} \right), 
\label{dhQdef}
\ee
and $dQ$ on its strictly upper triangular part. These matrices $dA$ and $dB$
are then ordered exactly as in eq. (\ref{CDlabel}), or most explicitly as:
\bea
&&dH_{N+\nu,1}\ldots dH_{N+\nu,N}dH_{N+\nu-1,1} \ldots dH_{N+\nu-1,N}
\ldots dH_{N+1,1}\ldots dH_{N+1,N}\nn\\
&&
dH_{N,1}dJ_{N,1}\ldots dH_{N,N-1}dJ_{N,N-1}d\hat{T}_{N,N}d\hat{Q}_{N,N}
dQ_{N,N+1}\ldots dQ_{N,N+\nu}\nn\\
&&
dH_{N-1,1}dJ_{N-1,1}\ldots dH_{N-1,N-2}dJ_{N-1,N-2}d\hat{T}_{N-1,N-1}
d\hat{Q}_{N-1,N-1}
dT_{N-1,N}dQ_{N-1,N}\ldots \nn\\
&&
dQ_{N-1,N+\nu}\ldots dH_{2,1}dJ_{2,1}d\hat{T}_{2,2}
d\hat{Q}_{2,2}\ldots dT_{2,N}dQ_{2,N}
dQ_{2,N+1}\ldots dQ_{2,N+\nu}\nn\\
&&d\hat{T}_{1,1}
d\hat{Q}_{1,1}\ldots dT_{1,N}dQ_{1,N}
dQ_{1,N+1}\ldots dQ_{1,N+\nu}\ .
\label{dHJTQlabel}
\eea
Next we write out explicitly the differentials in eq. (\ref{CDdiff}) 
in terms of all independent matrix elements of $\{dT,dQ,dH,dJ\}$:
\bea
\dc_{k,l}&=& dT_{k,l}+i\sum_{n=1}^{k-1}dH_{k,n}T_{n,l}+idh_{k,k}T_{k,l}
-i\sum_{n=k+1}^{N+\nu}\overline{dH_{n,k}}T_{n,l}\nn\\
&&+i\sum_{n=1}^{l-1}T_{k,n}\,\overline{dJ_{l,n}}-iT_{k,l}\,dj_{l,l}
-i\sum_{n=l+1}^{N}T_{k,n}\,dJ_{n,l}
\label{dcexpl}\\
\dd_{k,l}&=& dQ_{k,l}+i\sum_{n=1}^{k-1}dJ_{k,n}\,Q_{n,l}+idj_{k,k}\,Q_{k,l}
-i\sum_{n=k+1}^{N}\overline{dJ_{n,k}}\,Q_{n,l}\nn\\
&&+i\sum_{n=1}^{l-1}Q_{k,n}\,\overline{dH_{l,n}}-iQ_{k,l}\,dh_{l,l}
-i\sum_{n=l+1}^{N+\nu}Q_{k,n}\,dH_{n,l}\ .
\label{ddexpl}
\eea
For the matrix elements related through anti-self duality we have used the
relation $dH_{n,k}=-\overline{dH_{k,n}}$, where $\ \bar{}\ $
denotes the conjugate
quaternion \cite{Mehta}. Of course it holds that
$\overline{dH_{k,n}}/dH_{k,n}$ is 
non-vanishing. However, we will not need to compute these matrix elements as
they appear on the off-diagonal of the Jacobi matrix only. 
Let us distinguish 5 cases to show the block-triangularity of the Jacobi
matrix. The differentiation of quaternion elements by quaternion elements used
here is to be understood in a symbolical sense, in order to see the block 
structure
of the Jacobi matrix. In a second step we will differentiate independent
matrix element by independent matrix element.
\begin{itemize}
\item[i)]\underline{$k>l;\ k=N+\nu,\ldots,N+1,\ l=1,\ldots,N$:}

In this case we only have differentials $\dc_{k,l}$ simplifying to 
\be
\dc_{k,l}\ =\ i\sum_{n=1}^{l}dH_{k,n}T_{n,l}\ .
\label{i}
\ee
The variable $\dc_{k,l}$ only depends on $dH_{k,l}$: 
\be
\dc_{k,l}/dH_{k,l}\ =\ it_{l,l}\ ,
\label{iblock}
\ee
and not 
on variables to the right of $dH_{k,l}$ in the ordering eq. (\ref{dHJTQlabel}),
that is $dH_{k,n>l}$, $dH_{n<k,m}$, or any of the $dJ,dQ,dT$.  
\item[ii)]\underline{$k>l;\ k=N\ldots,2,\ l=1,\ldots,N$:}

We have both differentials $\dc_{k,l}$ an $\dd_{k,l}$ simplifying to 
\bea
\dc_{k,l} &=&
i\sum_{n=1}^{l}dH_{k,n}T_{n,l}-i\sum_{n=k}^{N}T_{k,n}dJ_{n,l}\ ,\nn\\
\dd_{k,l}&=& i\sum_{n=1}^{l}dJ_{k,n}\,Q_{n,l}
-i\sum_{n=k}^{N+\nu}Q_{k,n}\,dH_{n,l}\ .
\label{ii}
\eea
This gives the following $2\times2$ entry on the diagonal:
\be
\begin{tabular}{l|cc}
             & $dH_{k,l}$  & $dJ_{k,l}$  \\ \hline
$\dc_{k,l}$  & $it_{l,l}$  & $-it_{k,k}$ \\
$\dd_{k,l}$  & $-iq_{k,k}$ & $iq_{l,l}$  \\
\end{tabular}
\label{iiblock}
\ee
None of the
differentials depends on variables further to the right of $dH_{k,l}$
and $dJ_{k,l}$ in
eq. (\ref{dHJTQlabel}). 
\item[iii)]\underline{$k=l;\ k,l=1,\ldots,N$:}
\bea
\dc_{k,l} &=&dt_{k,k}+
i\sum_{n=1}^{k-1}dH_{k,n}T_{n,k}+idh_{k,k}t_{k,k}-it_{k,k}dj_{k,k}
-i\sum_{n=k+1}^{N}T_{k,n}dJ_{n,l}\ ,\nn\\
\dd_{k,l}&=& dq_{k,k}+i\sum_{n=1}^{k-1}dJ_{k,n}\,Q_{n,l}+idj_{k,k}q_{k,k}
-iq_{k,k}dh_{k,k}
-i\sum_{n=k+1}^{N+\nu}Q_{k,n}\,dH_{n,l}\ .
\label{iii}
\eea
The diagonal block entry reads:
\be
\begin{tabular}{l|crrc}
             & $\{dt_{k,k}\ ,$& $dh_{k,k}\}$ & $\{dj_{k,k}\ ,$  & $dq_{k,k}\}$ 
\\ \hline
$\dc_{k,l}$  & 1              & $it_{k,k}$   & $-it_{k,k}$   & 0   \\
$\dd_{k,l}$  & 0              & $-iq_{k,k}$  & $iq_{k,k}$    & 1   \\
\end{tabular}\ ,
\label{iiiblock}
\ee
with no dependence on variables further to the right of $d\hat{T}_{k,k}$ and 
$d\hat{Q}_{k,k}$.
\item[iv)]\underline{$k<l;\ k,l=1,\ldots,N$:}
\bea
\dc_{k,l}&=& dT_{k,l}+i\sum_{n=1}^{k-1}dH_{k,n}T_{n,l}+idh_{k,k}T_{k,l}
-i\sum_{n=k+1}^{l}\overline{dH_{n,k}}T_{n,l}\nn\\
&&+i\sum_{n=k}^{l-1}T_{k,n}\,\overline{dJ_{l,n}}-iT_{k,l}\,dj_{l,l}
-i\sum_{n=l+1}^{N}T_{k,n}\,dJ_{n,l}\ ,
\\
\dd_{k,l}&=& dQ_{k,l}+i\sum_{n=1}^{k-1}dJ_{k,n}\,Q_{n,l}+idj_{k,k}\,Q_{k,l}
-i\sum_{n=k+1}^{l}\overline{dJ_{n,k}}\,Q_{n,l}\nn\\
&&+i\sum_{n=k}^{l-1}Q_{k,n}\,\overline{dH_{l,n}}-iQ_{k,l}\,dh_{l,l}
-i\sum_{n=l+1}^{N+\nu}Q_{k,n}\,dH_{n,l}\ ,
\eea
with little simplifications. We have 
\be
\dc_{k,l}/  dT_{k,l}\ =\ 1\ ,\ \ \dd_{k,l}/dQ_{k,l}\ =\ 1\ ,
\label{ivblock}
\ee
with no dependence on variables further to the right of $dT_{k,l}$ and 
$dQ_{k,l}$.
\item[v)]\underline{$k<l;\ k=1,\ldots,N,\ l=N+1,\ldots,N+\nu$:}
\bea
\dd_{k,l}&=& dQ_{k,l}+i\sum_{n=1}^{k-1}dJ_{k,n}\,Q_{n,l}+idj_{k,k}\,Q_{k,l}
-i\sum_{n=k+1}^{N}\overline{dJ_{n,k}}\,Q_{n,l}\nn\\
&&+i\sum_{n=k}^{l-1}Q_{k,n}\,\overline{dH_{l,n}}-iQ_{k,l}\,dh_{l,l}
-i\sum_{n=l+1}^{N+\nu}Q_{k,n}\,dH_{n,l}\ ,
\eea
with 
\be
\dd_{k,l}/dQ_{k,l}\ =\ 1\ ,
\label{vblock}
\ee
and no dependence on variables to the right of $dQ_{k,l}$.
\end{itemize}
This proves that the Jacobi matrix in the given orderings eqs. (\ref{CDlabel})
and (\ref{dHJTQlabel}) is indeed block-triangular. 
It remains to compute its determinant in terms of the $1\times1$ and
$2\times2$ quaternion blocks on the diagonal.

In the following we collect all the contributions from the diagonal blocks. 
In case i) we parametrise for each matrix element the contributing part in
eq. (\ref{i})
\bea
dH_{k,l}\equiv \left( \begin{array}{cc}
r& s\\
t& u
\end{array} \right)\Rightarrow\  \dc_{k,l}\equiv \left( \begin{array}{cc}
a& b\\
c& d
\end{array} \right)= i\left( \begin{array}{cc}
rx_l& sx_l^*\\
tx_l& ux_l^*
\end{array} \right)+\ \ldots\ ,
\eea
leading to 
\be
\det\left(\frac{\partial(a,b,c,d)}{\partial(r,s,t,u)}\right)=\
\det(\mbox{diag}(ix_l,ix_l^*,ix_l,ix_l^*))\ =\ |x_l|^4\ .
\ee
The contribution to the Jacobian from case i) for all matrix elements is thus
\be
\prod_{l=1}^N\prod_{k=N+1}^{N+\nu}|x_l|^4\ =\ \prod_{l=1}^N|x_l|^{4\nu}\ .
\label{Jaci}
\ee
In case ii) we parametrise in addition 
$\dd_{k,l}$ and $dH_{k,l}$ as above with primed
variables. We thus obtain for the contributing terms in eq. (\ref{ii})
\bea
 \dc_{k,l}&=& \left( \begin{array}{cc}
a& b\\
c& d
\end{array} \right)\ \ = i\left( \begin{array}{cc}
rx_l& sx_l^*\\
tx_l& ux_l^*
\end{array} \right)- i\left( \begin{array}{cc}
x_kr'& x_ks'\\
x_k^*t'& x_k^*u'
\end{array} \right)+\ \ldots\ ,\nn\\
 \dd_{k,l}&\equiv& \left( \begin{array}{cc}
a'& b'\\
c'& d'
\end{array} \right)= i\left( \begin{array}{cc}
r'y_l& s'y_l^*\\
t'y_l& u'y_l^*
\end{array} \right)- i\left( \begin{array}{cc}
y_kr& y_ks\\
y_k^*t& y_k^*u
\end{array} \right)+\ \ldots\ .
\eea
This leads to the following $8\times8$ block where we only display
non-vanishing elements
\bea
\det\left(\frac{\partial(a,b,c,d,a',b',c',d')}{\partial(r,s,t,u,r',s',t',u')}
\right)&=&
\det\left( \begin{array}{cccc|cccc}
 ix_l&       &       &       &-iy_k&      &       &       \\
     & ix_l^*&       &       &     &-iy_k &       &       \\
     &       & ix_l  &       &     &      &-iy_k^*&       \\
     &       &       & ix_l^*&     &      &       &-iy_k^*\\\hline
-ix_k&       &       &       & iy_l&      &       &       \\
     &-ix_k  &       &       &     &iy_l^*&       &       \\
     &       &-ix_k^*&       &     &      &iy_l   &       \\
     &       &       &-ix_k^*&     &      &       &iy_l^* \\
\end{array} \right)\nn\\
&=&|z_l^2-z_k^2|^2|z_l^{*\,2}-z_k^2|^2\ .
\eea
Here we have used $x_ky_k=-z_k^2$ and the following 
formula for determinants of
matrices $A',B',C',D'$: $\det\left( \begin{array}{cc}
A'& B'\\
C'& D'
\end{array} \right)=\det(A')\det(D'-C'A'^{-1}B')$.
The contribution to the Jacobian from case ii) is then given by
\be
\prod_{k>l=1}^N |z_l^2-z_k^2|^2|z_l^{*\,2}-z_k^2|^2.
\label{Jacii}
\ee
In case iii) we obtain from eq. (\ref{iii}) for the contributing terms 
\bea
 \dc_{k,l}&=& \left( \begin{array}{cc}
a& b\\
c& d
\end{array} \right)\ \ = \left( \begin{array}{cc}
dx_k                    & ids_kx_k^*-ix_kdp_k\\
ids_k^*x_k-ix_k^*dp_k^* & dx_k^*             \\
\end{array} \right) +\ \ldots\ ,\nn\\
 \dd_{k,l}&=& \left( \begin{array}{cc}
a'& b'\\
c'& d'
\end{array} \right)\,= 
\left( \begin{array}{cc}
dy_k                    & idp_ky_k^*-iy_kds_k\\
idp_k^*y_k-iy_k^*ds_k^* & dy_k^*             \\
\end{array} \right)\  +\ \ldots\ ,
\eea
leading to the matrix
\bea
\det\left(\frac{\partial(a,d,b,c,a',d',b',c')}
{\partial(dx_k,dx_k^*,ds_k,ds_k^*,dy_k,dy_k^*,dp_k,dp_k^*)} 
\right)&=&
\det\left( \begin{array}{cccc|cccc}
 1   &       &       &       &     &      &       &       \\
     & 1     &       &       &     &      &       &       \\
     &       & ix_k^*&       &     &      &-iy_k  &       \\
     &       &       & ix_k  &     &      &       &-iy_k^*\\\hline
     &       &       &       & 1   &      &       &       \\
     &       &       &       &     & 1    &       &       \\
     &       &-ix_k  &       &     &      &iy_k^* &       \\
     &       &       &-ix_k^*&     &      &       &iy_k   \\
\end{array} \right)\nn\\
&=&|z_k^{*\,2}-z_k^2|^2\ .
\eea
This result is obtained similarly to the previous case, with the contribution 
to the Jacobian now reading
\be
\prod_{k=1}^N|z_k^{*\,2}-z_k^2|^2.
\label{Jaciii}
\ee
It is easy to see that the remaining cases iv) and v) simply give a
contribution unity. Collecting eqs. (\ref{Jaci}), (\ref{Jacii}) 
and (\ref{Jaciii}) we obtain as a final result the conjectured Jacobian in
eq. (\ref{Jacobi}).

\end{appendix}


\end{document}